\documentclass[journal]{IEEEtran}
\usepackage[pdftex]{graphicx}
\usepackage[cmex10]{amsmath}
\usepackage{amssymb}

\usepackage{caption}
\usepackage[hidelinks]{hyperref}
\usepackage{subcaption}
\usepackage{cite}
\usepackage{psfrag}
\usepackage{epstopdf}
\usepackage{algorithm}
\usepackage{algorithmic}
\usepackage{widetext}
\usepackage{float}
\usepackage{color}
\usepackage{soul}
\usepackage{wrapfig}
\usepackage[normalem]{ulem}
\usepackage{upgreek}
\usepackage{multirow}
\usepackage[table,xcdraw]{xcolor}
\usepackage[justification=centerlast]{caption}
\usepackage{tikz,pgfplots}
\usepackage{dblfloatfix}
\usepackage{adjustbox}
\DeclareUnicodeCharacter{2005}{\hspace{0.01em}}
\usetikzlibrary{calc}
\usepackage{mathtools}
\usetikzlibrary{arrows}
\pgfplotsset{compat=newest}
\usepgfplotslibrary{fillbetween}
\usepackage{xcolor}
\usepackage{pgfplots}
\usetikzlibrary{patterns}

\definecolor{myBlue}{RGB}{72,125,215}
\definecolor{myOrange}{RGB}{118,54,45}
\definecolor{InfinBlue}{RGB}{72,72,51}
\usepackage{xcolor}
\ifCLASSINFOpdf
\else
\fi

\begin{document}

\title{Reducing Computational Complexity of Neural Networks in Optical Channel Equalization:\\ From Concepts to Implementation}

\author{Pedro J. Freire, Antonio Napoli, Diego Arg\"uello Ron,  Bernhard Spinnler, Michael Anderson, Wolfgang Schairer, Thomas Bex, Nelson Costa, Sergei K. Turitsyn, Jaroslaw E. Prilepsky
\thanks{This paper was supported by the EU  Horizon 2020 program under the Marie Sklodowska-Curie grant agreements 813144 (REAL-NET) and 860360 (POST-DIGITAL).  JEP is supported by Leverhulme Trust, Grant No. RP-2018-063. SKT acknowledges support of the EPSRC project TRANSNET. 
}
\thanks{Pedro J. Freire, Diego Arg\"uello Ron, Michael Anderson, Sergei K. Turitsyn and Jaroslaw E. Prilepsky are with Aston Institute of Photonic Technologies, Aston University, United Kingdom, p.freiredecarvalhosouza@aston.ac.uk.}
\thanks{Antonio Napoli, Wolfgang Schairer, and Bernhard Spinnler are with Infinera R\&D, Sankt-Martin-Str. 76, 81541, Munich, Germany. anapoli@infinera.com}
\thanks{Nelson Costa is with Infinera Unipessoal, Lda, Rua da Garagem nº1, 2790-078 Carnaxide, Portugal, ncosta@infinera.com.}
\thanks{Manuscript received xxx 19, zzz; revised January 11, yyy.}}

\maketitle
\begin{abstract}
In this paper, a new methodology is proposed that allows for the low-complexity development of neural network (NN) based equalizers for the mitigation of impairments in high-speed coherent optical transmission systems. In this work, we provide a comprehensive description and comparison of various deep model compression approaches that have been applied to feed-forward and recurrent NN designs. Additionally, we evaluate the influence these strategies have on the performance of each NN equalizer. Quantization, weight clustering, pruning, and other cutting-edge strategies for model compression are taken into consideration. In this work, we propose and evaluate a Bayesian optimization-assisted compression, in which the hyperparameters of the compression are chosen to simultaneously reduce complexity and improve performance.
Next, this paper presents four distinct metrics (RMpS, BoP, NABS, and NLGs) that are discussed here that can be used to evaluate the amount of computing complexity required by various compression algorithms. These measurements can serve as a benchmark for evaluating the relative effectiveness of various NN equalizers when compression approaches are used. In conclusion, the trade-off between the complexity of each compression approach and its performance is evaluated by utilizing both simulated and experimental data in order to complete the analysis. 
By utilizing optimal compression approaches, we show that it is possible to design an NN-based equalizer that is simpler to implement and has better performance than the conventional digital back-propagation (DBP) equalizer with only one step per span. This is accomplished by reducing the number of multipliers used in the NN equalizer after applying the weighted clustering and pruning algorithms. Furthermore, we demonstrate that an equalizer based on NN can also achieve superior performance while still maintaining the same degree of complexity as the full electronic chromatic dispersion compensation block.  We conclude our analysis by highlighting open questions and existing challenges, as well as possible future research directions.
\end{abstract}


\IEEEpeerreviewmaketitle

\section{Introduction}\label{sec:Introduction}

To achieve satisfactory optical performance in modern high-speed optical transmission systems, the detrimental impact of linear and, most importantly, nonlinear transmission impairments that cap the systems’ throughput \cite{akc16,winzer2018fiber}, has to be mitigated. Several digital signal processing (DSP) algorithms specifically addressing optical fiber channel nonlinearity mitigation have already been proposed \cite{Cartledge:17}. However, the ``conventional'' equalizers/soft-demappers, which are mostly based on deterministic algorithms, have recently started to lose their attractiveness in favor of designs incorporating machine learning (ML) techniques\cite{musumeci2018overview,hager2018nonlinear,hager2020physics,freire2021performance,freire2022neural,nevin2021machine,deligiannidis2020compensation,deligiannidis2021performance,sidelnikov2018equalization}. In the meantime, the possibility of using neural networks (NNs) in digital communication systems was already discussed over 20 years ago \cite{ibnkahla2000applications}. In general, various ML-based approaches and, more specifically, deep artificial NNs, are rapidly finding their way into the telecommunication sector. This is mainly due to NNs' being universal approximators with virtually unlimited approximation capabilities\footnote{The Universal Approximation Theorem \cite{HORNIK1989359} states that no matter what the function is (with some fairly relaxed constraints on the function properties), there exists a feed-forward NN that can approximate that function to any desired degree of accuracy; a similar statement can be proven for recurrent NNs~\cite{schafer2007recurrent}.}. Thus, NNs can successfully reverse the channel propagation function and, thereby, efficiently mitigate transmission- and devices-induced impairments. Also, data science-related approaches can flourish in optical communication applications since large datasets can be obtained in a short period of time, which makes the (typically) data-hungry learning process easier. However, despite several recognized advantages and benefits of ML and, particularly, NNs in optical transmission equalization, there are still many challenges that can seriously hinder their success. One major challenge is the typically high computational complexity of NN-based algorithms, resulting in prohibitively strong requirements on the speed and energy consumption of end devices performing the equalization (although a lot of ``traditional'' approaches, like digital back-propagation, are also deemed too complex). 

It was demonstrated in~\cite{freire2021performance} that, when the NN equalizer's complexity is not constrained, combining a convolutional layer with a bidirectional long-short term memory (biLSTM) layer yields the best performance (among several NN structures studied). This is a consequence of the optical fiber channel's involving significant memory-related effects, primarily due to chromatic dispersion (but optical line components can introduce memory as well), but the recurrent NN models (to which the LSTM belongs) are apt for efficient memory handling \cite{Goodfellow-et-al-2016}. It was also shown in ~\cite{freire2021performance} that reducing the overall computational complexity by limiting the number of neurons, hidden units, filters, etc., of an NN may lead to significantly worse optical performance. This can be attributed to the infamous underfitting phenomena, i.e., the case when the reduced-structure NN loses the capacity to reverse the fairly complicated channel propagation function\cite{jabbar2015methods}. To address this performance-complexity trade-off, two well-known approaches can generally be considered. First, we can modify the original NN equalizer architecture, which recovers just one symbol at a time from a multisymbol input, so that multiple symbols can be recovered at a time~\cite{deligiannidis2021performance,sang2022low}. This may be achieved by using multidimensional regression predictive modeling (or a multidimensional classification when a soft demapper is coupled to the NN equalizing structure \cite{sang2022low}). In the case when the resulting multi-output NN architecture is similar to the original one (that recovered just one symbol at a time), the overall complexity per recovered symbol is reduced. This is the first method incorporated into our approach here. 
Second, we can use sophisticated NN model compression techniques to reduce the number of multiplications and, afterward, diminish the hardware complexity by allowing low bitwidth precision on the NN arithmetic operations. In this work, we describe how to design a NN equalizer based on the use of the aforementioned strategies, combining the multidimensional regression approach with advanced model compression techniques, namely pruning, weight clustering, and quantization. It is shown that the resulting NN-based equalizer is less complex than a standard (deterministic) and non-optimized digital back-propagation (DBP) equalizer with just 1 step-per-span (STpS). Furthermore, NN-based equalizers can achieve better optical performance than multi-step DBP-based ones with similar complexity. In this work, we:

\begin{itemize}
    \item Compare various pruning approaches that use \textit{recurrent} layers. Our results are then used to prune the optical channel equalizer model. We are not aware of such a comparison being done even in ML literature.
    \item Enhance existing compression strategies by utilizing Bayesian optimization (BO). BO enables improving optical performance while also reducing computational complexity.
    \item Investigate the potential of weight clustering to reduce the NN model's complexity (studied for the specific case of optical channel equalization), and calculate the achieved reduction in the number of multiplications in the equalizer model.
    \item Compare quantization strategies in the context of optical channel equalization (using recurrent layers).
    \item Provide four metrics for evaluating the computational complexity and explain when each one is adequate for carrying out the models' comparative analysis. 
\end{itemize}

This paper is organized as follows. Sec.~\ref{sec:nonlinearities} introduces the physical layer problem that we aim to mitigate using a post-equalizer. In Sec.~\ref{sec:NN_Design}, we describe the steps used to design the combined biLSTM+CNN equalizer that recovers multiple symbols following a multidimensional complex-valued regression predictive modeling approach configuration. Sec.~\ref{sec:Compression} presents the compression techniques that we address in this work and describes how we can use the BO method to optimize the trade-off between complexity and performance. Sec.~\ref{sec:NN} describes the experimental and simulated setup used. It also includes a description of the considered computational complexity metrics and explains how to compute them when compression techniques are used. Sec.~\ref{sec:Results} contains the main results, including the comparison between optical performance and computational complexity achieved when employing the different proposed strategies to reverse the channel propagation function. Our findings are described in the conclusions, which also include a discussion of open problems, challenges, and research opportunities.

\section{The Nonlinearity Problem in Optical Fiber Communications}\label{sec:nonlinearities}
\subsection{Propagation of Light in the Fiber}

The fundamental equation used to describe the propagation of light along an optical fiber is commonly referred to as the nonlinear Schr\"{o}dinger equation (NLSE)~\cite{agrawal21} and can be derived directly from the Maxwell equations, which describe the foundations of electricity and magnetism\cite{kodama1985optical}. The NLSE reads as:
\begin{equation}\label{eq:NLSE}
    \frac{\partial E}{\partial z} = (\hat{L} + \hat{N}) E,
\end{equation}
where $E$ is the electrical field as a function of the propagation distance $z$ and time $t$. $\hat{D}$ and $\hat{N}$, describe the linear and nonlinear parts of the NLSE, which are given by:
\begin{align}\label{eq:lin_nl_terms}
    \hat{L} &= \underbrace{-\frac{\alpha}{2}}_{\rm loss} - \underbrace{\frac{j\beta_2}{2}\frac{\partial^2}{\partial t^2}}_{\rm GVD} + \underbrace{\frac{j\beta_3}{6}\frac{\partial^3}{\partial t^3}}_{\textrm{GVD slope}}, \\ \nonumber
    \hat{N} &= \underbrace{j\gamma |E|^2}_{\textrm{Kerr effect}},
\end{align}
where $\alpha$, $\beta_{2,3}$, and $\gamma$ are the attenuation, the group velocity dispersion (GVD), and the nonlinear coefficient, respectively. If we substitute $\hat{L}$ and $\hat{N}$ from Eq.~(\ref{eq:lin_nl_terms}) into Eq.~(\ref{eq:NLSE}), it provides the explicit form of the NLSE:
\begin{equation}\label{eq:NLSEcomplete}
    \frac{\partial E}{\partial z} + \frac{\alpha}{2}E + \frac{j \beta_2}{2} \frac{\partial^2 E}{\partial t^2} - \frac{j\beta_3}{6}\frac{\partial^3 E}{\partial t^3} = j \gamma |E|^2 E .
\end{equation}
Eq.~(\ref{eq:NLSEcomplete}) is suitable to model optical fiber transmission when transmission along a single-polarization only is explored, e.g., intensity-modulation with direct-detection systems~\cite{agrawal21}. 
However, a coherent transceiver employs advanced digital signal processing (DSP) which enables detecting a dual-polarization signal, thus doubling the spectral efficiency of the system. In this context, the linear and non-linear interactions between the two signal polarizations must be taken into account. Consequently, the NLSE of Eq.~(\ref{eq:NLSEcomplete}) is extended in a vectorized form:
\begin{eqnarray}\label{eq:Manakov}
 \frac{\partial E_X}{\partial z} &=& \underbrace{-\frac{\alpha}{2}E_X + \frac{j \beta_2}{2} \frac{\partial^2}{\partial
t^2} E_X - \frac{j\beta_3}{6}\frac{\partial^3}{\partial t^3}E_X}_{\textrm{linear part}}\nonumber\\
				 & &  \underbrace{-j\gamma \frac{8}{9} \left(|E_X|^2 + |E_Y|^2
\right)E_X}_{\textrm{nonlinear part}},\nonumber\\
 \frac{\partial E_Y}{\partial z} & = & -\frac{\alpha}{2}E_Y + \frac{j \beta_2}{2} \frac{\partial^2}{\partial t^2}E_Y - \frac{j\beta_3}{6}\frac{\partial^3}{\partial t^3} E_Y \nonumber\\
				 &&  -j\gamma \frac{8}{9} \left(|E_X|^2 + |E_Y|^2 \right)E_Y.
\end{eqnarray}
This pair of equations is commonly referred to as ``the Manakov equation'', and it involves both polarization states. Here, $E_X$ and $E_Y$ represent the two orthogonal polarization components of the electric field $E$. In addition to the two polarizations, Eq.~\ref{eq:Manakov} properly averages the impact of residual birefringence that leads to fast polarization changes. Since the polarization state of the electric field changes rapidly, the resulting nonlinearities do not correspond to the ones from a linearly or circularly polarized field but to an average over the entire Poincar$\acute{e}$ sphere. The previous equations do not take into account, for example, stimulated Raman scattering (SRS). The SRS is a nonlinear effect that leads to the depletion of power from short to long wavelengths, achieving its maximum efficiency when the signals are separated by $\sim$100~nm. The Raman effect has mainly been explored to design distributed Raman amplifiers. Indeed, the SRS impact is usually negligible in C-band only systems, which occupy $\sim$35~nm.  However, with the advent of ultra-wideband optical systems, SRS will become the main transmission impairment in optical networks~\cite{ferrari2020assessment}.
\subsection{Channel Capacity Limitations caused by Nonlinear Kerr Effect}
The non-linear part of Eq.~(\ref{eq:Manakov}) imposes a severe limitation on the maximum achievable throughput in an optical fiber. In fact, the information theory indicates that the capacity of a linear channel increases monotonically by raising the transmitted signal power (or rather signal-to-noise ratio, SNR)\cite{shannon1948mathematical}. This theoretical limit is also commonly referred to as Shannon's limit. However, in fiber optics, this tendency does not hold because the term $\left(|E_{X}|^2 + |E_Y|^2 \right)E_{X,Y}$  becomes progressively more important as the transmitted signal power increases, thus causing phase distortions that limit the maximum throughput in the network\cite{essiambre2010capacity}. Consequently, there is an optimal optical signal power that balances the achievable maximum SNR and the signal distortion induced by the optical fiber's nonlinear behavior. 

These peculiar aspects of fiber propagation have been widely investigated, together with mitigation techniques, in both the optical and digital domains. The next subsection provides a brief overview of some studies carried out to mitigate the nonlinear Kerr effect in the digital domain. Nevertheless, a more complete review can be found in, e.g., Ref.~\cite{Cartledge:17}.

\subsection{Mitigation of Fiber Propagation Effects}
Eq.~(\ref{eq:Manakov}) is a multi-domain differential equation that does not have a closed-form solution. A possible way to solve it is to apply the ``Split-step Fourier method'' (SSFM). This method assumes that the linear ($\hat{L}$) and Kerr nonlinear ($\hat{N}$) effects can be separated and solved independently when a propagation step-size small enough is considered, alternating between them along the optical fiber. A more detailed description of this approach can be found in Refs.~\cite{sinkin2003optimization, millar2010mitigation}.
The absence of an analytical solution for the Manakov equations makes the perfect compensation of transmission effects very difficult. Additionally, and as an example, the loss of the phase information severally limits the compensation of transmission effects in direct-detection-based receivers (RXs). However, thanks to coherent detection, the amplitude and phase of the transmitted signal can be simultaneously detected at the RX input, which enables applying enhanced DSP algorithms to at least partially compensate for transmission effects.
Indeed, the linear effects, such as GVD and polarization mode dispersion (PMD), can be fully compensated for in the electronic domain by using a frequency domain equalizer in conjunction with a multiple-input multiple-output (MIMO) equalizer.
On the other hand, the compensation of the Kerr nonlinear effects that induce a self- and cross-phase modulation (SPM and XPM, respectively) on the transmitted signal is much more difficult.

The full compensation of the Kerr effect is troublesome as the equalizer would require complete knowledge of the propagation channel itself (for the SPM compensation), of the neighboring channels (for the XPM compensation), and of the amplified spontaneous emission (ASE) noise (intertwining with both SPM and XPM). Nevertheless, several methods have been proposed to digitally mitigate nonlinearities. Among them, the most relevant ones that are worth to be explicitly mentioned and described are: 1) maximum likelihood sequence estimation (MLSE); 2) Volterra-series based equalizers; 3) DBP; 4) NN-based techniques (we provide some respective references below).

 MLSE is the optimal method as long as there is no limitation on the number of states of the trellis code, as shown by~\cite{agazzi2004impact} for coherent- and by~\cite{savory2007imdd} for direct-detection systems. However, complying with this limitation means that it may become too complex and its potential commercial application ended with 10~Gb/s systems~\cite{kupfer2008measurement}, where it has been mainly used to compensate for GVD. At current high symbol rates, it seems unrealistic to implement a sufficiently low-power-consumption MLSE equalizer. 

Volterra equalizers were proposed in the '70s for satellite communications~\cite{benedetto1979modeling}, and provide a nonlinear version of the widely used finite impulse response (FIR) filters. They are based on the mathematical technique developed by Vito Volterra, which is an extension of Taylor's series but for a general function. Volterra equalizers can result in significant improvements in transmission quality~\cite{guiomar2012mitigation,cho2022volterra} but, like in the case of MLSE, their complexity is too high for realistic implementation.

DBP\footnote{Not to be confused with the backpropagation through the NN layers used for the training of NNs.} gained momentum about a decade ago when the article by Ip and Kahn~\cite{ip2008compensation} was published. The main idea behind DBP is to extend the MIMO equalizer by adding a nonlinear part, so that DBP would invert the nonlinear and linear parts of Eq.~\ref{eq:Manakov} by applying the SSFM and solving the propagation equation (\ref{eq:Manakov}) backward at the RX. However, DBP is effective only when combined with coherent detection and is deemed as being relatively complex for realistic implementation. Several methods have been proposed to simplify the DBP concept~\cite{napoli2014reduced, zhu2012nonlinearity, rafique2011compensation}, but its complexity is still considered to be high. 

NNs are intrinsically nonlinear and, therefore, match well with the type of effects we want to mitigate. Moreover, NNs can still be employed even in the absence of link information or in cases where the system configuration has changed as they obtain the required information directly from the received signal. However, NNs can be quite complex, often even more complex than DBP\cite{freire2020complex,freire2021performance}. As this limitation is the most relevant blocking point for the implementation of ASICs, this work specifically addresses this paramount issue, covering several hardware simplification techniques.
 
\section{Low Complexity Neural Network Design}\label{sec:NN_Design}
As described in~\cite{freire2021performance}, the bidirectional LSTM equalizer in a configuration of many-to-one (1D regression task), i.e., when a window of symbols is used to recover just the central one, leads to a computational complexity in terms of real multiplication per recovered symbol (RMpS) given by:
\begin{equation}\label{Eq_LSTMfull}
\begin{split}
  C_{\text{biLSTM}}= 2
  n_{s} \big( \underbrace{4n_{h}n_i}_{\text{$a$}}+\underbrace{4n_{h}^2}_{\text{$b$}}+\underbrace{3n_{h}}_{\text{$c$}}+\underbrace{n_on_{h}}_{\text{$d$}} \big),
  \end{split}
\end{equation}
where $n_s$ is the size of the input sequence in the time-domain, $n_{i}$ is the number of input features, $n_o$ is the output dimension (which is equal to $2$  - the real and imaginary parts of the symbol), and $n_h$ is the number of hidden units in the LSTM cell. In Eq.~(\ref{Eq_LSTMfull}), the addend $a$ is attributed to matrix multiplication of input and weights; $b$ to matrix multiplication of hidden states and weights; $c$ to pointwise multiplications occurring internally within the LSTM cell; and $d$ to matrix multiplication of hidden states and output weights\footnote{Here, we consider that a flatten layer was applied to achieve a many-to-one configuration. Instead, if the output comes from just one cell, the complexity of the term $d$ instead of 2$n_s n_o n_h$, would read as: 2$n_h n_o$}. During this investigation of computational complexity reduction, we found that simply applying compression techniques would not be enough to reduce the complexity beyond DBP level, because such compression strategies reduce the multiplications between input and weights, as in $a$, $b$, and $d$, but do not impact internal multiplications as in $c$. As a result, the multiplication $n_s n_h$ would become the bottleneck to achieving a reduction of complexity. To mitigate this effect, we can follow two different strategies. As suggested in~\cite{deligiannidis2021performance}, we can utilize basic vanilla recurrent neural networks (RNNs) in our equalizers insofar as they lack an intrinsic point-wise multiplier and, therefore, do not suffer from this issue. The second possibility, again indicated in Ref.~\cite{deligiannidis2021performance}, is to recover several symbols at a time rather than just the central one, which allows for eliminating some of the $n_s$ multiplications. 

Firstly, we tried using the vanilla RNN and optimizing it using the BO. However, while comparable performance was shown in Ref.~\cite{deligiannidis2021performance} when using the LSTM and vanilla RNNs, we observed quite different performances when using these NNs (with both tuned using the BO). Indeed, and using the standard DBP as the reference comparison scenario, as is typically done in the literature, the vanilla RNN barely outperformed the 1 STpS DBP, while the LSTM-based architecture showed better performance than a 3 STpS DBP. In this case, the BO showed substantially low ($\approx 5.10^{-5}$) learning rates in the vanilla RNN scenario, in an attempt to reduce the impact of exploding gradients (that such layers are known to have). Consequently, the training process got stuck in the local minima of the loss function landscape, which limited the optical performance improvement attainable by the equalizer. The LSTM cell, an enriched variant of the vanilla RNN cell with several gating units that help propagate the gradient and govern the flow of information through the NN, solves this gradient problem. However, at the cost of additional complexity. 

\begin{figure}[ht!]
\centering\includegraphics[width=0.49\textwidth]{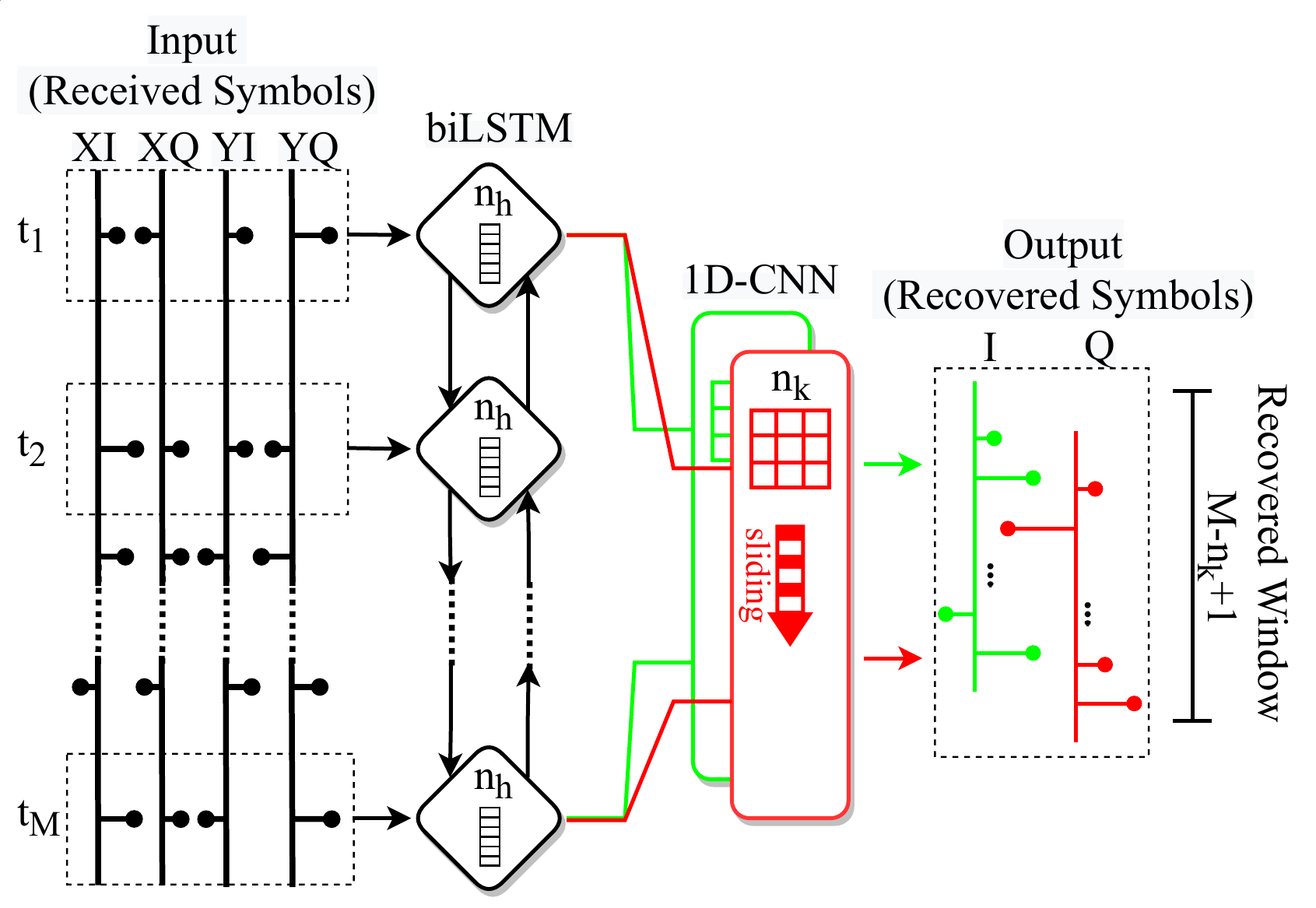}
\caption{Schematic of the biLSTM+CNN equalizer: the input consists of $M$ real (I) and imaginary (Q) parts of the symbols. The LSTM cells are indicated by lozenges containing $n_h$ hidden units. The LSTM output is sequentially processed by the convolutional layer with two filters to compute the I and Q components of the symbols.}
\label{fig:MODEL}
\end{figure}

\begin{figure*}[htbp]
  \centering
\begin{subfigure}{.244\textwidth}
  \centering
\includegraphics[width=\textwidth]{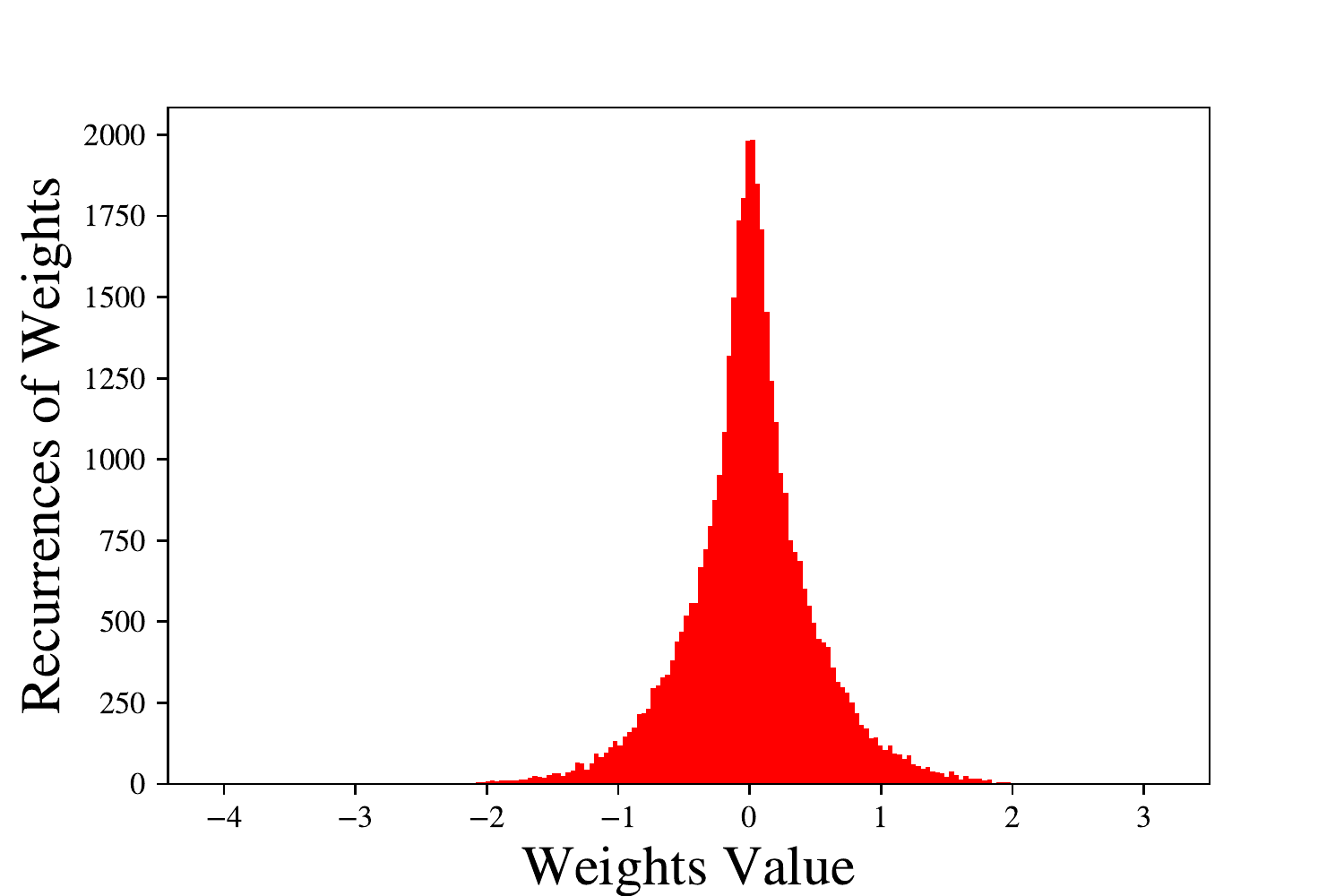}
\caption{Original.}  
\end{subfigure}
\hfill
\begin{subfigure}{.244\textwidth}
    \centering
\includegraphics[width=\textwidth]{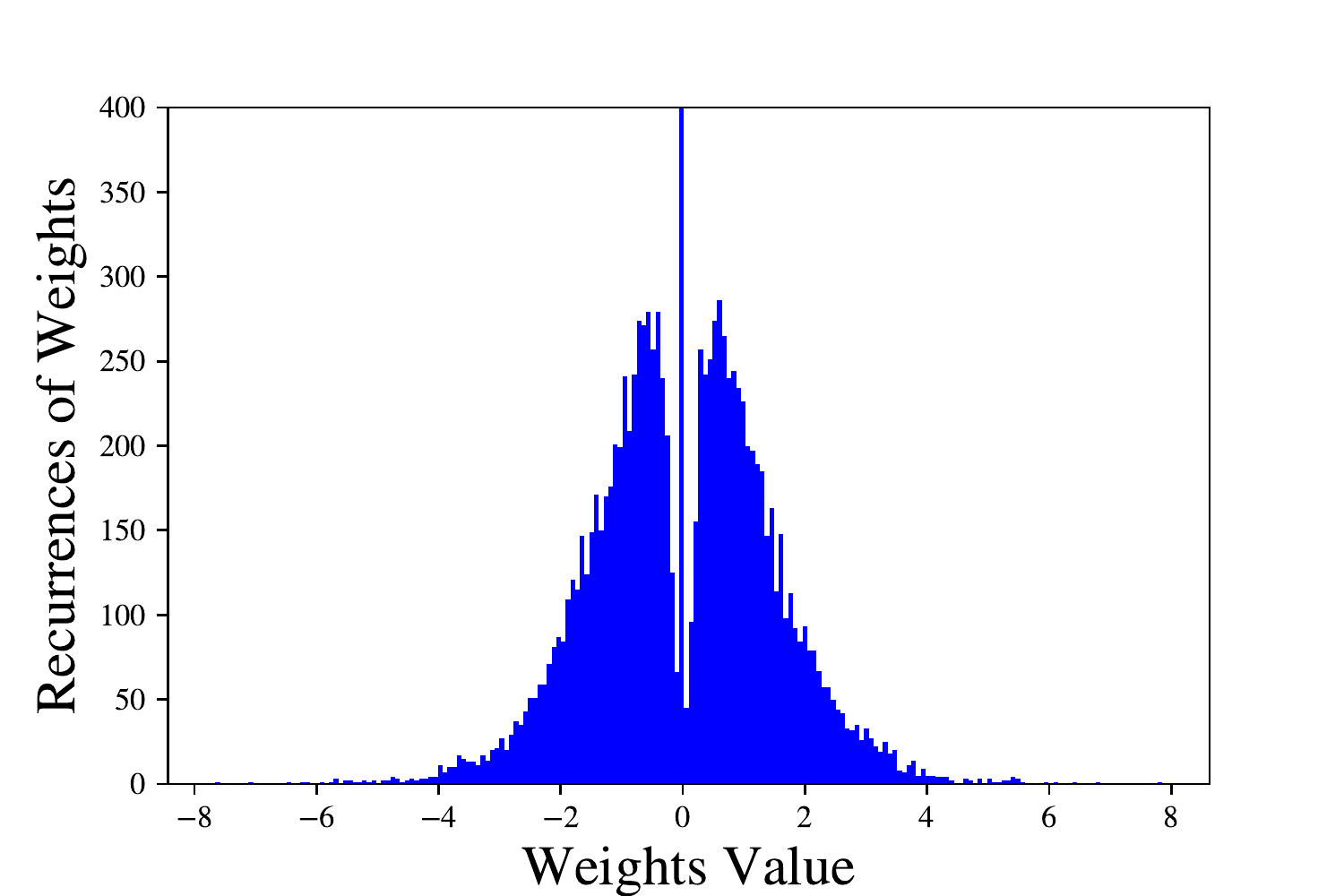}
\caption{After Pruning.}
\end{subfigure}
\hfill
\begin{subfigure}{.244\textwidth}
    \centering
\includegraphics[width=\textwidth]{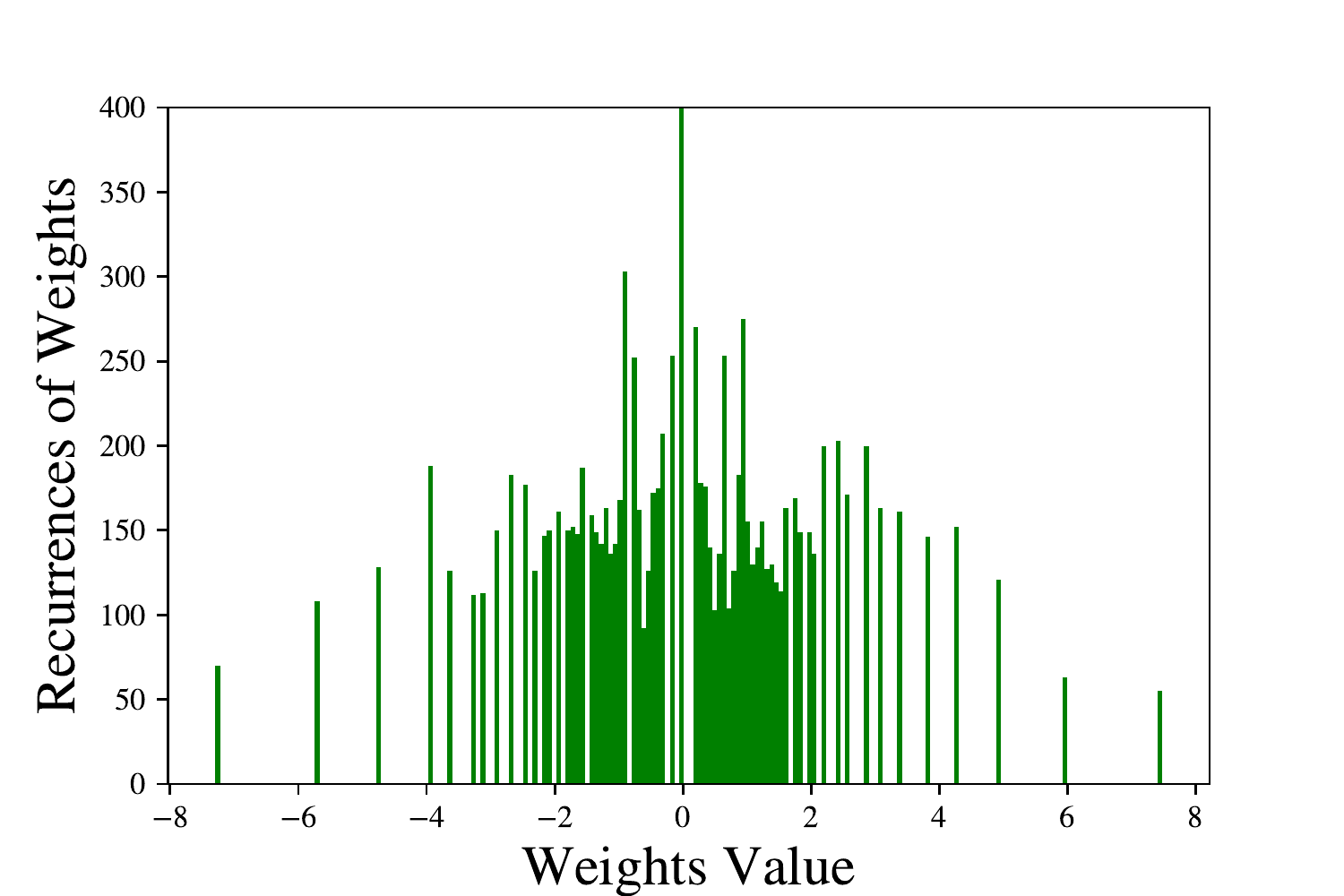}
\caption{After clustering.}
\end{subfigure}
\hfill
\begin{subfigure}{.244\textwidth}
    \centering
\includegraphics[width=\textwidth]{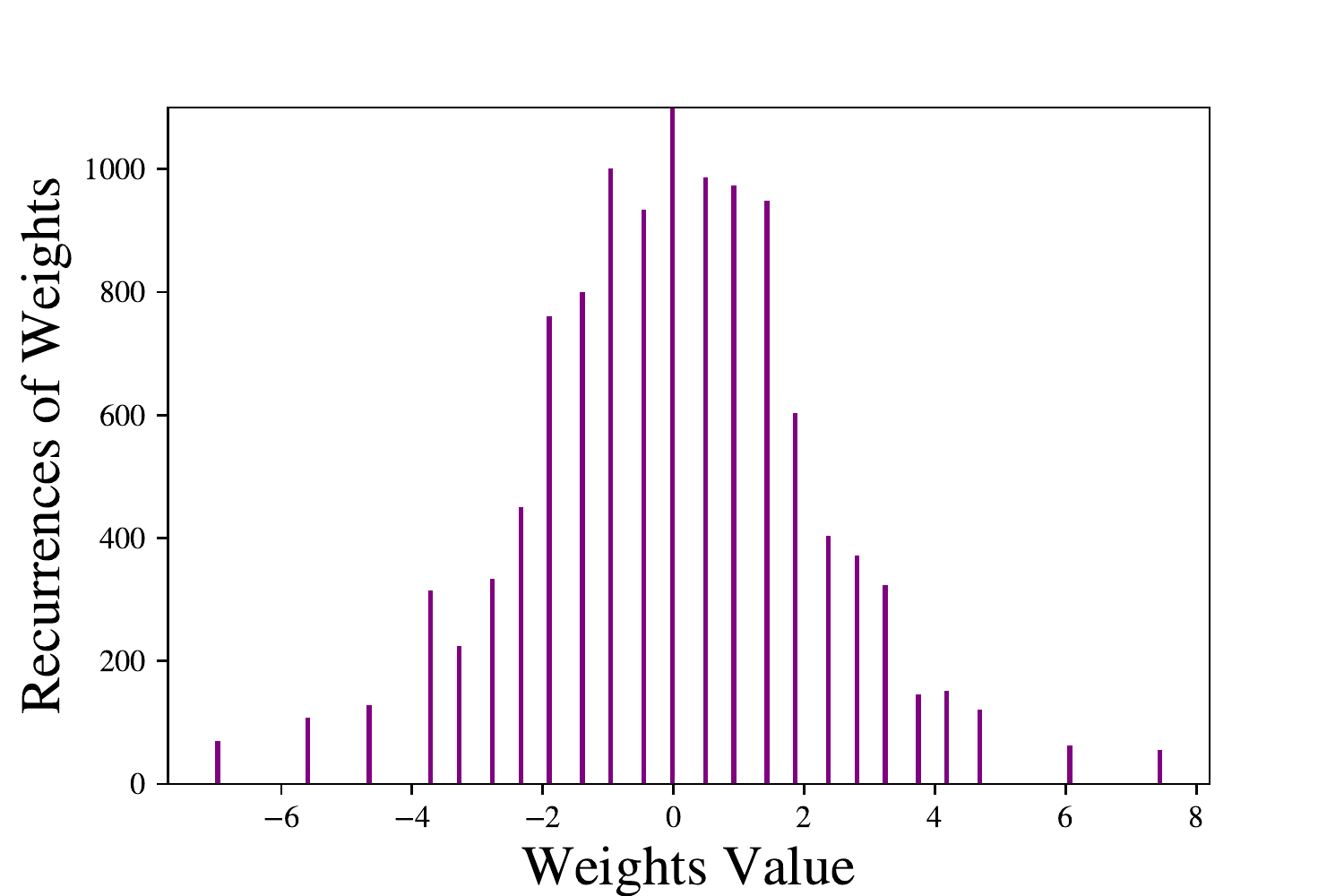}
\caption{After quantization.}
\end{subfigure}
\caption{Illustration of the weight distribution of the recurrent kernel in the LSTM layer of our NN equalizer from the different phases of the NN design in this paper. (a) Original trained NN; (b) The pruning phase; (c) The weight clustering phase; (d) The quantization phase. }
\label{fig:weightdistribuition}
\end{figure*}

The better performing LSTM equalizer is considered henceforth. However, we have enhanced it by recovering multiple symbols with the same NN structure instead, as proposed in Ref.~\cite{deligiannidis2021performance}. To recover multiple symbols, we need to consider that, since chromatic dispersion plays an important role in fiber perturbation, if the NN equalizer processes a window of $M$ symbols as input, we will be able to recover $\text{M} - x$ symbols only, where $M$ is the number of input symbols and $x$ depends on the system memory length. Since the initial and final symbols of the window will lack important information from their neighbors (due to dispersion-induced memory), they may not be recovered properly. The simplest way to reduce the dimensionality of the time window tensor without losing information is by using a 1D convolutional layer. For this purpose, we use a 1D-CNN with kernel size $n_k$, the padding set to zero, the dilation and stride to 1, and just two filters to represent the real and imaginary parts of each symbol. In this case, the number of recovered symbols (the recovery window) is $\text{M}-n_k+1$.  We have used the BO to estimate the appropriate values for $M$, $n_h$, $n_k$, learning rate, and mini-batch size, limiting the number of hidden units to at most 150 for complexity constraint reasons. The equalizer scheme is shown in Fig.~\ref{fig:MODEL}.

The computational complexity  of the bidirectional LSTM + 1D-CNN equalizer, in terms of RMpS, can be represented using the formulae in Ref.~\cite{freire2021performance}, but this time taking into consideration the parallel recovery of $n s-n k+1$ symbols as:

\begin{equation}
\label{Eq_original}
 \mathrm{RMpS}_{\text{NN}}= \frac{2
  n_{s}n_{h}(4n_i+4n_{h}+3)}{n_s-n_k+1} + 2n_h n_o n_k,
\end{equation}
The analysis of Eq.~(\ref{Eq_original}) shows that the number of multiplications has decreased when compared to that of the initial biLSTM equalizer, but compression techniques are still required to further reduce the number of multiplications to a level at least below 1 STpS DBP without affecting the resulting model's performance.

\section{Overview of Deep Compression Techniques}\label{sec:Compression}
Generally, the subject of deep NNs compression is vast~\cite{alqahtani2021literature,deng2020model}, and new compression methods emerge almost continuously. This section presents some chosen compression strategies that can be efficiently used to overcome the constraints limiting the real-time deployment of NNs. The strategies to reduce the high processing resources as well as to cut down on energy consumption will also be discussed. According to Ref.~\cite{liang2021pruning, 9043731}, the compression can often be accomplished with little loss of accuracy and, in some situations, the accuracy may even rise\cite{ron2022experimental}. Three methods of network compression are discussed below: pruning, weight clustering, and quantization. Fig.~\ref{fig:weightdistribuition} illustrates how the weight distribution of the NN changes after applying each of these compression techniques.

\subsection{Pruning}
Pruning is the process of removing parameters, neurons, or even layers or parts of a NN that do not significantly impact its performance to reduce its computational complexity. The area of NN pruning is wide and encompasses several subcategories: (a) static or dynamic; (b) one-shot or iterative; (c) structured or unstructured; (d) magnitude-based or information-based; (e) global or layer-wise. Detailed information on the different types of pruning can be found in, e.g., Refs.~\cite{blalock2020state,liang2021pruning,liu2018rethinking, augasta2013pruning, vadera2020methods,han2015deep}. In our work, we prune the lowest magnitude weights globally throughout the NN~\cite{blalock2020state}. This low complexity, traditional unstructured global magnitude pruning has already proven to be quite effective~\cite{frankle2018lottery,blalock2020state,tung2018deep,renda2020comparing,9043731}. To be more specific, we consider a static, iterative, unstructured, global magnitude-based pruning. In this case, we remove weights offline from the network after training and before inference. Moreover, iterative pruning allows us to prune more weights while preserving accuracy.

\begin{figure*}[htbp]
    \centering
    \includegraphics[width=\textwidth]{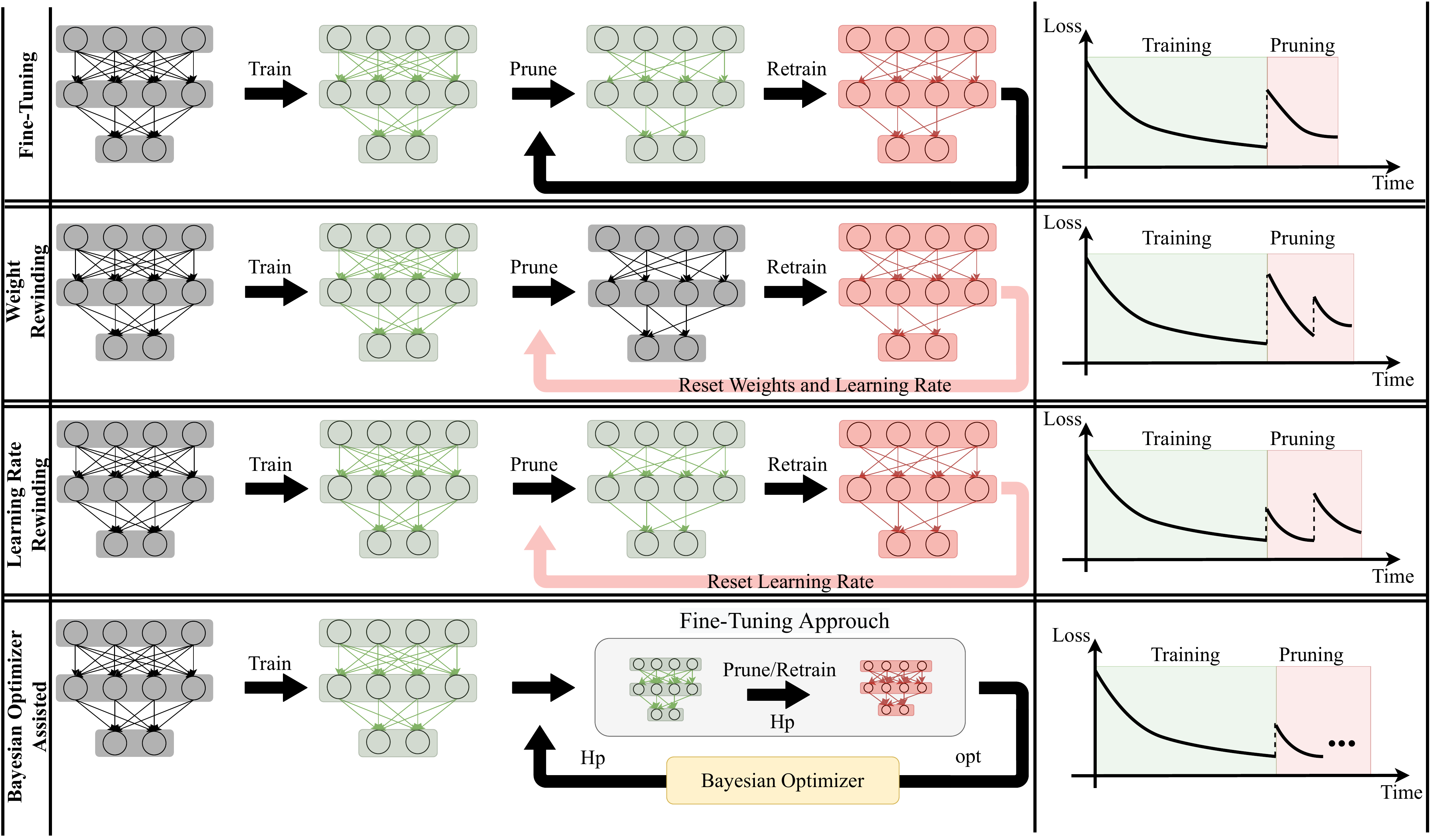}
    \caption{A schematic of the fine-tuning, weight rewinding, learning rate rewinding, and Bayesian optimizer-assisted pruning strategies is shown. A qualitative representation of the evaluation loss over the training time process is shown on the right-hand side. $H_p$ is the set of hyperparameters suggested by the BO for the pruning phase.}
    \label{fig:pruning_type}
\end{figure*}

The four (most promising) strategies for the iterative-pruning retraining process that are applied in our study are schematically depicted in Fig.~\ref{fig:pruning_type}. The four approaches are referred to as fine-tuning, weight rewinding, learning rate rewinding, and Bayesian optimizer assisted.
\subsubsection{Fine-Tuning approach} 
This method prunes the model once it has been trained. In a second step, it trains the weights that remain after pruning using a constant learning rate; the latter is usually the same as the final learning rate of the original training procedure. The first panel of Fig.~\ref{fig:pruning_type} shows how the fine-tuning scheme is implemented. After determining the fine-tuning period, we use the traditional gradual pruning method (a polynomial decay)~\cite{zhu2017prune}. The pruning polynomial decay approach quickly prunes the network at the beginning when there are many redundant connections, and gradually reduces the number of weights pruned each time. This procedure results in a smooth loss function during the fine-tuning period, which is beneficial for the learning process to maintain an accuracy close to the original NN model.  

This approach can be used when employing the other ``deterministic'' methods for the equalization of optical fiber nonlinearities. For example, this pruning technique can be used (in a simpler way) to eliminate the less relevant coefficients of the Volterra equalizers~\cite{chuang2019sparse,huang201893,6647643} and to trim the unimportant triplets (making the triplet feature vector more sparse). It can be used also when employing perturbation approaches~\cite{zhang2019field, melek2020nonlinearity, kumar2021deep}. Several papers investigated the use of fine-tuning, mainly in short-reach intensity-modulated systems \cite{li2021high, wan2018nonlinear, zhang2020compressed, wang2021low, ge2020compressed, reza2018nonlinear}, to reduce the complexity of the model. So far, the analysis of pruning in optical channel equalization has been restricted to the case of the feed-forward NN models only. In our work, we will also present such an analysis for a recurrent equalizer and deal with the case of coherent optical transmission.

\subsubsection{Weight rewinding approach}
This method was introduced in Ref.~\cite{frankle2018lottery} dealing with the lottery ticket hypothesis. The main idea supporting it is that a dense NN with random initialization contains a subnetwork that, when trained in isolation, can match the test precision of the original network after training. This approach is separated into three parts, as shown in the second line of Fig.~\ref{fig:pruning_type}. First, during the initial training process, the weights of each epoch are saved. Then, at the end of the initial training, a percentage of the connections are pruned and the remaining weights and the learning rate are reset to their prior values (that we had at the $k$-th epoch of the initial training); the choice of the particular epoch number $k$, used in our work, is explained below. Subsequently, the retraining restarts from the $k$-th epoch and goes up to the last epoch, followed by a fresh round of pruning using the remaining weights. The cycle of resetting weights and learning rates is repeated until a specific degree of sparsity is achieved. In this approach, the loss function oscillates over the pruning time since the loss increases every time the weights are reset, but the process tends to converge to the reference before pruning. 

In the context of channel equalization, to the best of our knowledge, the first and only paper that applied this method is the recent work by Koike-Akino et al.~\cite{Koike2021}, where such a technique has been tested in the feedforward model called ResMLP. It was shown that this approach can give a sparsity of 99\% compared to the initial overparameterized solution with 6 layers and more than $10^6$ parameters. In this work, and similarly to Ref.~\cite{Koike2021}, we use rewinding of the first epoch, meaning that $k=1$. 

\subsubsection{Learning rate rewinding approach} 
This method was introduced in~\cite{renda2020comparing} and combines fine-tuning with weight rewinding. The third panel of Fig.~\ref{fig:pruning_type} shows how this method operates. While the weight rewinding, described above, rewinds both the weights and the learning rate, the learning rate rewinding simply rewinds the learning rate, leaving the weights to be re-trained after pruning from their values at the end of the initial training phase (like in the fine-tuning approach described above). In a nutshell, after initial training, a percentage of connections are pruned and re-trained while just the learning rate schedule is rewinded. This cycle is repeated until the network sparsity is at the desired level. To the best of our knowledge, this approach has not yet been evaluated in the optical equalization task.

\subsubsection{Bayesian optimizer assisted approach} \label{sec:BO_sparsity}
The two previous approaches were proposed because just fine-tuning the initial hyperparameters of the NN does not guarantee that the performance of the equalizer remains similar. A possible explanation for this effect is that, once the pruning of the NN starts, the optimization problem's target changes. Consequently, the hyperparameters of the new architecture may also need to be adjusted. Indeed, and as stated in Ref.~\cite{weerts2020importance}, we may lose performance if the hyperparameters are set to a default value when fine-tuning the NN's weights after compression. Solutions as in~\cite{he2018amc} leverage reinforcement learning to provide the model compression policy, determining layer-wise pruning rates. Alternatively, we use the BO-based approach to not only define the pruning policy, but also other important hyperparameters of the model (the number of tuning epochs, learning rate, batch size, and initial/final sparsity) thus optimizing the trade-off between performance and computational complexity. We note that this is a completely new approach that has not been tested in any other applications. 

Let us briefly specify the BO approach\footnote{The hyperparameter optimization can be done using methods other than the BO, although, as mentioned in Ref.~\cite{cho2020basic,snoek2012practical}, the hyperparameter optimization strategy, the BO offers numerous advantages over search algorithms in terms of finding good candidates with fewer interactions.} that seeks the global optimum $\mathbf{x}^{*}$ of a black-box function $opt$, where $opt(\mathbf{x})$ can be evaluated for any arbitrary ${x} \in \mathcal{X}$. That is, ${x}^{*}=\arg \min _{{x} \in \mathcal{X}} opt({x})$, where $\mathcal{X}$ is a hyperparameter space that can contain categorical, discrete, and continuous variables \cite{cho2020basic}. For solving the problem formulated above, the BO assumes that the function $opt$ was sampled from a Gaussian process. The BO maintains a posterior distribution for this function when observations are made\cite{snoek2012practical}. The observations, for our application, are the outcomes of our performing the NN-based equalization trials with different hyperparameters. To choose the hyperparameters for the next trial, in this work we have optimized the expected improvement over the current best result, see more in Ref.~\cite{frazier2018tutorial}.

In our case, the optimization process involves the following procedure. After the initial training phase, the NN model with hyperparameters $H_i\in \mathcal{X}$, has a total computational complexity (say, expressed in terms of real multiplications) $C_i$, and a certain performance $P_i$ (the $P_i$ is evaluated using a testing dataset). Then, we use the BO to minimize the following objective function:
\begin{equation} \label{BO_equation}
    opt = \begin{dcases}
      (P_i-P_p)\frac{C_p}{C_i},              & P_i > P_p\\
  -\frac{C_i}{C_p},& \text{if } P_p \geq P_i
\end{dcases},
\end{equation}
where $P_p$ and $C_p$ are the performance and computational complexity observed when using a set of hyperparameters $H_p\in \mathcal{X}$ in the pruning and fine-tuning process, respectively. The two possible scenarios that may occur when pruning is applied are covered by Eq.~(\ref{BO_equation}): i) the first one corresponds to the usual case where $P_i$ is better than the pruned performance $P_p$. In such a situation, the goal of minimizing $opt$ is equivalent to minimizing the $P_i - P_p$ gap and, at the same time, reducing the number of multiplications $C_p$ when compared to the initial ones, $C_i$. ii) the second case takes place when the pruned NN improves the performance, $P_p > P_i$. This case occurs when pruning enables escaping a local minimum thus improving the NN performance. The focus is then to reduce the computational complexity. According to Eq.~(\ref{BO_equation}), this means that the reduction of $opt$ can only be achieved by reducing $C_p$, (since $C_i$ is constant).
To the best of our knowledge, this procedure is a new approach (even in the ML science): it aims at identifying the best balance between the model's performance and computational complexity, by selecting a good candidate for the parameters set $H_p$.

\subsection{Weights clustering}\label{Compression_cluester}
The weights clustering also referred to as the weight-sharing compression approach, is another method that can be explored to reduce the NN model's complexity by reducing the number of effective weights used by the model. This approach takes into account that several connections may share the same weight value, and then fine-tunes those shared weights. In the case of feedforward structures, this strategy was already successfully employed to minimize the complexity of NN models~\cite{han2015deep,wang2020compressing,son2018clustering,wu2018deep}. In this paper, we use the same method as in~\cite{han2015deep}, but modify it for the recurrent layers as well. Following the selection of a centroids initialization technique, \cite{han2015deep}, a minimal distance from each weight to such centroids is used to determine the shared weights for each layer of a trained network so that all weights in the same cluster share the same weight value. The weights are not shared between the layers to prevent further performance loss and because sharing weights between sequential layers does not lower computing complexity.
\begin{figure}[htbp]
\centering
\includegraphics[width=0.45\textwidth]{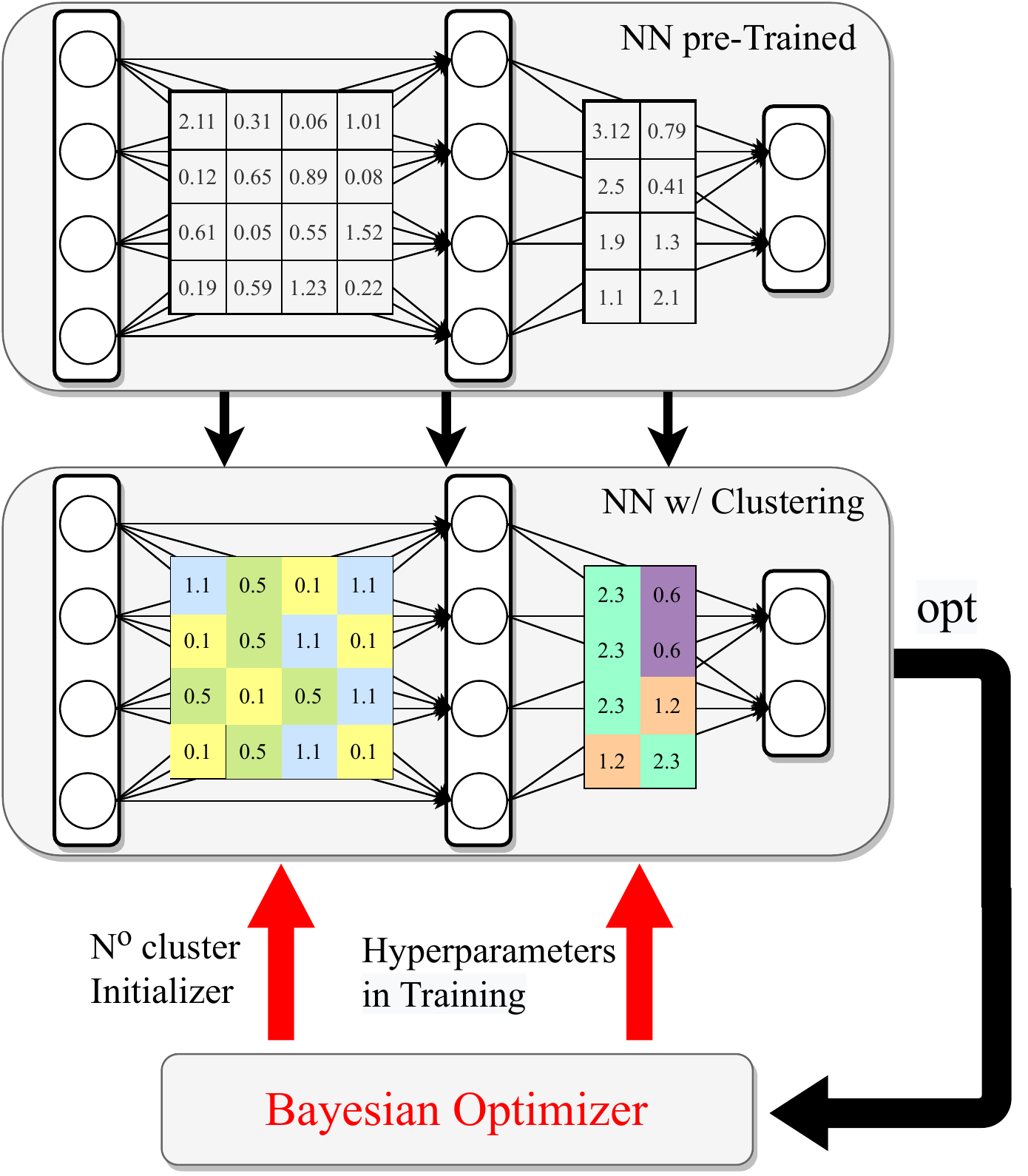}
\caption{Scheme of weight clustering over dense layers using the BO of its design. Once the NN weights are trained, a selection of weights per layer is forced to be in the closest centroid learned using stochastic gradient descent.}
\label{fig:clustering_example}
\end{figure}
Fig.~\ref{fig:clustering_example} illustrates how this strategy is applied jointly with the BO. To apply the weight clustering, we need to define three parameters: i) the number of clusters, ii) the centroids initialization technique, and iii) the weights fine-tuning process. The BO is used to select these parameters so that the performance degradation is minimized. The objective function depicted in Eq.~(\ref{BO_equation}) was also used for the BO. Four possible centroid initializers to choose from were provided to the BO: linear-, random-, density-, and K-means-based. Using the weight clustering approach has the advantage of reducing the number of distinct multipliers in matrix multiplication to at least the number of clusters per input element. Then, the results of the multipliers are sent to the different adders. To illustrate the weights clustering operation, consider the first matrix in Fig.~\ref{fig:clustering_example}. Suppose that the input vector $I$, the output vector $O$, and the weight matrix $W$ before clustering are linked as follows (to explain the method, we explicitly use 4-dimensional vectors and respective matrices):
\begin{equation}
      O \! = \! W \times  I \! =  \!
\begin{bmatrix}
           w_{11} & w_{12} & w_{13}  & w_{14}\\
           w_{21} & w_{22} & w_{23}  & w_{24}\\
           w_{31} & w_{32}& w_{33}  & w_{34}\\
           w_{41} & w_{42} & w_{43}  & w_{44}\\
         \end{bmatrix}\begin{bmatrix}
           i_{1} &
           i_{2} &
           i_{3} &
           i_{4}
         \end{bmatrix} .
  \end{equation}
In Fig.~\ref{fig:clustering_example}, we cluster this matrix with 3 centroids, $c_1$, $c_2$, and $c_3$, so the new equation connecting input and output, becomes:
  \begin{equation}
      \begin{bmatrix}
           o_{1} \\
           o_{2} \\
           o_{3} \\
           o_{4}
         \end{bmatrix} = 
\begin{bmatrix}
           c_1 & c_3 & c_2  &  c_1\\
           c_2 & c_3 &  c_1  & c_2\\
           c_3& c_2& c_3 &  c_1\\
           c_2 & c_3 &  c_1  & c_2\\
         \end{bmatrix}\begin{bmatrix}
           i_{1} &
           i_{2} &
           i_{3} &
           i_{4}
         \end{bmatrix} .
  \end{equation}
This result shows that, in the worst case scenario, the new number of multiplications would decrease from 16 (input size $\times$ output size = 4$\times$4)  to  (input size $\times$ number of clusters = 4$\times$3), because in this case we can carry out all possible unique multiplications ($i_1c_1$, $i_1c_2$, ... ,$i_4c_3$), and the rest of the operations are additions. However, by properly designing such a matrix multiplication, the number of multiplications can even be further reduced. In the same example, we may define the output $O$ as follows:
\begin{equation}\label{eq:weight-c2}
      \begin{bmatrix}
           o_{1} \\
           o_{2} \\
           o_{3} \\
           o_{4}
         \end{bmatrix} = 
\begin{bmatrix}
           i_1c_1  + (i_2+i_4)c_2  +i_3c_3\\
            (i_1 + i_2+i_4)c_3 + i_3c_2\\
           i_1c_2  + (i_2+i_4)c_1  + i_3c_3\\
           (i_1 + i_3)c_1  + (i_2+i_4)c_2  \\
         \end{bmatrix}.
\end{equation}
Notice that the number of multiplications is reduced to 8 unique multiplications in Eg. (\ref{eq:weight-c2}), which is half of its original value (the number of additions remains the same). It is important to note that the benefit resulting from using this technique depends on the lengths of the input vector and the weight matrix, as well as on how the learned weight pattern is spread over the weight matrix. In addition, weight clustering is also used as a form of heterogeneous quantization. In this sense, when assuming a quantization of, for instance, 3 bits, the weight clustering approach will try to identify the 8 unique weights that can best describe the original weight distribution of the NN model. In this case, this type of nonuniform quantization is implemented by maintaining a codebook structure that stores the shared weights, and the weights are grouped by index after calculating the gradient of each layer~\cite{han2015deep, lee2021cluster}. Importantly, in our current problem, the weight clustering contributes the most to the NN-based equalizer complexity reduction. Moreover, we note that clustering has never been used in the NN-based optical channel equalization. 

Finally, it is worth clarifying how the learning process occurs, when backpropagation is used to update the clusters of centroids and the original weights. The TensorFlow implementation used in this paper works with a lookup table to hold the centroid values during the model training, as described in Ref.~\cite{Clsteringtensorflow}. The weights array is populated with a ``gather'' operation so that, during the backpropagation, the gradients can be calculated in the usual way.  The lookup table is then adjusted using the cumulative gradient values for the weights that correspond to the same centroid. The original weights are also updated by using a straight-through estimator to overwrite the non-differential structure of clustering with an identity function, which allows all upstream gradients to be used in the updated original non-clustered weights of the layer~\cite{cho2021dkm,Clsteringtensorflow}.

\subsection{Quantization}\label{sec:quant-theor}
Quantization is used to lower the bitwidth of the numbers participating in arithmetic operations along the signal processing, which typically helps to significantly reduce the computation complexity of the processing. This means that a quantized model can use, for example, integers, instead of floating-point numbers for some or all operations. Therefore, quantization allows representing the model using less memory and doing high-performance vectorized operations on a variety of hardware platforms \cite{gholami2021survey}. 

Quantization has demonstrated excellent and consistent results when used during the training and inference using different NN models \cite{gholami2021survey,liang2021pruning,cheng2017survey, weng2021neural}. Particularly, it is especially effective during inference because it saves computing resources without significantly decreasing accuracy. NNs benefit from quantization because NNs are remarkably robust to aggressive quantization and extreme discretization.  This robustness emerges from the large number of parameters involved in the NN, meaning that they are typically working with over-parameterized models. In this subsection, we present the categories of quantizations addressed in this work, in terms of their mode (post-training quantization\cite{bai2021towards} or quantization-aware training \cite{alvarez2016efficient}) and quantization approach (homogeneous \cite{duarte2018fast} or heterogeneous \cite{coelho2021automatic}).

\begin{figure}[htbp]
\centering
\includegraphics[width=0.45\textwidth]{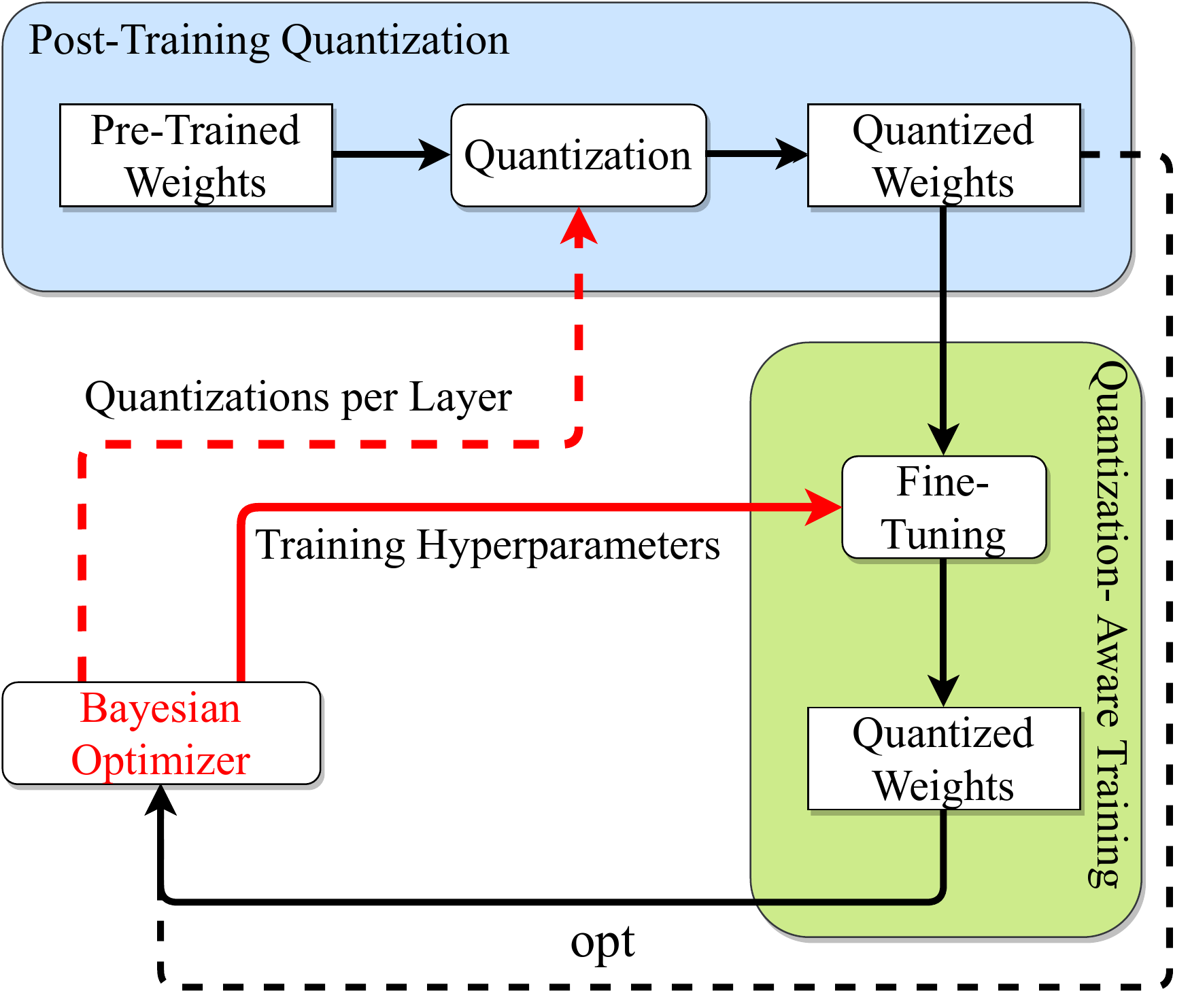}
\caption{Schematic of the NN quantizer. BO can help with the post-training quantization by finding the best bitwidth precision per layer and with the quantization-aware training by finding the best hyperparameters for the fine-tuning training after quantization.}
\label{fig:quantization}
\end{figure}

\subsubsection{Homogeneous or Heterogeneous Quantization}

Homogeneous is the most common quantization approach. The homogeneous quantization consists of reducing the precision of all NN weights to the same number of bits.  In this case, we use the same type of quantization and number of bits across the entire NN model. However, because the layered structure of multilayered NN models offers high quantization flexibility, it is natural to assume that different layers may impact the loss function differently, which favors a mixed-precision quantization approach. The process of quantizing the layers differently across the NN is known as heterogeneous quantization, and it can be a critical step toward improving the complexity-performance trade-off.  In this case, we quantize distinct layers with varying bitwidths into their fixed-point representation, as in Ref.~\cite{coelho2021automatic}.

 There are several different types of quantization that may be used. In this work we focus on the uniform quantization\cite{goyal2021fixed}, the power of two quantization (PoT) \cite{zhou2017incremental}, and the additive power of two quantization (APoT) \cite{li2019additive}, since they have a wide range of applications and usually deliver quite good results. Regarding those types of quantization, they usually convert the floating-point representation to a fixed-point representation, thus using integer mathematics instead of a floating-point one. This approach reduces both the memory and computing requirements for the realization of a particular solution.

Uniform quantization is the most common and simplest quantization approach. The uniform quantization applied to the NN elements can be expressed as\cite{goyal2021fixed}:
\begin{equation} \label{eq:quantization_uni}
    Q(x)_{BW} =  R(x, \alpha L_{uni}(BW)),
\end{equation}
\begin{equation} \label{eq:list_uni}
    L_{uni}(BW)= \left[-1, 0, \underbrace{\pm\frac{1}{2^{BW-1}},..., \pm\frac{2^{BW-1} -1}{2^{BW-1}}}_{2^{BW-1}-1 \, \,  \text{amplitudes}}\right],
\end{equation}
where $Q(\ldots)_{\cdots}$ is the quantization operator, $x$ is a real-valued input (it can be weights or an activation function), $BW$ is the quantized bitwidth value,  $R(x,\alpha L_{uni})$ is the function that rounds $x$ to the nearest element on the list $L_{uni}$ that contains all quantization levels, and $\alpha$ is a scaling level that guarantees that the largest weight in the NN will not be clipped. The quantization error is introduced by the rounding functions, depending on the $BW$ precision. Note that, in this paper, we use a representation format that besides the ``-1'' and ``''0 values involves $2^{BW-1}-1$ additional amplitudes, defining a total of $2^{BW}-2$ positive or negative levels.  The int8 quantization ($BW=8$) is one of the most widely used uniform quantization schemes, not only for the ML frameworks such as TensorFlow and PyTorch, but also for the hardware toolchains such as NVIDIA TensorRT~\cite{prasanna2019deep} and Xilinx DNNDK~\cite{DNNDK}. The int8 quantization has the advantage of typically not leading to relevant performance degradation (as can be observed from our results as well - see Fig.~\ref{fig:result_PTQ}). In this work, we will not restrict ourselves to int8 quantization only, but, in contrast, will also use the BO to determine the best number of bits for the quantization process. 

The PoT quantization is a logarithmic quantizer \cite{miyashita2016convolutional} designed to approximate the weights to the closest power of two in the range defined by the considered number of bits. Mathematically, we can represent the PoT quantization considering $2^{BW}$ elements as~\cite{miyashita2016convolutional,li2019additive, przewlocka2022power}:
\begin{equation} \label{eq:quantization_PoT}
    Q(x)_{BW} =  R(x, \alpha L_{Pot}(BW)),
\end{equation}
\begin{equation} \label{eq:list_PoT}
    L_{Pot}(BW)= \left[-1, 0,  \underbrace{\pm \frac{ 1}{2}, ...,   \pm \frac{1}{2^{(2^{BW-1} - 1)}} }_{2^{BW-1}-1 \, \,  \text{amplitudes}} \right].
\end{equation}
The POT quantization leads to much smaller computational complexity when compared to the uniform quantization because all multiplications can be represented in terms of bit-shift operations (since we have power-of-two values only). However, as pointed out in several works~\cite{nayak2019bit,li2019additive, zhou2017incremental,miyashita2016convolutional}, the performance of the PoT-quantized system can degrade compared to the uniform quantized one due to this scheme's rigid resolution problem.

The APoT quantization was recently proposed to encompass the benefits of PoT and uniform quantization types. As stated in the original paper~\cite{li2019additive}, the PoT and uniform quantizations are special cases of APoT with specific design parameters. The goal of APoT is to have fewer shift-adds than the uniform quantization, but at the same time to take the advantage of its non-uniform quantization levels as the PoT does.  Mathematically, we can represent the APoT quantization \cite{li2019additive} considering $2^{BW}$ levels as:
\begin{equation} \label{eq:quantization_APoT}
    Q_{BW}(x) =  R \big(x, \alpha L_{\text{APot}}(BW) \big),
\end{equation}
\begin{equation} \label{eq:list_APoT}
    L_{APot}(BW)= \left[-1, \left\{ \sum_{i=0}^{n-1}p_i \right\} \right],
\end{equation}
\begin{equation} \label{eq:list_APoT2}
    p_i \in \left[0,  \underbrace{\pm \frac{ 1}{2^{i+1}},\pm \frac{ 1}{2^{n+i+1}}, ...,   \pm \frac{1}{2^{[(2^{K} - 2)*n + i+1}]} }_{2^{K}-1 \, \,  \text{amplitudes}}\right],
\end{equation}
where $n$ is the number of additive terms, $k$ is defined as $k=(BW-1)/n$, and $\{ \ldots \}$ is the set containing  all possible combinations of $n$ additions from the $2^{K}$ different elements in the list $p_i$\footnote{To explain the notations better, consider the case where $n=2$ and $k=3$, such that we have two sets, $p_0=\{p_0^1, \, p_0^2, \, p_0^3\}$, and $p_1 = \{p_1^1, p_1^2, p_1^3 \}$. 
Then, $\{ \sum_{i=0}^1 p_i\} = \{p_0^1+p_1^1, \, p_0^1+p_1^2,$ \,
$p_0^1 + p_1^3, \, p_0^2 +p_1^1, \, \ldots, \, p_0^3+p_1^3 \}$.
}.  
In this description, by setting $n=1$ we have the PoT case, whereas setting $k=1$ leads to the uniform case\footnote{Eqs. (\ref{eq:list_uni}), (\ref{eq:list_PoT}), (\ref{eq:list_APoT}) are valid for$b_w>1$; when $b_w=1$, we have the same set of values, 1 or 0, independently of the type of quantization.}. In this work, we have considered $n$ fixed and equal to 2, 3, or 4. We have also addressed the case from the original paper \cite{li2019additive}, where $k$ was fixed equal to 2.  Importantly, we will show the drawback of using the APoT quantization with $k=2$ when the model has already been pruned.

\subsubsection{Post-Training Quantization}

The post-training quantization (PTQ) \cite{goyal2021fixed,gholami2021survey,habi2021hptq,bai2021towards} is a conversion technique in which all trained weights and activations of the NN model are converted to some fixed point representation, following some quantization precision established after the training phase. As indicated in Fig.~\ref{fig:quantization}, blue box,  a quantization approach is applied after training the neural network weights, and the quantized weights are saved for future use. As a result, the PTQ is an extremely fast method of quantizing NN models. Moreover, we found that, when using the PTQ, a quick grid search was already enough to analyze all possibilities and to get a satisfactory result. Thus, we decided not to use the BO to determine the optimal precision (bitwidth per layer). However, this approach usually leads to a small degradation of the model's performance, independent of the selected quantization approach.  

\subsubsection{Quantization Aware Training}

As stated previously, the inference performance of the quantized integer models is generally worse than the one of the floating-point models due to the information loss induced by quantization. To address this limitation, a method known as quantization-aware training (QAT)\cite{gholami2021survey,zhang2021training,hawks2021ps} was proposed. QAT accounts for the loss of information during the training phase, resulting in a smaller performance degradation during inference. In this work, we use the QAT approach proposed in Ref.~\cite{jacob2018quantization}, where the quantized weight levels are optimized. Afterward, the quantization is reversed, but the final forward-propagated values also include the errors aggregated by the weight quantization scheme. 

The implementation of the QAT is illustrated in the green box in Fig.~\ref{fig:quantization}. In the QAT case, the fine-tuning block operates as follows: 1) it receives the weights quantized via the chosen quantization strategy; 2) then, it performs the forward propagation; 3) afterward, it converts all variables to float precision; 4) finally, it does the backpropagation. This cycle is repeated until the weights are definitively quantized. The inference giving the quantized structure performance is then completed. 

\begin{figure}[htbp]
\centering
\includegraphics[width=0.5\textwidth]{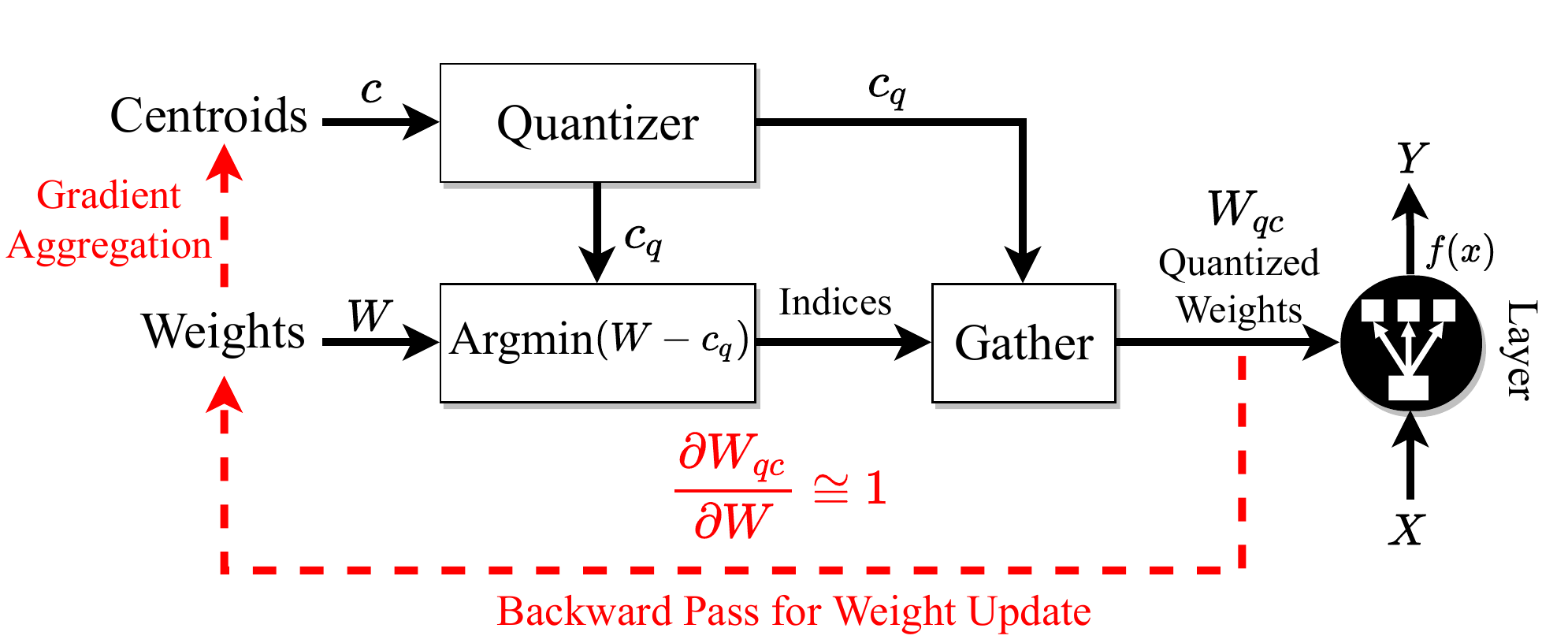}
\caption{Quantization-aware training scheme for forward and backpropagation. The forward propagation uses the quantized alphabet $c_q$ to generate the quantized weights $W_{qc}$ and, in the backpropagation, such a weight is ``skipped" by imposing its gradient to be one (straight-through estimator, Ref.~\cite{przewlocka2022power}).}
\label{fig:qat}
\end{figure}

For practical reasons, the QAT scheme for learning depicted in Fig.~\ref{fig:qat} is similar to the one used with weight clustering.  In the case of weight clustering, the quantizer box is an identity since the cluster centroids are the alphabet that the NN is training to learn (a nonuniform quantizer), and both centroids and weights are updated in the training process. We can instead force the centroids to be fixed into a defined alphabet (e.g. uniform, Eq.~(\ref{eq:list_uni}); POT, Eq.~(\ref{eq:list_PoT}); APOT, Eq.~(\ref{eq:list_APoT}), and update the weights only (the centroids will fall into one of the possibilities in the quantization alphabet). For the forward propagation, all weights in the NN, $W$, are quantized to the nearest element of the quantized alphabet $c_q$, resulting in the quantized weights $W_{qc}$ that will be used to compute the loss function. However, in the backpropagation stage, we compute the gradients using the floating-point values ($W$). This is possible because the backpropagation engine is forced to ``ignore" the quantization step used in the forward propagation. The latter is done by assuming that  $\dfrac{\partial W_{qc}}{\partial W}=1$. As stated in~\cite{przewlocka2022power}, this process is known as a Straight-Through Estimator, and it results in a smoother transition between consecutive quantization levels in the learning process.

Finally, BO was used to fine-tune, i.e., find the best hyperparameters, in the QAT. Note that, differently from the other two compression approaches where the computational complexity $C$ is measured in terms of the number of real multiplications, it is now measured in terms of the number of bit operations.

\subsubsection{Quantization Applications in Optical Channel Equalization: the Current State of the Art}
Several quantization strategies have already been proposed to equalize optical channels. Regarding the post-training quantization, the authors of Ref.~\cite{kaneda2020fpga} implemented an MLP-based equalizer with two hidden layers in an FPGA (XCZU9EG FFVC900) using post-training quantization with traditional uniform int8 precision; the quantized equalizer was tested in an experimental setup of a 50Gb/s  PON with a 30~km SSMF link. Next, this time using a recurrent NN-based equalizer, Ref.~ \cite{huang2022low} tested the equalizer in a PAM4-based 100-Gbps PON signal transmission over a 20~km SSMF fiber testbed and applied post-training quantization changing the bitwidth of the weights from 8 to 2 bits, to evaluate the BER degradation resulting from the quantization. Also,  the authors of Ref.~\cite{huang2022low} implemented such an equalizer in an FPGA using the Xilinx Vivado toolset for high synthesis.
The authors of Ref.~\cite{he2021fiber} focused on coherent transmission. In this case, a complex-valued dimension-reduced triplet input neural network was proposed and experimentally tested with a 16-QAM 80 Gbps single polarization signal transmitted along 1800 km of SSMF (100 km SSMF loop). In this study, to validate the robustness of such a NN equalizer on the quantization, they reduced the bit precision of weights to up to 2 bits, observing mostly only minor performance degradation. Finally, in \cite{ron2022experimental}, an MLP equalizer was used to mitigate the impairments in a 30 GBd 1000 km system. In this case, the PTQ strategy together with the traditional uniform 8 bits quantization was demonstrated using low-performing hardware (Raspberry Pi and Jetson nano). 

Regarding the QAT strategies description, an important discussion on the quantization of NN weights was held in Ref.~ \cite{aoudia2019towards} where it was emphasized that the equalizer inference should be performed by a fixed-point system to address a more hardware-friendly situation. An MLP-based equalizer was used, and its weights were quantized with a PoT quantization strategy. The authors incorporated the quantization error in the training of the equalizer by using the Learning-Compression (LC) algorithm, which is a possible QAT strategy. The authors of Ref.~\cite{xu2019efficient} used a deep CNN equalizer to assess their proposed quantization strategy, which combines QAT and post-equalization to find the most appropriate number of bits for uniform quantization. Considering a theoretical dispersive channel with AWGN noise and ISI, the CNN equalizer achieved performance comparable to the one of the full-precision model when using only 5-bit weights. More recently, the paper \cite{Koike2021} demonstrated that the APoT strategy could provide much higher resilience to quantization than the ordinary PoT.  In this case, a ResMLP equalizer was tested ( using simulation results) considering the transmission of a  dual-polarization 64/256QAM, 34~GBd 11Ch-WDM signal over 22 spans of 80~km of SSMF.

\section{Assessment of performance of Neural Network based Equalizers}\label{sec:NN}

\subsection{Experimental and Numerical Setups}\label{Sec:exper}

\begin{table*}[htbp]
\caption{The best hyperparameters were found by the BO for the two transmission setups considered in this work. The same NN configuration is employed for the numerical and experimental setups.}
    \centering
    \resizebox{\textwidth}{!}{
\begin{tabular}{|c|c|c|c|c|c|c|}
\hline
Transmission Case & Input Window & Output Window & Hidden Units & Kernel Size & Mini-Batch size & Learning Rate \\ \hline\hline
\begin{tabular}[c]{@{}c@{}}20$\times$50km SSMF link\\  64QAM 30GBd\end{tabular} & $221$ & $171$ & $100$ & $51$ & $3502$ & $0.001$ \\ \hline
\begin{tabular}[c]{@{}c@{}}9$\times$110km SSMF link\\  64QAM 34GBd\end{tabular} & $221$ & $195$ & $117$ & $27$ & $2153$ & $0.0013$ \\ \hline
\end{tabular}
}\label{NN_configurations}
\end{table*}

\begin{figure}[ht!]
\centering\includegraphics[width=0.5\textwidth]{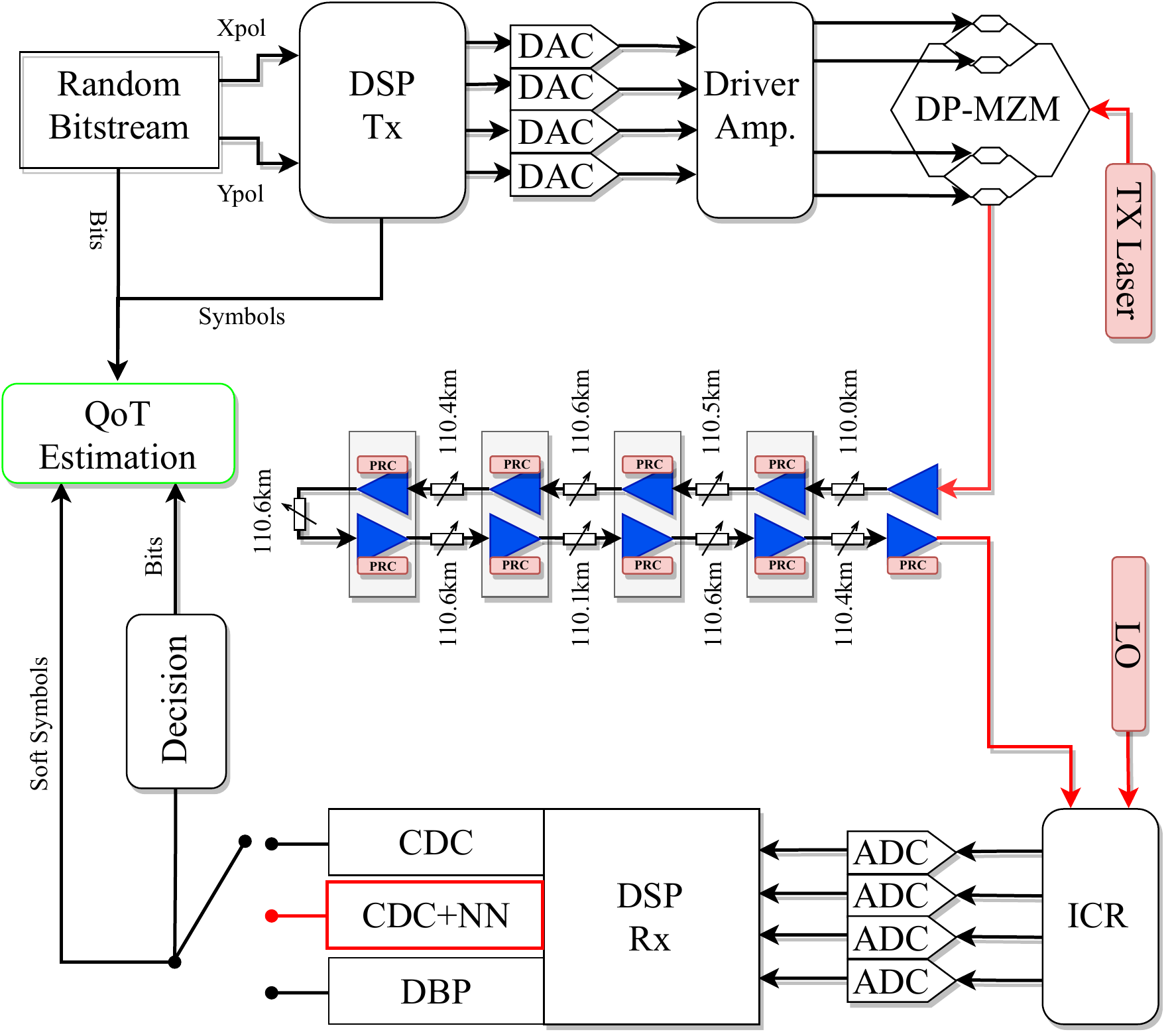}
\caption{Experimental setup. The input of the NN (shown as the red rectangle after DSP RX) is the soft output of the regular DSP before the decision unit.}
\label{setup}
\end{figure}

The performance of the NN-based equalizers with reduced complexity is assessed using data not only from numerical simulations but also from a real experimental setup to make the analysis as complete as possible. The setup used in our experiment is depicted in Fig.~\ref{setup}. At the transmitter side, a dual-polarized probabilistic shaped 64QAM (8bits/4D symbol)\footnote{We address in the experiment the PS case to show that the equalizer works in a variety of different scenarios.} 34.4 Gbaud symbol sequence was mapped out of data bits generated by a Marsenner twister generator \cite{matsumoto1998mersenne}. Then, a digital root-raised cosine (RRC )filter with a roll-off factor 0.1 was applied to limit the channel bandwidth to 37.5~GHz. The resulting filtered digital samples were resampled and uploaded to a digital-to-analog converter (DAC) operating at 88~GSamples/s. The DAC outputs were amplified by a four-channel electrical amplifier that drove a dual-polarization in-phase/quadrature Mach–Zehnder modulator, modulating the continuous waveform carrier produced by an external cavity laser at $\lambda = 1.55 \, \mu m$. 
The resulting optical signal was transmitted along 9$\times$110~km spans of SSMF with lumped (EDFA) amplification. The EDFA noise figure was in the 4.5 to 5~dB range. The SSMF is characterized at$\lambda = 1.55 \, \mu m$ by an  attenuation coefficient $ \alpha = 0.21$~dB/km, a dispersion coefficient $D = 16.8$~ps/(nm·km), and an  effective nonlinear coefficient $\gamma$ = 1.14 (W$\cdot$ km)$^{-1}$.

At the Rx side, the optical signal was converted to the electrical domain using an integrated coherent RX. The resulting signal was sampled at 50~Gsamples/s with a digital sampling oscilloscope and processed by an offline DSP based on the algorithms described in~\cite{kuschnerov2010data}. First, the bulk accumulated dispersion was compensated using a frequency domain equalizer, which was followed by the mitigation of the carrier frequency offset. A constant-amplitude zero autocorrelation-based training sequence was then located in the received frame and the equalizer transfer function was estimated from it. Afterward, the two polarizations were demultiplexed and the signal was corrected for clock frequency and phase offsets. The carrier phase estimation was then done with the help of pilot symbols. Subsequently, the resulting soft symbols were used as input for the NN equalizer. Finally, the pre-FEC BER was evaluated from the signal at the NN output.

The experimental transmission setup was mimicked by simulation. In this case, the transmission of a DP-64QAM, single-channel (SC)  34.4~Gbaud signal pre-shaped by an RRC filter with 0.1 roll-off, with an upsampling rate of 8 samples per symbol (275.2~GSamples/s) over the same fiber link is assumed. We have also tested an additional simulated setup consisting in the transmission of a DP-64QAM signal (but with a symbol rate of 30~Gbaud) along $20\times50$~km SSMF spans. The propagation of the optical signal along the optical fiber was simulated by solving the Manakov equations (\ref{eq:Manakov}) using the split-step Fourier method (with a resolution of $1$~km per step). Each fiber span was followed by an EDFA with the noise figure $\text{NF}=4.5$~dB, which fully compensates for fiber losses and adds amplified spontaneous emission noise. At the RX, after the full electronic chromatic dispersion compensation (CDC) by the frequency-domain equalizer and downsampling of the signal to the symbol rate, the received symbols are normalized to the transmitted ones. The performance of the system was evaluated in terms of the Q-factor, defined as: $Q = 20 \: \mathrm{log_{10}} \left[\sqrt{2} \: \mathrm{erfc^{-1}}(2\,\rm BER)\right]$.

\begin{figure}[ht!]
    \centering
    \includegraphics[width=\linewidth]{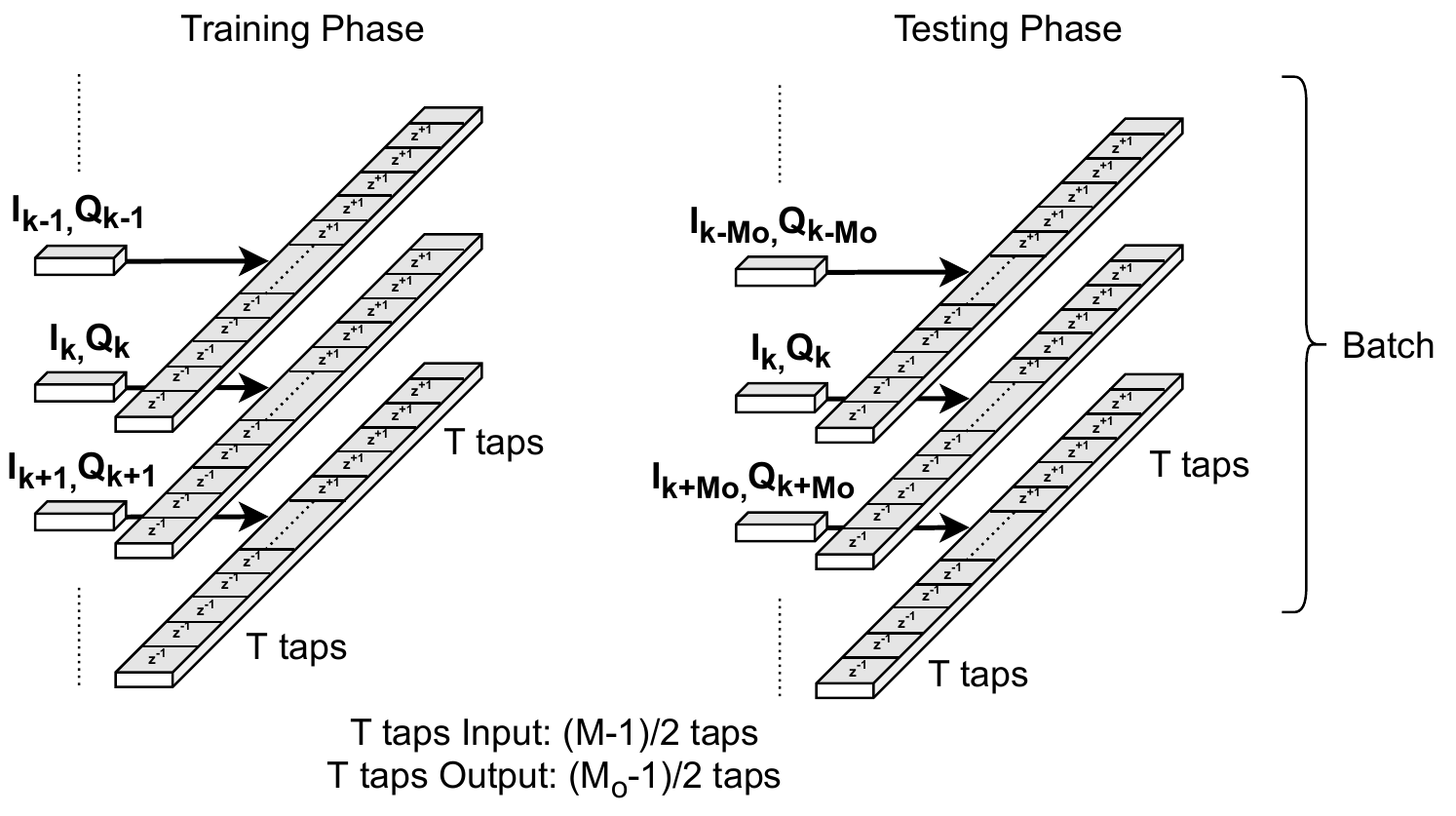}
    \caption{Data flow generation for training and testing the NN-based equalizer}
    \label{fig:data_gen}
\end{figure}

Focusing now on the biLSTM + CNN NN implemented in this work, the mean square error (MSE) loss estimator and the classical Adam algorithm for the stochastic optimization step~\cite{gulli2017deep} were used when training the weights and bias of the NN. The training hyperparameters (mini-batch size and learning rate) and the NN design hyperparameters (output window, hidden units of the LSTM, and kernel size of the 1D-CNN) were found using the BO procedure described in~\cite{freire2020complex}. An input window with 221 symbols was selected because it allows recovering a large number of symbols simultaneously, thus reducing the computational complexity. The BO optimization cycle starts with the training of the NN via backpropagation for 1000 epochs with a fixed set of hyperparameters. The BER is evaluated after each training epoch. For training, we used a fixed dataset with $2^{20}$ data points (vector of symbols), and, at every epoch, we picked $2^{18}$ random input data points from this dataset.  For testing, we used a never-seen-before dataset with $2^{18}$ data points. Here we recap in more detail the data generation for the training and testing phases. Our multi-symbol equalizer, as described in Section III, takes $M$ symbols as input data point and recovers $M_o$ symbols as output. This produces a level of parallelization of the solution, which reduces the computational complexity per recovered symbol. For training purposes, we need as much data as we can produce to train our NN model. In this sense, each input data point in the training dataset corresponds to the vector of M symbols (I and Q for both X and Y pol) for every available time $k$ in the transmission, which at the end produces many vectors with overlapped output symbols. In the testing phase, the data was generated to simulate the real benefits of such output parallelization. In this sense, for each input data point created in time $k$, we will skip $M_o$ times before generating the next point to ensure that all recovered symbols are unique and BER can be calculated with distinct symbols, avoiding any metric miscalculations. Fig.~\ref{fig:data_gen} summarizes the data creation procedure in both the training and testing phases.

 Following the training phase, the best BER was fed to the BO as an optimization target~\cite{freire2020complex} (the optimizer assumes a Gaussian conditional distribution of BERs). Using this input, the optimizer updates the process model and generates a new set of hyperparameters to be tested. After 20 Bayesian optimizer cycles, we selected the set of hyperparameters leading to the lowest BER. The BO grid space considered was: mini-batch size [32 to 5000], learning rate [0.0001 to 0.002], hidden units [1 to 150], and kernel size [1 to 200]. In this case, the output window is directly defined by the input window and the kernel size of the 1D-CNN. The BO was used to learn the best hyperparameters for the two different transmission setups considered in this paper. The results of the optimization process are summarized in Table~\ref{NN_configurations}.  Here, we also emphasize that the automated kernel acquired by the BO method can be explained by the fact that shorter links are predicted to have bigger nonlinear memory; hence, the BO discovered a larger kernel memory for the 20$times$50km ($n_k=51$) link than for the 9$times$110km link ($n_k=27$).

Finally, we would like to mention about the two benchmark lines used in this paper. two benchmarks: (ii) one for the complexity provided by the CDC and (ii) one for nonlinear mitigation given by DBP, where we used the implementation described in~\cite{napoli2014reduced}. Our primary goal was to assess the complexity of NN with respect to CDC, while guaranteeing a level of nonlinear compensation comparable to the one of the widely used DBP \footnote{The CDC benchmark is the most important because our primary goal is to show the readiness of NN with respect to the already available algorithm in commercial transponders. In contrast, none of the existing DBP versions has reached the hardware level of implementation.}. 

The CDC block was designed using a frequency domain equalizer (FDE). FDE gets rid of dispersion by multiplying the signal by the opposite of the transfer function for dispersion. After the transmission, the amount of dispersion that has built up is estimated, and based on that, the FDE changes its parameters on the fly. In terms of computational complexity, the CDC block corresponds to two linear steps of the DBP method with 2 and 1 samples per symbol, respectively, and its computational complexity in terms of the number of real multiplications per transmitted symbol is \cite{sidelnikov2021advanced}:
\begin{equation}
    C_{C D E}=4 \cdot\left(\frac{2 N\left(\log _2 N+1\right)}{\left(N-N_{D_2}+1\right)}+\frac{N\left(\log _2 N+1\right)}{\left(N-N_{D_1}+1\right)}\right) \text {, }
\end{equation}

where $N$ is the FFT size and $N_{D_q}=q \tau_D / T$, where $\tau_D$ corresponds to the dispersive channel impulse response and $T$ is the symbol interval. Factor 4 in the expression corresponds to the fact that one complex multiplication can be expressed through four real ones. 

The DBP used in this paper was also used in multiple papers, as previously reported in Ref \cite{lin2014adaptive,napoli2014performance,napoli2014reduced} . In summary, this DBP is implemented using the symmetric split-step Fourier method. In this case, both the linear filter parameters and the nonlinear operators are optimized to reduce the equalized BER, and the nonlinear step was thought to be completely static.  Also, the DBP is implemented in the frequency domain with oversampling factor equal to 2\footnote{All NN-based equalizers presented in this work operates with 1 samples per symbol.}, and the FFT size is set to 256, which was not optimized any further.  In the case of a single channel, the computational complexity of the DBP method in terms of the number of required real multiplications per transmitted symbol can be estimated as \cite{sidelnikov2021advanced} :
\begin{equation}
    C_{D B P-1 c h}=4 N_{S p} N_{S t p S p}\left(\frac{N\left(\log _2 N+1\right) q}{\left(N-N_{D_q}+1\right)}+q\right) \text {, }
\end{equation}

where $N_{S p}$ is the total number of spans, $N_{S t p S p}$ is the number of propagation steps per span, and $q$ is the oversampling factor.

\subsection{Computational Complexity Evaluation Metrics} \label{sec:complexity metric}
\begin{figure}[htbp]
\centering
\includegraphics[width=0.45\textwidth]{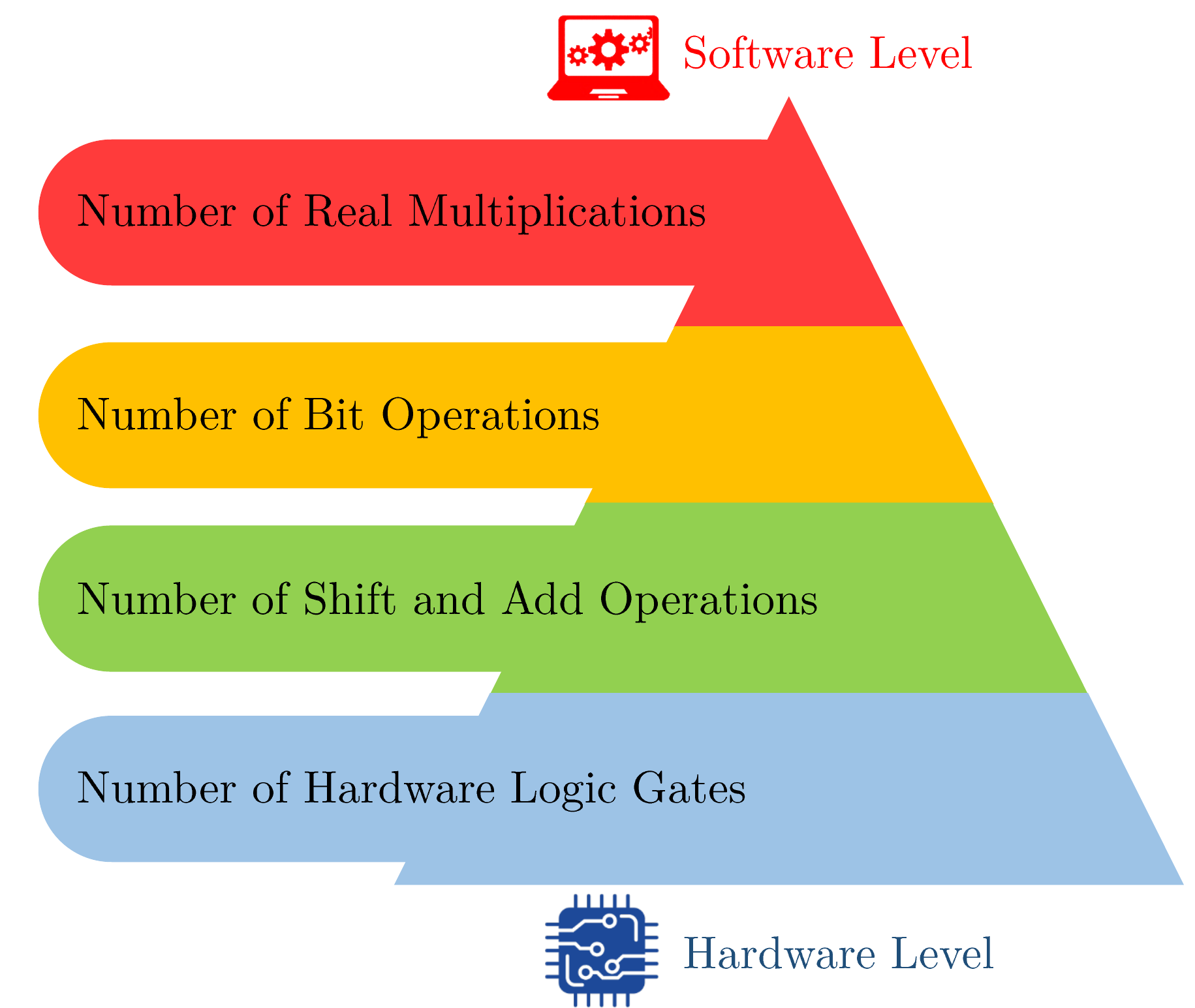}
\caption{Computational complexity metrics diagram illustrating the various levels of complexity measurement from software to hardware.}
\label{fig:complexity_metrics}
\end{figure}

The accurate evaluation of computational complexity is critical when designing a DSP (to assess the potential for hardware implementation)~\cite{spinnler2010equalizer}.
Fig.~\ref{fig:complexity_metrics} summarizes the four most commonly used criteria for assessing computational complexity, from the software to the hardware level.

\begin{table*}[ht!]
\caption{Overview of Main Metrics for Evaluating Computational Complexity in NN Models.}
    \centering
    \resizebox{\textwidth}{!}{
\begin{tabular}{cll}
\hline\hline
\rowcolor[HTML]{D3D3D3} 
\textbf{\normalsize Metric} & \multicolumn{1}{c}{\cellcolor[HTML]{D3D3D3}\textbf{ \normalsize Description}} & \multicolumn{1}{c}{\cellcolor[HTML]{D3D3D3}\textbf{\normalsize Usage}} \\ \hline\hline
\textbf{Number of Real Multiplications} & \begin{tabular}[c]{@{}l@{}}

Number of float point multiplications used in the\\NN model.
\end{tabular} & 
\begin{tabular}[c]{@{}l@{}}

When comparing with reference solutions utilizing float point\\  operations, this metric is advised as an initial estimate.

\end{tabular} \\ \hline
\textbf{Number of Bit Operations} & \begin{tabular}[c]{@{}l@{}}
Number of multiply-and-accumulate operations where\\ the number of multiplications is weighted with\\ the contribution of the  quantization used by the input,\\activation function and weights  in the NN model.
\end{tabular} & 
\begin{tabular}[c]{@{}l@{}}
This metric is recommended when the same quantization is employed,\\ but the bitwidth precision varies. These differences in bitwidth precision\\ need to be taken into consideration when comparing a solution's\\ complexity and performance to a reference solution.
\end{tabular} \\ \hline
\textbf{Number of Add and Shift Operations} & \begin{tabular}[c]{@{}l@{}}

Number of fixed-point operations used in the NN\\ model when multiplications are implemented with \\bit shifters and adders.
\end{tabular} & 
\begin{tabular}[c]{@{}l@{}}

This metric is recommended when different bitdwidth precision\\ and  quantization strategies are used. For example, when the PoT \\quantization is used and we want to show its complexity advantage\\  against the Uniform quantization. If different quantization strategies \\are not used, the number of bit operations metric should be used.

\end{tabular} \\ \hline
\textbf{Number of Hardware Logic Gates} & \begin{tabular}[c]{@{}l@{}}

Number of logic gates required to implement the\\ NN model in hardware.
\end{tabular} & 
\begin{tabular}[c]{@{}l@{}}

This metric is recommended when the size of hardware needs to\\ be assessed. It is highly dependent on the selected implementation\\ approach. 
As an example, it can be used to compare the size of \\the NN model with other traditional blocks in ASIC.

\end{tabular} \\  \hline\hline
\end{tabular}
    }
    \label{tab:overview}
\end{table*}

We have introduced in~\cite{freire2022computational} a general way of estimating the computational complexity for different-type of NN layers. The proposed metrics were the number of multiplications\cite{jacobsen2007fast,spinnler2010equalizer}, the number of bit operations\cite{baskin2021uniq, hawks2021ps,wu2018deep}, the number of shift and add operations \cite{freire2022computational} and the number of hardware logic gates \cite{sahin2006neural, 5280233}. A brief description of each of these metrics is presented in Table.~\ref{tab:overview}. In this subsection, we focus on the expressions of computational complexity when combining the biLSTM and 1D-CNN layers, and on how the compression techniques impact the computational complexity of NN equalizers.

Traditionally, the simplest estimation of complexity refers to the number of real multiplications of the algorithm only. This metric is also known as the number of real multiplications per recovered symbol (RMpS)~\cite{freire2021performance}. The RMpS corresponding to the bidirectional LSTM + CNN equalizer is given by Eq.~(\ref{Eq_original}). Since in this work we consider the unstructured pruning, which prunes all layers in the same way, the number of multiplications is:
\begin{equation}
\label{Eq_sparsity}
\begin{split}
     \mathrm{RMpS}_{\text{NN}}= \left(\frac{2
  n_{s}n_{h}(4n_i+4n_{h})}{n_s-n_k+1} + 2n_h n_o n_k \right) (1-\mu) \\ + \frac{6
  n_{s}n_{h}}{n_s-n_k+1},
\end{split}
\end{equation}
where we assume that the achieved sparsity level is equal to $\mu$. The explanation for the variables entering Eq.~(\ref{Eq_sparsity}) can be found below Eq.~(\ref{Eq_LSTMfull}).  Note that the pruning coefficient reflects the multiplications' reduction only for the weight multiplications. Since we are interested in the recurrent layers, the number of pointwise multiplications that occur internally in the recurrent cell is not affected by pruning. Aside from pruning, we can also use weight clustering to reduce the RMpS. As indicated in subsection \ref{Compression_cluester}, and also taking into account the equations that describe the LSTM cell and the 1D-CNN layer in Ref.~\cite{freire2021performance}, each LSTM cell depends on four input kernel matrices, $W^{i,f,o,c}$, and four recurrent kernel matrices, $U^{i,f,o,c}$. These four matrices for input and recurrent kernel are usually treated as two matrices (say, $W$ and $U$) with shapes $[n_i,4n_h]$ and $[n_h, 4n_h]$, respectively. Now, consider that in the input matrix, $W$, we can identify some number of clusters, $c_i$, and for the recurrent kernel matrix, $U$, we can identify $c'_h$ clusters. Therefore, the number of unique real multipliers would be $n_i * c_i$ for the multiplications involving the matrix $W$, and $n_h * c'_h$ for the multiplication with the matrix $U$\footnote{Note that, once those multiplications are performed, a synthesis data-flow / routing algorithm \cite{sun2007fpga} would be needed to distribute the result of such a multiplier to the correct adders, but here we do not account for the complexity of this design step.}.  Then, the contribution of unique multiplications to the overall complexity of the LSTM, is:
%
\begin{equation}
\label{LSTM_cluster1}
 \mathrm{C}_{\text{LSTM}}= 
  n_{s} \left(n_i c_i + n_hc'_h + 3n_h\right),
\end{equation}

For the complexity analysis, we also need to include the contribution of the 1D-CNN layer.  Since the 1D-CNN layer receives the output of the biLSTM layer, the 1D-CNN layer with kernel size $n_k$ and the number of filters $n_o$ will possess a CNN kernel tensor with the shape [$n_k$, $2n_h$, $n_o$]. Suppose that we have identified  $c''_j$ clusters in each of $n_o$ the filter and the biLSTM the output has the shape [$n_s$, $2n_h$]. Now we can split the operation for the convolution of the biLSTM output with a CNN filter as  $c''_j$  multiplications between each of the clustered kernel values to the sum of the selected input elements which share the same clustered weight value. This operation has to be repeated $n_s- n_k +1$ times (the output shape of the CNN layer) for one filter, and then for the total number of operations we multiply this value by the number of filters $n_o$. The remaining operations are just additions. Therefore, we can eventually represent the 1D-CNN layer complexity contribution as:
\begin{equation}
\centering
\label{out_cluster}
  \mathrm{C}_{\text{CNN}}= (n_s-n_k+1) \left(n_o c''_j \right).
\end{equation}
And now, the ultimate complexity for the biLSTM + CNN equalizer with clustering and pruning, becomes:
\begin{equation}
\label{Eq_cluster}
 \mathrm{RMpS}_{\text{NNp+c}}=  \frac{\mathrm{C}_{\text{LSTM}}^{forward}+\mathrm{C}_{\text{LSTM}}^{backward}+\mathrm{C}_{\text{CNN}}}{n_s-n_k+1},
\end{equation}
We notice that the pruning simplifies the clustering method since we have to group less weights.  In summary, by doing clustering first, we observed that the training was not that efficient because the structural change was too abrupt, and so the learning afterward was more complex. However, when we prune first, the set of weights to be clustered and be fine-tuned, drops to around 70\% in its size, which, in our tests, helped decrease the learning complexity of this new structure. Also, when defining the number of clusters in each layer ($c_i$, $c'_h$, and $c''_j$), one of the clustered values is almost always zero. Consequently,  this cluster does not add multiplications, and we have even lower computation complexity.

When comparing solutions that use floating-point arithmetic with the same bitwidth precision, the RMpS is usually a meaningful metric for comparative estimates. When moving to fixed-point arithmetic, a second metric known as the number of bit-operations (BOPs) should be adopted to understand the impact of changing the bitwidth precision on the complexity\footnote{Two assumptions are made in the definition of the BoPs. First,  we assume that each parameter is only fetched once from an external memory; second, the cost of fetching a $b$-bit parameter is assumed to be equal to $b$-BOPs~\cite{hawks2021ps,baskin2021uniq}. Also, the bias is supposed to be quantized in the same way as the weights.}. Because we can readily find the number of bit operations required by the additions and multiplications, we can calculate the BOPs associated with the NN inference process, expressed in terms of multiply-and-accumulate operations (MACs)\cite{baskin2021uniq, hawks2021ps,wu2018deep,freire2022computational}.  As described in Ref.~\cite{freire2022computational}, the BOP complexity for the LSTM and 1D-CNN layers can be expressed as:

\begin{equation} \label{BOP.lstm}
\begin{split}
        \mathrm{BOP}_{\text{LSTM}}
={}& 4n_{s}n_{h}\text{Mult}(n_{i}, b_{w}, b_{i})\\
+{}& 4n_{s}n_{h}\text{Mult}(n_{h}, b_{w}, b_{a})\\
+{}& 3n_{s}n_{h}b_{a}^2\\
+{}& 9n_{s}n_{h}\text{Acc}(n_{h}, b_{w}, b_{a}),
\end{split}
\end{equation}
\begin{equation} \label{BOP.cnn}
\begin{split}
        \mathrm{BOP}_{\text{CNN}}
={}& OutputSize \cdot n_{f}\text{Mult}(n_{i}n_{k}, b_{w}, b_{i})\\
+{}& n_{f}\text{Acc}(n_{i}n_{k}, b_{w}, b_{i}),
\end{split}
\end{equation}
where, in the context of NNs, $b_w$ is the number of bits used to represent the weights of the NN, $b_i$ is the number of bits used to represent the input, and $b_a$ is the number of bits used to represent the NN's activation functions. For the convenience of further presentation, we have used short notations, Mult and  Acc:
\[ \text{Mult}(n_{i}, b_{w}, b_{i}) = n_{i} b_{w} b_{i} + (n_{i}\!-\!1)\big(b_{w} \!+ \!b_{i} \!+\lceil \log_2(n_{i}) \!\rceil \big),\] 
and 
\[ \text{Acc}(n_{i}, b_{w}, b_{i}) = b_{w} + b_{i} +\lceil \log_2(n_{i}) \rceil.\] 
The Acc expression represents the actual bitwidth of the accumulator\footnote{The accumulator is the register in which the intermediate arithmetic logic unit results are stored. For more detailed explanation, see \cite{freire2022computational}.} required for MAC operations. Therefore, for our NN-based equalizer (biLSTM+1D-CNN) with 4 input features, 2 output features, $n_h$ hidden units in the LSTM cell, $n_k$ convolutional kernel size, and $n_s = M$ memory time window, we can represent the required number of BoPs considering that the output of biLSTM is the input of 1D-CNN, as follows:
\begin{equation} \label{BOP.lstm2}
\begin{split}
        \mathrm{BOP}_{\text{LSTM}}^{for/backward}
={}& 4Mn_{h}\text{Mult}(4, b_{w}, b_{i})\\
+{}& 4Mn_{h}\text{Mult}(n_{h}, b_{w}, b_{a})\\
+{}& 3Mn_{h}b_{a}^2\\
+{}& 9Mn_{h}\text{Acc}(n_{h}, b_{w}, b_{a}),
\end{split}
\end{equation}
\begin{equation} \label{BOP.cnn2}
\begin{split}
        \mathrm{BOP}_{\text{CNN}}
={}& (M-n_k+1) \cdot 2\text{Mult}(2n_{h}n_{k}, b_{w}, b_{i})\\
+{}& 2\text{Acc}(2n_{h}n_{k}, b_{w}, b_{i}),
\end{split}
\end{equation}
\begin{equation}
\label{boptotap}
 \mathrm{BoP}_{\text{NN}}=  \frac{\mathrm{BoP}_{\text{LSTM}}^{forward}+\mathrm{BoP}_{\text{LSTM}}^{backward}+\mathrm{BoP}_{\text{CNN}}}{M-n_k+1}.
\end{equation}

Most real DSP implementations use a dedicated logic macros (e.g., DSP Slice in FPGAs or MAC in ASIC), where the BoP metric fits as a good complexity estimation/comparison metric. However, with the advances in new quantization techniques\cite{li2019additive,Koike2021,elhoushi2021deepshift,you2020shiftaddnet}, those multiplications can also be implemented using just bit shifter- and adder-based algorithms \cite{gentili1995efficient, evans1994efficient, lee2003frequency}, when the fixed-point multiplications are used\footnote{Note that the translation from multiplications to additions and shift operations adds some quantization noise/error since we round the original coefficients when converting them from a float representation to a fixed representation. However, in the context of NNs, this can be partially mitigated by training the NN with the quantized weights, as it was done in Refs.~\cite{Koike2021,elhoushi2021deepshift, you2020shiftaddnet} and in this current work.}. Therefore, to account for the impact of our using different quantization strategies, we can utilize the metric that evaluates the number of additions and bit shift (NABS) operations. 
 
 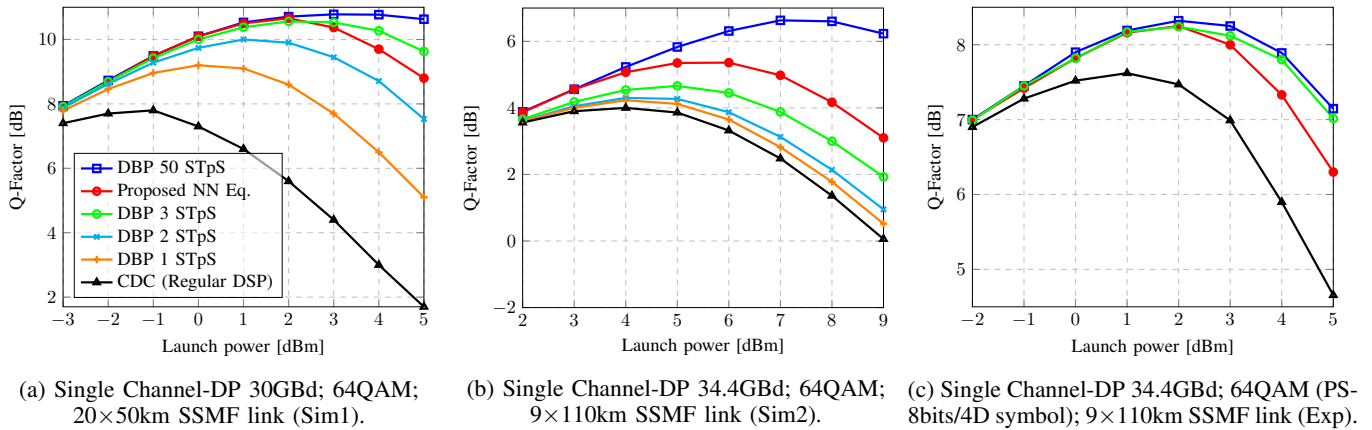
\begin{figure*}[ht!] 
\begin{subfigure}{.33\textwidth}
    \centering
 \begin{tikzpicture}[scale=0.7]
    \begin{axis} [ylabel={BER}, 
        xlabel={Launch power [dBm]},
        ylabel={Q-Factor [dB]},
        grid=both,  
        xmin=-3, xmax=5,
    	xtick={-3, ..., 5},
    	ymin=1.7, ymax=11,
        legend style={legend pos=south west, legend cell align=left,fill=white, fill opacity=0.6, draw opacity=1,text opacity=1},
    	grid style={dashed}]
        ]
        \addplot[color=blue, mark=square, very thick]     coordinates {
    (-3,  7.94)(-2, 8.73)(-1, 9.49)(0, 10.1)(1,10.53)(2,10.71)(3,10.78)(4,10.77)(5,10.63)
    };
    \addlegendentry{DBP 50 STpS};
    
    \addplot[color=red, mark=*, very thick]   
    coordinates {
    (-3, 7.88)(-2, 8.69)(-1, 9.49)(0, 10.1)(1,10.5)(2,10.66)(3,10.37)(4,9.7)(5,8.8)
    };
    \addlegendentry{Proposed NN Eq.};

    \addplot[color=green, mark=o, very thick]     coordinates {
        
    (-3, 7.92)(-2, 8.69)(-1, 9.42)(0, 10)(1,10.38)(2,10.56)(3,10.53)(4,10.27)(5,9.63)
    };
    \addlegendentry{DBP 3 STpS};

            \addplot [color=cyan, mark=x, very thick]    coordinates {
           (-3, 7.88)(-2, 8.62)(-1, 9.28)(0,9.74)(1,10)(2,9.9)(3,9.45)(4,8.71)(5,7.53)
    };
    \addlegendentry{DBP 2 STpS};

        \addplot[color=orange, mark=+,very thick]    coordinates {
    (-3,  7.79)(-2, 8.46)(-1, 8.96)(0, 9.2)(1,9.1)(2,8.6)(3,7.7)(4,6.5)(5,5.1)
    };
    \addlegendentry{DBP 1 STpS};
        \addplot[color=black,mark=triangle, very thick]    coordinates {
    (-3, 7.4)(-2, 7.7)(-1, 7.8)(0, 7.3)(1,6.6)(2,5.6)(3,4.4)(4,3)(5,1.7)
    };
    \addlegendentry{CDC (Regular DSP)};
    
    \end{axis}
    \end{tikzpicture}
    \caption{Single Channel-DP 30GBd; 64QAM;\\ 20$\times$50km SSMF link (Sim1).}
    \label{fig:res_a} 
\end{subfigure}\hfill
\begin{subfigure}{.33\textwidth}
    \centering
 \begin{tikzpicture}[scale=0.7]
    \begin{axis} [ylabel={BER}, 
        xlabel={Launch power [dBm]},
        ylabel={Q-Factor [dB]},
        grid=both,  
        xmin=2, xmax=9,
    	xtick={2, ..., 9},
    	ymin=-2, ymax=7,
        legend style={legend pos=south west, legend cell align=left,fill=white, fill opacity=0.6, draw opacity=1,text opacity=1},
    	grid style={dashed}]
        ]
        \addplot[color=blue, mark=square, very thick]     coordinates {
    (2,      3.88)(3, 4.56)(4, 5.23)(5, 5.83)(6,6.31)(7,6.63)(8,6.6)(9,6.23)
    };
    
    \addplot[color=red, mark=*, very thick]   
    coordinates {
     (2,      3.87)(3, 4.56)(4, 5.07)(5, 5.35)(6,5.36)(7,4.98)(8,4.17)(9,3.1)
    };

    \addplot[color=green, mark=o, very thick]     coordinates {
    (2,      3.66)(3, 4.18)(4, 4.54)(5, 4.66)(6,4.45)(7,3.88)(8,3)(9,1.93)
    };


            \addplot [color=cyan, mark=x, very thick]    coordinates {
    (2,      3.61)(3, 4.05)(4, 4.3)(5, 4.27)(6,3.87)(7,3.13)(8,2.14)(9,0.95)
    };

        \addplot[color=orange, mark=+,very thick]    coordinates {
    (2,      3.6)(3, 4)(4, 4.23)(5, 4.12)(6,3.65)(7,2.82)(8,1.78)(9,0.52)
    };
        \addplot[color=black,mark=triangle, very thick]    coordinates {
    (2,      3.56)(3, 3.9)(4, 4)(5, 3.86)(6,3.32)(7,2.48)(8,1.36)(9,0.06)
    };
    
    \end{axis}
    \end{tikzpicture}
    \caption{Single Channel-DP 34.4GBd; 64QAM;\\ 9$\times$110km SSMF link (Sim2).}    \label{fig:res_b} 
\end{subfigure}\hfill
\begin{subfigure}{.33\textwidth}
    \centering
 \begin{tikzpicture}[scale=0.7]
    \begin{axis} [ylabel={BER}, 
        xlabel={Launch power [dBm]},
        ylabel={Q-Factor [dB]},
        grid=both,  
        xmin=-2, xmax=5,
    	xtick={-2, ..., 5},
    	ymin=4.5, ymax=8.5,
        legend style={legend pos=south west, legend cell align=left,fill=white, fill opacity=0.6, draw opacity=1,text opacity=1},
    	grid style={dashed}]
        ]
        \addplot[color=blue, mark=square, very thick]     coordinates {
    (-2, 7)(-1, 7.45)(0, 7.9)(1, 8.19)(2,8.32)(3,8.25)(4,7.89)(5,7.146)
    };
    
    \addplot[color=red, mark=*, very thick]   
    coordinates {
    (-2,  7)(-1, 7.42)(0, 7.82)(1, 8.16)(2,8.25)(3,8)(4,7.33)(5,6.3)
    };

    \addplot[color=green, mark=o, very thick]     coordinates {
 (-2,  6.99)(-1, 7.45)(0, 7.82)(1, 8.17)(2,8.24)(3,8.12)(4,7.8)(5,7.016)
    };


        \addplot[color=black,mark=triangle, very thick]    coordinates {
    (-2,  6.9)(-1, 7.28)(0, 7.52)(1, 7.62)(2,7.47)(3,6.99)(4,5.9)(5,4.656)
    };
    
    \end{axis}
    \end{tikzpicture}
    \caption{Single Channel-DP 34.4GBd; 64QAM (PS-8bits/4D symbol); 9$\times$110km SSMF link (Exp).}    \label{fig:res_c} 
\end{subfigure}

  \caption{Performance of the proposed NN equalizer, benchmarked against DBP~ \cite{Kahn:01} for three different  transmission scenarios.}
  \label{fig:result_powers} 
\end{figure*}

Next, we discuss how to translate the complexity of the NN equalizer from RMpS into the NABS metric, in the cases when we utilize uniform quantization, PoT quantization, and APoT quantization (see also Section \ref{sec:Compression}.C). According to Ref.~\cite{freire2022computational}, the NABSs metric takes into account the conversion of all multipliers into adders and shifters, and computes the complexity of the total number of adders (including the pre-existing adders of the NN structure) based on bit precision while ignoring the cost of the bit shift operations. The NABSs complexity for the LSTM and 1D-CNN layers can be expressed as~\cite{freire2022computational}:
\begin{equation} \label{nabs.lstm}
\begin{split}
    \mathrm{NABS_{LSTM}} 
={}& 4n_{s}n_{h}\big[n_{i}(X_{w} +1)-1\big]\text{Acc}(n_{i}, b_{w}, b_{i})\\
+{}& 4n_{s}n_{h}\big[n_{h}(X_{w} +1) +1\big]\text{Acc}(n_{h}, b_{w}, b_{a})\\
+{}& 6n_{s}n_{h}b_{a}.
\end{split}
\end{equation}
\begin{equation} \label{nabs.cnn}
\begin{split}
    \mathrm{NABS_{CNN}} 
={}& OutputSize \cdot n_{f}\big[n_{i}n_{k}(X_{w} + 1 )-1\big]\\ 
{}&\cdot\text{Acc}(n_{i}n_{k}, b_{w}, b_{i})\\
+{}& n_{f}\text{Acc}(n_{i}n_{k}, b_{w}, b_{i}).\\
\end{split}
\end{equation}
In these expressions, $X$ represents the number of adders required at most, to perform the multiplication when considering that the first bit represents the sign and the remaining ones contain the magnitude of the weight. For the uniform quantization, we have: $X = b_{w}-2$, whereas in the case of POT quantization, we have: $X = 0$, because each multiplication costs only a shift \cite{gentili1995efficient, przewlocka2022power}. Lastly, for the APOT quantization, we have: $X = \mathrm{n}$, where n denotes the number of additive terms. These equations are in line with the expected complexity behavior from Ref.~\cite{li2019additive}, where it is stated that by using the APOT with $k=2$ (which means that $\mathrm{n} = (b_{w} -2)/2$), the multiplication would be approximately 2 times faster than when using the uniform quantization. Thus, for the biLSTM + 1D-CNN equalizer considered in our work, we have the following expressions for the NABSs complexity per recovered symbol:
\begin{equation}\label{nabs.lstm2}
\begin{split}
    \mathrm{NABS_{LSTM}^{for/backward}} 
={}& 4Mn_{h}\big[4(X_{w} +1)-1\big]\text{Acc}(4, b_{w}, b_{i})\\
+{}& 4Mn_{h}\big[n_{h}(X_{w} +1) +1\big]\text{Acc}(n_{h}, b_{w}, b_{a})\\
+{}& 6Mn_{h}b_{a},
\end{split}
\end{equation}
\begin{equation}\label{nabs.cnn2}
\begin{split}
    \mathrm{NABS_{CNN}} 
={}& (M-n_k+1) \cdot 2\big[2n_{h}n_{k}(X_{w} + 1 )-1\big]\\ 
{}&\cdot\text{Acc}(2n_{h}n_{k}, b_{w}, b_{i})\\
+{}& 2\text{Acc}(2n_{h}n_{k}, b_{w}, b_{i}),
\end{split}
\end{equation}
\begin{equation}
\label{NABStotap}
 \mathrm{NABS}_{\text{NN}}=  \frac{\mathrm{NABS}_{\text{LSTM}}^{forward}+\mathrm{NABS}_{\text{LSTM}}^{backward}+\mathrm{NABS}_{\text{CNN}}}{M-n_k+1}.
\end{equation}

Notably, the BoPs and NABSs expressions given above do not take into account the effects of pruning and weight clustering, but they can be corrected, similarly to how the RMpS metric at the beginning of this subsection.

Finally, the metric that is even closer to the hardware level is the number of logic gates (NLGs) that are used for the hardware (e.g. ASIC or FPGA) implementation of a signal processing device. It is different from the NABSs metric because it indicates the real cost of implementation. Within this metric, the cost of activation functions, represented by look-up tables (LUTs), is also taken into account. However, this metric is not used in this work since it already depends on the particular hardware type that we do not consider here. 
 
\section{Results} \label{sec:Results}

\subsection{Multi-Symbol Equalizer Performance}

We start by presenting the benchmark scenario obtained using the nonlinear equalization, i.e. using the equalizers without compression. In this case, we can see the increase in optimum launch power after equalization and the corresponding Q-factor improvement concerning the case without nonlinear equalization. To speed up the training process and the acquisition of results, we have trained our model at the highest launch power and applied the transfer learning strategy \cite{freire_transfer} for the remaining lower launch powers. For these lower power levels, we fine-tuned the NNs for around five epochs. Fig.~\ref{fig:result_powers} shows the results of Q-factor over launch power dependence for three transmission scenarios.  For the simulated transmission with $20\times50$~km, the NN equalizer enabled increasing the optimum launch power from -1~dBm to 2~dBm. Furthermore, the maximum Q-factor increased by about 2.8 dB, showing a similar maximum performance as that achieved by 3 STpS DBP. For the transmission over $9\times110$~km system, which has a similar total transmission length but more than doubled span length, we have the enhanced impact of ASE noise. This effect can be observed in the results depicted in Fig.~\ref{fig:result_powers}~(b). In this case, the optimum power increased from 4~dBm to 6~dBm and the optimum Q-factor improved by around 1.3~dB due to the equalization. Although the performance improvement enabled by the NN equalizer was lower than in the previous case, it was still higher than the one enabled by the 3 STpS DBP (about 0.7~dB performance improvement). Finally, Fig.~\ref{fig:result_powers}~(c) shows the results obtained in the experimental setup described in Sec.~\ref{Sec:exper}. In this case, we observe that the NN equalizer leads to an increase in the optimum launch power of about 1~dB, and an increase in the maximum Q-factor of about 0.7~dB. In this case, we observed that compared to the results of the numerical modeling of a similar system (shown in Fig.~\ref{fig:result_powers}~(b)), the NN allows us to approach the performance of 50 StPS DBP closer, because the probabilistic shaping (considered in the experimental setup, but not in the numerical modeling of Fig.~\ref{fig:result_powers}~(b)) modifies both signal-signal and signal-noise interactions at higher signal powers [116]~\cite{cho2019probabilistic}. Also, the limited performance of DBP and NN-based equalization within the experimental conditions can be attributed to non-ideal conditions, such as the presence of polarization effects and transceiver noise. 

Following this initial analysis, we chose the best launch power in each transmission setup to evaluate the performance degradation resulting from using the different compression approaches.
\begin{figure*}[t!] 
\begin{subfigure}{.33\textwidth}
    \centering
 \begin{tikzpicture}[scale=0.7]
    \begin{axis} [ylabel={BER}, 
        xlabel={Sparsity (\%)},
        ylabel={Q-Factor [dB]},
        grid=both,  
        xmin=20, xmax=80,
        ymax=11,
        legend style={legend pos=south west, legend cell align=left,fill=white, fill opacity=0.6, draw opacity=1,text opacity=1},
    	grid style={dashed}]
        ]
        \addplot[color=blue, mark=square, very thick]     coordinates {
    (20,10.64)(40,10.4)(60,8.89)(80,7.26)     
    };
    \addlegendentry{Fine Tune};
    
    \addplot[color=orange, mark=*, very thick]   
    coordinates {
    (20,10.41)(40,10)(60,9.4)(80,8.7)     
    };
    \addlegendentry{Weight Rewinding};

    \addplot[color=green, mark=triangle, very thick]     coordinates {
    (20,10.64)(40,10.4)(60,9.7)(80,9)     

    };
    \addlegendentry{Learning Rate Rewinding};
                \addplot [color=red, mark=*, width=6pt,mark size=4pt,smooth]    coordinates {
   (72,10.47) 
    };
    \addlegendentry{BO+FT};
    
            \addplot [color=purple, dashed, very thick]    coordinates {
    (20,10.77)(40,10.77)(60,10.77)(80,10.77)     
    };
    \addlegendentry{Original (No Pruning)};
    
    \end{axis}
    \end{tikzpicture}
    \caption{Single Channel-DP 30GBd; 64QAM;\\20$\times$50km SSMF link (Sim1).}
    \label{fig:res_a4} 
\end{subfigure}\hfill
\begin{subfigure}{.33\textwidth}
    \centering
 \begin{tikzpicture}[scale=0.7]
    \begin{axis} [ylabel={BER}, 
        xlabel={Sparsity (\%)},
        ylabel={Q-Factor [dB]},
        grid=both,  
        xmin=20, xmax=80,
        ymax=5.5,
        legend style={legend pos=south west, legend cell align=left,fill=white, fill opacity=0.6, draw opacity=1,text opacity=1},
    	grid style={dashed}]
        ]
        \addplot[color=blue, mark=square, very thick]     coordinates {
    (20,5.31)(40,5.22)(60,4.75)(80,3.93)     
    };
    \addlegendentry{Fine Tune};
    
    \addplot[color=orange, mark=*, very thick]   
    coordinates {
    (20,5.15)(40,5.13)(60,5.13)(80,5.1)     
    };
    \addlegendentry{Weight Rewinding};

    \addplot[color=green, mark=triangle, very thick]     coordinates {
    (20,5.31)(40,5.29)(60,5.14)(80,4.9)     

    };
    \addlegendentry{Learning Rate Rewinding};
    \addplot [color=red, mark=*, width=6pt,mark size=4pt,smooth]    coordinates {
   (70, 5.28) 
    };
    \addlegendentry{BO+FT};
    
            \addplot [color=purple, dashed, very thick]    coordinates {
    (20,5.35)(40,5.35)(60,5.35)(80,5.35)     
    };
    \addlegendentry{Original (No Pruning)};

    
    \end{axis}
    \end{tikzpicture}
    \caption{Single Channel-DP 34.4GBd; 64QAM;\\9$\times$110km SSMF link (Sim2).}    \label{fig:res_b4} 
\end{subfigure}\hfill
\begin{subfigure}{.33\textwidth}
    \centering
 \begin{tikzpicture}[scale=0.7]
    \begin{axis} [ylabel={BER}, 
        xlabel={Sparsity (\%)},
        ylabel={Q-Factor [dB]},
        grid=both,  
        xmin=20, xmax=80,
        ymax=8.5,
        legend style={legend pos=south west, legend cell align=left,fill=white, fill opacity=0.6, draw opacity=1,text opacity=1},
    	grid style={dashed}]
        ]
        \addplot[color=blue, mark=square, very thick]     coordinates {
    (20,8.22)(40,8.16)(60,7.91)(80,7.22)     
    };
    \addlegendentry{Fine Tune};
    
    \addplot[color=orange, mark=*, very thick]   
    coordinates {
    (20,8.17)(40,8.1)(60,8)(80,7.9)     
    };
    \addlegendentry{Weight Rewinding};

    \addplot[color=green, mark=triangle, very thick]     coordinates {
    (20,8.22)(40,8.19)(60,8.05)(80,7.95)     

    };
    \addlegendentry{Learning Rate Rewinding};
                \addplot [color=red, mark=*, width=6pt,mark size=4pt,smooth]    coordinates {
   (61,8.14) 
    };
    \addlegendentry{BO+FT};
    
            \addplot [color=purple, dashed, very thick]    coordinates {
    (20,8.22)(40,8.22)(60,8.22)(80,8.22)     
    };
    \addlegendentry{Original (No Pruning)};

    
    \end{axis}
    \end{tikzpicture}
    \caption{Single Channel-DP 34.4GBd; 64QAM (PS-8bits/4D symbol); 9$\times$110km SSMF link (Exp).}    \label{fig:res_pruning} 
\end{subfigure}

  \caption{Optical performance when using different pruning techniques for several NN sparsity level. The optimum launch power (without pruning) is set for each case in this study: (Sim 1) 2dBm; (Sim 2) 6dBm; (Exp) 2dBm }
  \label{fig:result_pruning} 
\end{figure*}
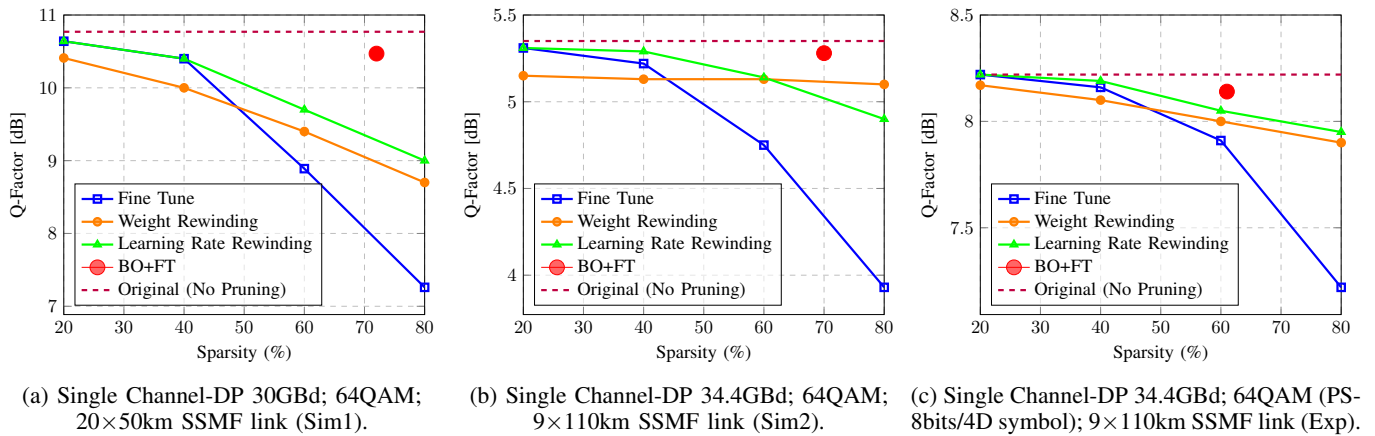
\subsection{Pruning Study}
We start our comparative study by doing an analysis similar to what was done in Ref.~\cite{renda2020comparing} for image classification, but this time in the context of coherent optical channel post-equalization. Additionally, besides the fine-tuning approach, the weight rewinding, and the learning rate rewinding, in our work, the fine-tuning assisted by the Bayesian optimizer is also considered.

The results for the three transmission setups obtained after pruning are depicted in Fig.~\ref{fig:result_pruning}. Similarly to the results shown in Ref.~\cite{renda2020comparing},  the weight rewinding and the learning rate rewinding outperform the fine-tuning when we have a high level of compression ($\geq 50\%$). As an example, for the $60\%$ sparsity and when employing the fine-tuning for retraining, the Q-factor is reduced by $1.9$~dB, $0.6$~dB, and $0.3$~dB for the three considered transmission scenarios, as compared to the original performance. If instead, the rewinding approaches are used, the Q-factor degradation of only $1.1$~dB, $0.2$~dB, and $0.2$~dB, and of $1.4$~dB, $0.2$~dB, and $0.2$~dB, are observed for the learning rate and weight rewinding, respectively, when considering the same three transmission scenarios. However, when the fine-tuning is assisted by the BO to select the hyperparameters (as described in Sec.~\ref{sec:BO_sparsity}), we observe that the performance can be significantly improved compared to the other approaches. This approach enabled reaching high sparsity (even higher than the $60\%$ example mentioned above), leading to a Q-factor degradation not exceeding $0.3$~dB, $0.1$~dB, and $0.1$~dB for the three considered transmission scenarios.  This result shows the potential of the BO-assisted fine-tuning approach, to outperform the previous model compression techniques.
In our view, the superior performance of the BO-assisted pruning comes from the ability of this approach to cope with the dimensionality changes in multidimensional trainable parameters' space when the NN architecture is pruned. Therefore, the training hyperparameters to achieve a good local minimum may differ from the initial ones, and the BO is capable of identifying this new set of training hyperparameters, while the other methods use their previous values obtained before pruning. However, we emphasize that the BO requires significant computational effort, which means that this method is appropriate mainly for offline applications. When we are interested in achieving the result in the fastest way, the learning rate rewinding is the recommended approach.

Interestingly, the weight rewinding approach performed worse than the fine-tuning approach in cases where the sparsity was lower than $50\%$, while the learning rate rewinding led to similar or even better performance as compared to fine-tuning. This result can be explained by recalling that the original model was learned using the transfer learning approach, which aids in the learning process by improving generalization and avoiding local minima. When fine-tuning and learning rate rewinding are used, the original weights are the starting point of the pruning process, preserving the good initialization provided by transfer learning. However, in the case of weight rewinding, the weights are reinitialized randomly after the pruning, which can be detrimental to training, thus leading to higher performance degradation.

Regarding the computational complexity reduction in terms of RMpS when using the BO plus fine-tuning (BO+FT) approach, in the result depicted in Fig.~\ref{fig:result_pruning}~(a), the BO+FT approach achieved a sparsity of $72\%$, which represents a reduction from $1.29\mathrm{e+}5$ to $3.66\mathrm{e+}4$ in the RMpS value. In the case of Fig.~\ref{fig:result_pruning}~(b), the achieved sparsity was $70\%$, which represents a reduction from $1.42\mathrm{e+}5$ to $4.31\mathrm{e+}4$ in RMpS. Finally, in the case depicted in Fig.~\ref{fig:result_pruning}~(c), the attained sparsity was $61\%$, which gives a reduction from $1.42\mathrm{e+}5$ to $5.58\mathrm{e+}4$ in RMpS number.

\subsection{Clustering Study}
In this section, we evaluate the weight clustering compression technique. To the best of our knowledge, this is the first time that the trade-off between optical performance and computational complexity when using such a technique in optical communications has been assessed. Note that quantization and clustering can be implemented by maintaining a codebook structure that stores the shared weights for each layer. However, in this work, we have also used weight clustering as a pre-step to simplify the problem for the next step, where the traditional quantization techniques are used. The first goal of this subsection is to assess if the weight clustering can reduce the number of multiplications without impacting the performance significantly. The BO described in Sec.~\ref{Compression_cluester} is used in this work to find the new training hyperparameters and the number of $k$-weight clusters throughout the NN-structure, so that the RMpS is given by Eq.~(\ref{Eq_cluster}).

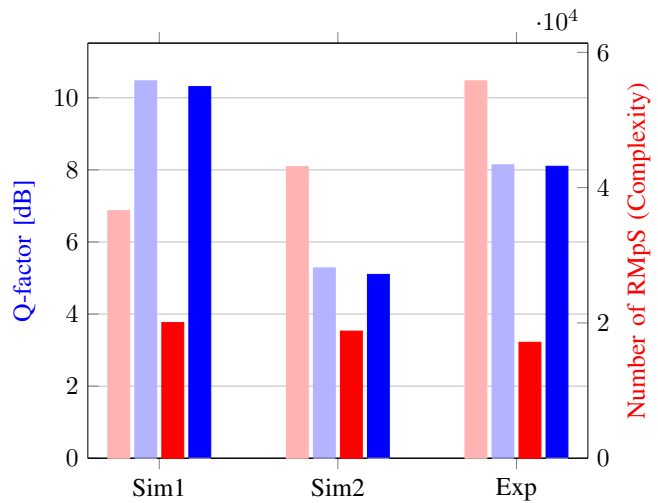
\begin{figure}[ht!]
    \centering
\begin{tikzpicture}[scale=0.97]

  \begin{axis}[
    axis y line*=left,
    ybar=5*\pgflinewidth,
    symbolic x coords={Sim1, Sim2, Exp},
    bar width=0.3cm,
    enlarge x limits=0.2,
        ymin=0,
    xtick=data,
    axis y line*=left,
    ymajorgrids = true,
    ylabel = {\textcolor{blue}{Q-factor [dB]}} ,
    ]
      \addplot[style={blue!30,fill=blue!30,mark=none}]
            coordinates {(Sim1, 0) (Sim2, 0) (Exp, 0)};
    \addplot[style={blue!30,fill=blue!30,mark=none}]
          coordinates {(Sim1, 10.47) (Sim2, 5.28) (Exp, 8.14)};
  \addplot[style={blue!30,fill=blue!30,mark=none}]
            coordinates {(Sim1, 0) (Sim2, 0) (Exp, 0)};
    \addplot[style={blue,fill=blue,mark=none}]
          coordinates {(Sim1, 10.31) (Sim2, 5.1) (Exp, 8.1)};

  \end{axis}
  \begin{axis}[
    ybar,
    bar width=0.3cm,
        enlarge x limits=0.2,
        ymin=0,
    symbolic x coords={Sim1, Sim2, Exp},
    xtick=\empty,
    axis y line*=right,
    ylabel=axis2,
    ylabel = {\textcolor{red}{Number of RMpS (Complexity)}},
    ]
    \addplot[style={red!30,fill= red!30,mark=none}]
          coordinates {(Sim1, 36595) (Sim2, 43093) (Exp, 55782)};
             \addplot[style={blue!30,fill=blue!30,mark=none}]
            coordinates {(Sim1, 0) (Sim2, 0) (Exp, 0)};
    \addplot[style={red,fill= red,mark=none}]
          coordinates {(Sim1, 20048) (Sim2, 18792) (Exp, 17135)};
            \addplot[style={blue!30,fill=blue!30,mark=none}]
            coordinates {(Sim1, 0) (Sim2, 0) (Exp, 0)};

 \end{axis}
\end{tikzpicture}
    \caption{Optical performance (blue) and complexity (red) evaluation of the pruning + clustering  (darker colors) and pruning only (lighter colors) approaches for the three considered transmission systems (Sim1, Sim2, and Exp).}
    \label{fig:clustering}
\end{figure}

Fig.~\ref{fig:clustering} depicts the impact of weight clustering on the performance and on computational complexity in the three considered transmission scenarios. This Fig. demonstrates that weight clustering leads to a small degradation in the Q-factor while still allowing us to lower the computational complexity considerably. In Sim1 in Fig.\ref{fig:clustering},  when $74$ clusters were used, we see a Q-factor degradation of $0.2$~B and a reduction in complexity from $36$k to $20$k RMpS, when compared to the pruned architecture results. In Sim2 in the same figure, we observe a similar degradation of the Q-factor and a reduction of complexity from $43$k to $19$k RMpS when $68$ clusters are used. Finally, for our experimental data, and using $62$ clusters only, we observe that the Q-factor remains mostly unchanged, and the complexity is reduced from $55$k to $17$k RMpS. We observed that clustering the weights after pruning leads to better results than clustering the original weights. Moreover, the training time is also improved in the former case since fewer parameters need to be learned during the training phase.

Now we focus our analysis on the complexity part of the weight clustering technique, i.e., how much can the number of weight clusters be reduced while still enabling relevant optical performance improvement? Only the Sim1 transmission scenario is considered in the analysis, as this is the case where nonlinear mitigation shows the most noticeable improvement. 

\begin{figure}
    \centering
\begin{tikzpicture}[scale=0.9]

  \begin{axis}[
    axis y line*=left,
    ybar=5*\pgflinewidth,
    symbolic x coords={CDC,NN(2W),NN(3W),NN(4W),DBP},
    bar width=0.15cm,
    enlarge x limits=0.1,
        ymin=0,ymax=10,
    xtick=data,
    axis y line*=left,
    ymajorgrids = true,
    ylabel = {\textcolor{black}{Q-factor [dB]}} ,
        legend style={at={(0.5,-0.25)},
anchor=north,legend columns=-1},
    ]

    \addplot[style={brown!100,fill=brown!100,mark=none}]
          coordinates {(CDC, 7.7) (NN(2W), 7.97)(NN(3W), 9)(NN(4W), 9.12) (DBP, 8.96)};
    \addplot[style={gray!100,fill=gray!100,mark=none}]
          coordinates {(CDC, 7.3) (NN(2W),8.35)(NN(3W), 9.31)(NN(4W), 9.53) (DBP,9.2)};
    \addplot[style={green!100,fill=green!100,mark=none}]
          coordinates {(CDC, 6.57) (NN(2W), 8.32)(NN(3W), 9.36)(NN(4W), 9.66) (DBP,9.1)};
    \addplot[style={blue!100,fill=blue!100,mark=none}]
          coordinates {(CDC, 5.56) (NN(2W), 7.16)(NN(3W), 8.9)(NN(4W), 9.38) (DBP, 8.6)};
\legend{-1dBm,0dBm,1dBm,2dBm}
  \end{axis}
  \begin{axis}[
    ybar,
    bar width=0.15cm,
        enlarge x limits=0.1,
        ymin=0, ymax=2000,
    symbolic x coords={CDC,NN(2W),NN(3W),NN(4W),DBP},
    xtick=\empty,
    axis y line*=right,
    ylabel=axis2,
    ylabel = {\textcolor{red}{Number of RMpS (Complexity)}},
    ]
   \addplot[draw=red, thick, dashed, smooth,mark=*,mark size=3pt, color = red]
          coordinates {(CDC, 113) (NN(2W),549)(NN(3W),820)(NN(4W),1091) (DBP,1673)};

 \end{axis}
\end{tikzpicture}
    \caption{Optical performance and complexity results when employing very aggressive weight clustering in the Sim1 transmission scenario: 2 weights clustering [NN(2W)], 3 weights clustering [NN(3W)], and 4 weights clustering [NN(4W)]. The traditional CDC and reference 1 STpS DBP results are shown as benchmark.}
    \label{fig:clustering_lowcompelxity}
\end{figure}
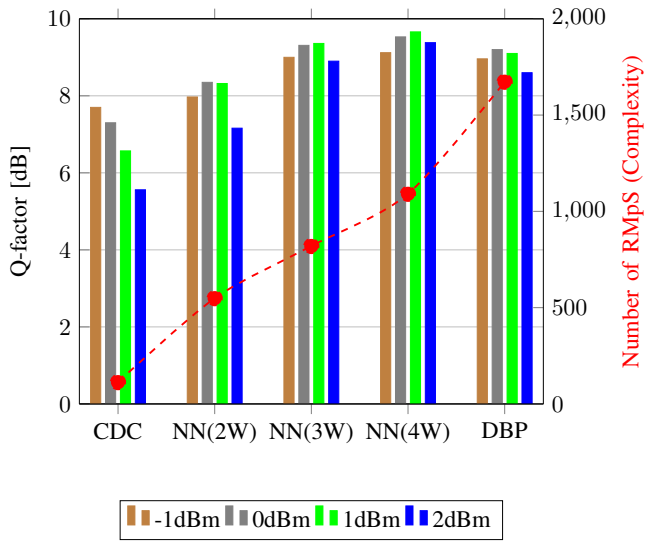

We assess the potential of the weight clustering technique when using up to four distinct weights. Launch powers in the range from -1~dBm to 2~dBm are tested to assess if the optimum launch power changes when using such an aggressive compression approach. The achieved results are compared to the ones obtained when using linear equalization only (CDC) or 1 STpS DBP. 
Fig.~\ref{fig:clustering_lowcompelxity} depicts the Q-factor for each equalization approach, as well as the number of RMpS\footnote{The sparsity of the NN structure was not preserved while doing the clustering, because we observed that by allowing the zero-value weights, where pruning removes the nodes, Fig.~\ref{fig:weightdistribuition}~(b), to acquire a different (small, but non-zero) value helped in improving the overall performance when an ultra-low number of weight clusters is used. Additionally, the training phase, in this case, took much longer (10k epochs).}. Fig.~\ref{fig:clustering_lowcompelxity}  shows that, when using the CDC, the optimum launch power is -1~dBm, leading to the Q-factor of 7.8~dB (113~RMpS are used in this case). If the reference 1~STpS DBP is used,  the optimum launch power changes to 0~dBm, enabling the Q-factor of 9.2dB but requiring 1673~RMpS. We notice that the NN-based equalizer enables outperforming the 1~STpS DBP, which is often used as a benchmark. Indeed, the NN with 4 clustered weights per layer [NN(4W) case in the figure], enables achieving a Q-factor of 9.7~dB (at 1~dBm optimum launch power) using 1091~RMpS. Instead, if only 3 clustered weights are used [NN(3W) case in the figure], a Q-factor of 9.4~dB (at 1~dBm) can be reached, requiring 820~RMpS only. As expected, using 2 clustered weights [NN(2W) case in the figure] leads to the worst performance, where we can achieve the Q-factor of 8.4~dB (at 0~dBm), thus still outperforming the CDC, but at the expense of 549~RMpS complexity.

Finally, if we consider that the multiplier complexity is proportional to $b_i b_w$ (as described in Sec.~\ref{sec:complexity metric}), the savings in complexity enabled by the clustering technique can be even higher than the ones indicated above. Indeed, considering that $b_i=8$ bits for all cases, the coefficients of the CDC and DBP filters are also represented by $8$ bits. However, for the NN with 3 and 4 clustered weights, we can encode the weights using just a 2-bit format. So, for the cases of 3W and 4W NN, which performed better than 1 STpS DBP, the complexity calculated as $ \mathrm{RMpS}*b_i*b_w$, is just 1.82 and 2.42 times higher than the CDC one (and 8.15 and 6.13 times lower than the 1~STpS DBP), respectively.  Here we note that the CDC benchmark is the most important because our primary goal is to show the readiness of NN with respect to the already available algorithm in commercial transponders. In contrast, none of the existing DBP versions has reached the hardware level of implementation. In this context, Fig.~\ref{fig:clustering_lowcompelxity} and the previous analysis shows that an NN-based equalizer achieves a performance close to that obtained with the ``DBP''~\cite{napoli2014reduced}, while approaching the complexity of the CDC block.

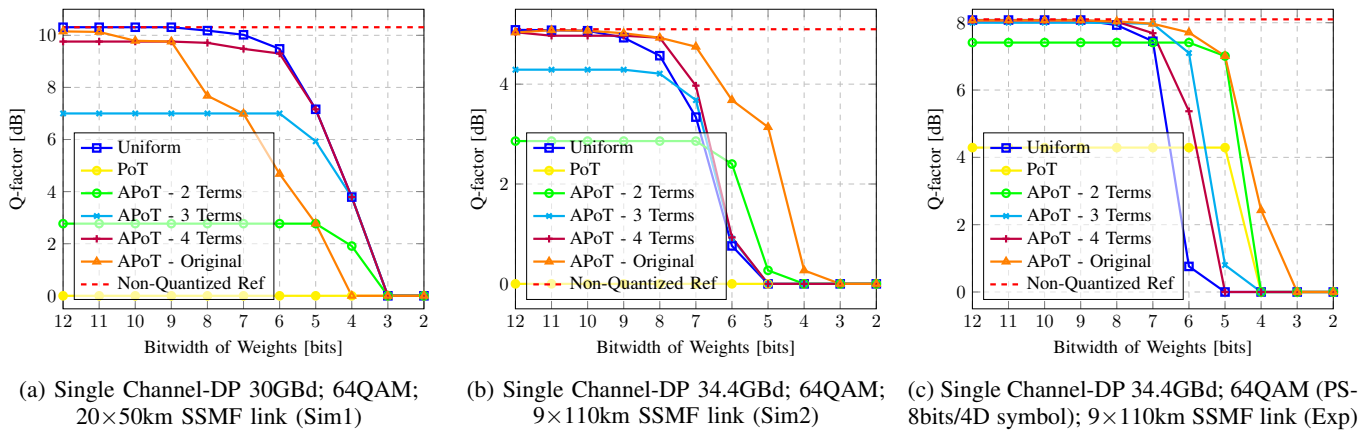
\begin{figure*}[ht!] 
\begin{subfigure}{.33\textwidth}
    \centering
 \begin{tikzpicture}[scale=0.7]
    \begin{axis} [ylabel={BER}, 
        xlabel={Bitwidth of Weights [bits]},
        ylabel={Q-factor [dB]},
        grid=both,  
        xmin=2, xmax=12,
    	xtick={12, ..., 2},
          x axis line style={-},
          x post scale=-1,
    	ymin=-0.5, ymax=11,
        legend style={legend pos=south west, legend cell align=left,fill=white, fill opacity=0.6, draw opacity=1,text opacity=1},
    	grid style={dashed}]
        ]
        \addplot[color=blue, mark=square, very thick]     coordinates {
    (2,0)(3,0)(4,3.79)(5,7.16) (6,9.48) (7,10.02) (8,10.18) (9,10.31) (10,10.31) (11,10.31) (12,10.31)    
    };
    \addlegendentry{Uniform};
    
    \addplot[color=yellow, mark=*, very thick]   
    coordinates {
    (2,0)(3,0)(4,0)(5,0) (6,0) (7,0) (8,0) (9,0) (10,0) (11,0) (12,0) 
    };
    \addlegendentry{PoT};

    \addplot[color=green, mark=o, very thick]     coordinates {
    (2,0)(3,0)(4,1.91)(5,2.77) (6,2.77) (7,2.77) (8,2.77) (9,2.77) (10,2.77) (11,2.77) (12,2.77) 

    };
    \addlegendentry{APoT - 2 Terms};

    \addplot [color=cyan, mark=x, very thick]    coordinates {
    (2,0)(3,0)(4,3.79)(5,5.93) (6,7) (7,7) (8,7) (9,7) (10,7) (11,7) (12,7) 

    };
    \addlegendentry{APoT - 3 Terms};
    
    \addplot [color=purple, mark=+, very thick]    coordinates {
    (2,0)(3,0)(4,3.79)(5,7.16) (6,9.3) (7,9.48) (8,9.71) (9,9.76) (10,9.76) (11,9.76) (12,9.76) 

    };
    \addlegendentry{APoT - 4 Terms};

    \addplot [color=orange, mark=triangle, very thick]    coordinates {
    (2,0)(3,0)(4,0)(5,2.77) (6,4.68) (7,6.99) (8,7.68) (9,9.76) (10,9.79) (11,10.13) (12,10.15) 

    };
    \addlegendentry{APoT - Original};
    
    \addplot[color=red, dashed, very thick]    coordinates {
    (2,10.31)(3,10.31)(4,10.31)(5,10.31) (6,10.31) (7,10.31) (8,10.31) (9,10.31) (10,10.31) (11,10.31) (12,10.31) 
    };
    \addlegendentry{Non-Quantized Ref};
    
    \end{axis}
    \end{tikzpicture}
    \caption{Single Channel-DP 30GBd; 64QAM;\\20$\times$50km SSMF link (Sim1)}
    \label{fig:res_a2} 
\end{subfigure}\hfill
\begin{subfigure}{.33\textwidth}
    \centering
 \begin{tikzpicture}[scale=0.7]
    \begin{axis} [ylabel={BER}, 
        xlabel={Bitwidth of Weights [bits]},
        ylabel={Q-factor [dB]},
        grid=both,  
        xmin=2, xmax=12,
    	xtick={12, ..., 2},
          x axis line style={-},
          x post scale=-1,
    	ymin=-0.5, ymax=5.5,
        legend style={legend pos=south west, legend cell align=left,fill=white, fill opacity=0.6, draw opacity=1,text opacity=1},
    	grid style={dashed}]
        ]
        \addplot[color=blue, mark=square, very thick]     coordinates {
    (2,0)(3,0)(4,0)(5,0) (6,0.76) (7,3.34) (8,4.57) (9,4.93) (10,5.07) (11,5.08) (12,5.087265463) 
    };
    \addlegendentry{Uniform};
    
    \addplot[color=yellow, mark=*, very thick]   
    coordinates {
    (2,0)(3,0)(4,0)(5,0) (6,0) (7,0) (8,0) (9,0) (10,0) (11,0) (12,0) 
    };
    \addlegendentry{PoT};

    \addplot[color=green, mark=o, very thick]     coordinates {
    (2,0)(3,0)(4,0)(5,0.27) (6,2.4) (7,2.86) (8,2.86) (9,2.86) (10,2.86) (11,2.86) (12,2.86)

    };
    \addlegendentry{APoT - 2 Terms};

    \addplot [color=cyan, mark=x, very thick]    coordinates {
    (2,0)(3,0)(4,0)(5,0) (6,0.91) (7,3.68) (8,4.21) (9,4.29) (10,4.29) (11,4.29) (12,4.29) 
    };
    \addlegendentry{APoT - 3 Terms};
    
    \addplot [color=purple, mark=+, very thick]    coordinates {
    (2,0)(3,0)(4,0)(5,0) (6,0.93) (7,3.97) (8,4.93) (9,4.97) (10,4.97) (11,4.97) (12,5.04) 
    };
    \addlegendentry{APoT - 4 Terms};

    \addplot [color=orange, mark=triangle, very thick]    coordinates {
    (2,0)(3,0)(4,0.27)(5,3.14) (6,3.68) (7,4.75) (8,4.93) (9,5.02) (10,5.07) (11,5.08) (12,5.036) 

    };
    \addlegendentry{APoT - Original};
    
    \addplot[color=red, dashed, very thick]   coordinates {
    (2,5.1)(3,5.1)(4,5.1)(5,5.1) (6,5.1) (7,5.1) (8,5.1) (9,5.1) (10,5.1) (11,5.1) (12,5.1) 
    };
    \addlegendentry{Non-Quantized Ref};
    
    \end{axis}
    \end{tikzpicture}
    \caption{Single Channel-DP 34.4GBd; 64QAM;\\9$\times$110km SSMF link (Sim2)}    \label{fig:res_b2} 
\end{subfigure}\hfill
\begin{subfigure}{.33\textwidth}
    \centering
 \begin{tikzpicture}[scale=0.7]
    \begin{axis} [ylabel={BER}, 
        xlabel={Bitwidth of Weights [bits]},
        ylabel={Q-factor [dB]},
        grid=both,  
        xmin=2, xmax=12,
    	xtick={12, ..., 2},
          x axis line style={-},
          x post scale=-1,
    	ymin=-0.5, ymax=8.4,
        legend style={legend pos=south west, legend cell align=left,fill=white, fill opacity=0.6, draw opacity=1,text opacity=1},
    	grid style={dashed}]
        ]
        \addplot[color=blue, mark=square, very thick]     coordinates {
    (2,0)(3,0)(4,0)(5,0) (6,0.76) (7,7.45) (8,7.93) (9,8.08) (10,8.08) (11,8.08) (12,8.08) 
    };
    \addlegendentry{Uniform};
    
    \addplot[color=yellow, mark=*, very thick]   
    coordinates {
    (2,0)(3,0)(4,0)(5,4.29) (6,4.29) (7,4.29) (8,4.29) (9,4.29) (10,4.29) (11,4.29) (12,4.29)  
    };
    \addlegendentry{PoT};

    \addplot[color=green, mark=o, very thick]     coordinates {
    (2,0)(3,0)(4,0)(5,7.01) (6,7.41) (7,7.41) (8,7.41) (9,7.41) (10,7.41) (11,7.41) (12,7.41) 
    };
    \addlegendentry{APoT - 2 Terms};

    \addplot [color=cyan, mark=x, very thick]    coordinates {
    (2,0)(3,0)(4,0)(5,0.8) (6,7.1) (7,7.97) (8,7.99) (9,8) (10,8) (11,8) (12,8) 
    };
    \addlegendentry{APoT - 3 Terms};
    
    \addplot [color=purple, mark=+, very thick]    coordinates {
    (2,0)(3,0)(4,0)(5,0) (6,5.37) (7,7.69) (8,8.03) (9,8.07) (10,8.07) (11,8.07) (12,8.07) 
    };
    \addlegendentry{APoT - 4 Terms};

    \addplot [color=orange, mark=triangle, very thick]    coordinates {
    (2,0)(3,0)(4,2.43)(5,7.01) (6,7.71) (7,7.97) (8,8.04) (9,8.072) (10,8.072) (11,8.072) (12,8.072) 
    };
    \addlegendentry{APoT - Original};
    
    \addplot[color=red, dashed, very thick]    coordinates {
    (2,8.1)(3,8.1)(4,8.1)(5,8.1) (6,8.1) (7,8.1) (8,8.1) (9,8.1) (10,8.1) (11,8.1) (12,8.1) 
    };
    \addlegendentry{Non-Quantized Ref};
    
    \end{axis}
    \end{tikzpicture}
    \caption{Single Channel-DP 34.4GBd; 64QAM (PS-8bits/4D symbol); 9$\times$110km SSMF link (Exp)}    \label{fig:res_c2} 
\end{subfigure}

  \caption{Performance of the Post Training Quantization (Homogeneous Approach).}
  \label{fig:result_PTQ} 
\end{figure*}

\begin{figure*}[hb!] 
\begin{subfigure}{.3\textwidth}
    \centering
 \begin{tikzpicture}[scale=0.7]
    \begin{axis} [ylabel={BoP}, 
        xlabel={Bitwidth of Weights [bits]},
        grid=both,  
        xmin=2, xmax=12,
    	xtick={12, ..., 2},
          x axis line style={-},
          x post scale=-1,
        legend style={legend pos=south west, legend cell align=left,fill=white, fill opacity=0.6, draw opacity=1,text opacity=1},
    	grid style={dashed}]
        ]
    \addplot[color=blue, mark=square, very thick]     coordinates {
    (2, 7624068)(3, 9799092)(4, 11974116)(5, 14149139)(6, 16324163)(7, 18499187)(8, 20674210)(9, 22849234)(10, 25024258)(11, 27199282)(12, 29374305)  
    };
    \addlegendentry{20x50km SSMF link 64QAM 30GBd};
            
    \addplot[color=red, mark=square, very thick]     coordinates {
    (2, 8306735)(3, 10703876)(4, 13101017)(5, 15498157)(6, 17895298)(7, 20292439)(8, 22689580)(9, 25086721)(10, 27483862)(11, 29881002)(12, 32278143)
    };
    \addlegendentry{90x100km SSMF link 64QAM 34GBd};

    \end{axis}
    \end{tikzpicture}
    \caption{BoPs for the NN used in Sim1/Sim2.}    \label{fig:BOP_uniform} 
\end{subfigure}\hfill
\begin{subfigure}{.3\textwidth}
    \centering
 \begin{tikzpicture}[scale=0.7]
    \begin{axis} [ylabel={NABS}, 
        xlabel={Bitwidth of Weights [bits]},
        grid=both, 
        ymode=log,
        xmin=2, xmax=12,
    	xtick={12, ..., 2},
          x axis line style={-},
          x post scale=-1,
    	ymin=2500000, ymax=70000000,
        legend style={legend pos=north east, legend cell align=left,fill=white, fill opacity=0.6, draw opacity=1,text opacity=1},
    	grid style={dashed}]
        ]
    \addplot[color=blue, mark=square, very thick]     coordinates {
    (2, 3350228)(3, 6926390)(4, 10758407)(5, 14846278)(6, 19190005)(7, 23789587)(8, 28645023)(9, 33756315)(10, 39123461)(11, 44746463)(12, 50625320) 
    };
    \addlegendentry{Uniform};
    
    \addplot[color=yellow, mark=square, very thick]     coordinates {
    (2, 3350228)(3, 3478154)(4, 3606079)(5, 3734005)(6, 3861930)(7, 3989856)(8, 4117781)(9, 4245707)(10, 4373632)(11, 4501558)(12, 4629483)
    };
    \addlegendentry{PoT};
    
    \addplot[color=green, mark=square, very thick]     coordinates {
    (2, 3350228)(3, 6926390)(4, 7182243)(5, 7438096)(6, 7693949)(7, 7949802)(8, 8205655)(9, 8461508)(10, 8717361)(11, 8973214)(12, 9229067)
    };
    \addlegendentry{APoT - 2 Terms};
    
    \addplot[color=cyan, mark=square, very thick]     coordinates {
    (2, 3350228)(3, 6926390)(4, 10758407)(5, 11142187)(6, 11525968)(7, 11909748)(8, 12293529)(9, 12677309)(10, 13061090)(11, 13444870)(12, 13828650)
    };
    \addlegendentry{APoT - 3 Terms};
    
    \addplot[color=orange, mark=square, very thick]     coordinates {
    (2, 3350228)(3, 6926390)(4, 10758407)(5, 14846278)(6, 15357986)(7, 15869694)(8, 16381402)(9, 16893110)(10, 17404818)(11, 17916526)(12, 18428234)
    };
    \addlegendentry{APoT - 4 Terms};
    
    \addplot[color=red, mark=square, very thick]     coordinates {
    (2, 3350228)(3, 6926390)(4, 7182243)(5, 11142187)(6, 11525968)(7, 15869694)(8, 16381402)(9, 21108911)(10, 21748547)(11, 26859838)(12, 27627401)
    };
    \addlegendentry{APoT - Original};
    \end{axis}
    \end{tikzpicture}
    \caption{ NABS for the NN used in Sim1.}    \label{fig:NABS_11} 
\end{subfigure}\hfill
\begin{subfigure}{.3\textwidth}
    \centering
 \begin{tikzpicture}[scale=0.7]
    \begin{axis} [ylabel={NABS}, 
        xlabel={Bitwidth of Weights [bits]},
        grid=both, 
        ymode=log,
        xmin=2, xmax=12,
    	xtick={12, ..., 2},
          x axis line style={-},
          x post scale=-1,
    	ymin=2500000, ymax=70000000,
        legend style={legend pos=north east, legend cell align=left,fill=white, fill opacity=0.6, draw opacity=1,text opacity=1},
    	grid style={dashed}]
        ]
    \addplot[color=blue, mark=square, very thick]     coordinates {
    (2, 3610121)(3, 7471525)(4, 11614914)(5, 16040289)(6, 20747649)(7, 25736995)(8, 31008327)(9, 36561644)(10, 42396947)(11, 48514235)(12, 54913509)
    };
    \addlegendentry{Uniform};
    
    \addplot[color=yellow, mark=square, very thick]     coordinates {
    (2, 3610121)(3, 3751112)(4, 3892103)(5, 4033093)(6, 4174084)(7, 4315075)(8, 4456066)(9, 4597057)(10, 4738048)(11, 4879038)(12, 5020029)
    };
    \addlegendentry{PoT};
    
    \addplot[color=green, mark=square, very thick]     coordinates {
    (2, 3610121)(3, 7471525)(4, 7753508)(5, 8035492)(6, 8317475)(7, 8599459)(8, 8881443)(9, 9163426)(10, 9445410)(11, 9727394)(12, 10009377)
    };
    \addlegendentry{APoT - 2 Terms};
    
    \addplot[color=cyan, mark=square, very thick]     coordinates {
    (2, 3610121)(3, 7471525)(4, 11614914)(5, 12037890)(6, 12460867)(7, 12883843)(8, 13306819)(9, 13729796)(10, 14152772)(11, 14575749)(12, 14998725)
    };
    \addlegendentry{APoT - 3 Terms};
    
    \addplot[color=orange, mark=square, very thick]     coordinates {
    (2, 3610121)(3, 7471525)(4, 11614914)(5, 16040289)(6, 16604258)(7, 17168227)(8, 17732196)(9, 18296165)(10, 18860135)(11, 19424104)(12, 19988073)
    };
    \addlegendentry{APoT - 4 Terms};
    
    \addplot[color=red, mark=square, very thick]     coordinates {
    (2, 3610121)(3, 7471525)(4, 7753508)(5, 12037890)(6, 12460867)(7, 17168227)(8, 17732196)(9, 22862535)(10, 23567497)(11, 29120814)(12, 29966769)
    };
    \addlegendentry{APoT - Original};
    \end{axis}
    \end{tikzpicture}
    \caption{ NABS for the NN used in Sim2/Exp.}    \label{fig:NABS_12} 
\end{subfigure}
  \caption{Computational complexity when using uniform quantization (a) and when different types of quantization are used (b/c). $b_i$ and $b_a$ have 16 bit precision whereas $b_w$ has a value in the range of 12 to 2 bits.}
  \label{fig:bbbbb} 
\end{figure*}
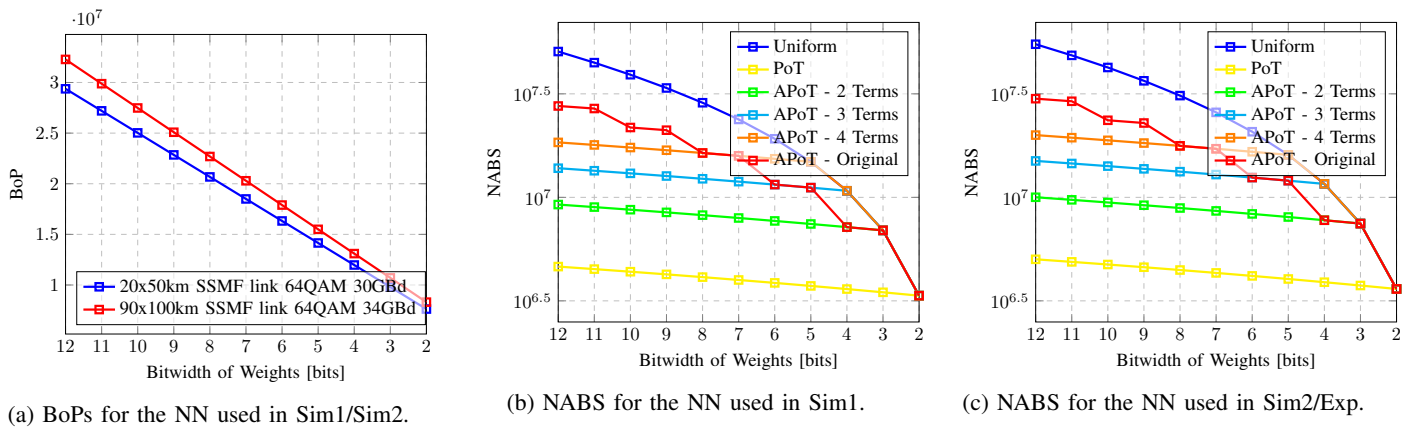

\subsection{Quantization Study}
Quantization is the other approach considered in this work to significantly reduce the computational complexity of equalizers. The PTQ homogeneous, PTQ heterogeneous, QAT homogeneous, and QAT heterogeneous approaches are considered in this subsection, see Sec. \ref{sec:quant-theor} for the approach details. 
In each case, a combined biLSTM+CNN equalizer whose weights have undergone the clustering procedure depicted in Fig.~\ref{fig:clustering} is quantized.
The performance and complexity in terms of BoPs and NABSs of the different quantization techniques are further assessed for different bit precisions.

\subsubsection{PTQ homogeneous approach} \label{sec:PTQ_Homog_subsect}
We start by assuming that all weights in the structure are quantized uniformly and with the same bit precision. 
Fig.~\ref{fig:result_PTQ} depicts the Q-factor as a function of the bit precision for the three considered transmission scenarios and using the APoT with 2, 3, and 4 additive terms quantization technique as well as the original version in \cite{li2019additive}, the uniform quantization and PoT\footnote{We have established a floor value of 0dB for the Q-factor since a lower Q-factor just means the information is completely corrupted.}. 

From the results depicted in Fig.~\ref{fig:result_PTQ}, we underline the noticeable impact of sparsity. The PoT and APoT techniques were purposely designed to have the majority of the quantization levels close to zero, since the weight distribution after training also shows a concentration of weights close to zero value, see Fig.~\ref{fig:weightdistribuition}. However, when pruning the NN structure, such weights are removed, Fig.~\ref{fig:weightdistribuition}~(b), the quantization levels above the pruning threshold are no longer used and, more importantly, the remaining weights are underrepresented. Consequently, the uniform quantization shows the best performance (for the reduced bitwidth of the weights) in the case shown in Fig.~\ref{fig:result_PTQ}~(a), where the sparsity is 72\%, whereas the APoT and POT reveal a better performance in the scenario depicted in Fig.~\ref{fig:result_PTQ}~(c), where the sparsity is 60\%. Interestingly, up to 8 bits precision, we could always find a quantization scheme that provides similar optical performance as when using the original 32 bits precision for the three considered transmission scenarios. 
For a high precision bitwidth, the uniform quantization always shows superior results, whereas for a lower bit precision (say, for less than 8 bits) and when the weight distribution is not heavily compromised by sparsity, the original APoT introduced in \cite{li2019additive} results in the best performance.  Here we highlight that, when doing with the PTQ strategy, the weight distribution must serve as the main indicator to select the best type of quantization. Also, note that, the POT has performed badly in all cases studied in this subsection. As described in Ref.~\cite{li2019additive}, the PoT quantization does not benefit even in the case from more bits, as we also observed in our work. The PoT quantization has a rigid resolution, in which by adding an extra bit, all new quantization levels concentrate around 0 and, thus, the PoT cannot increase the model’s expressiveness efficiently enough, as one would expect by addition more bits\footnote{This problem can be partially solved by training further the weights after approximating, as described next in the QAT section.}.
 
Now we assess the computational complexity of the different quantization techniques. In our analysis, we considered $b_i=b_a=16$ bits and $b_w$ changing between 12 and 2 bits. Fig.~\ref{fig:BOP_uniform} depicts the BoP metric for equally compressed models, showing that the BoP decreases almost linearly with the value of $b_{w}$. Since the Sim2 transmission scenario requires a NN structure with more hidden units and CNN filters than Sim1, the number of BoPs for this case is also higher than that for Sim1.  In this analysis, we are evaluating the total number of operations needed, as it is usually done in the literature \cite{hawks2021ps}, and therefore, do not account for the benefits stemming from weight clustering. 

\begin{figure*}[ht!]
  \centering
\begin{subfigure}{.32\textwidth}
  \centering
\includegraphics[width=0.99\textwidth]{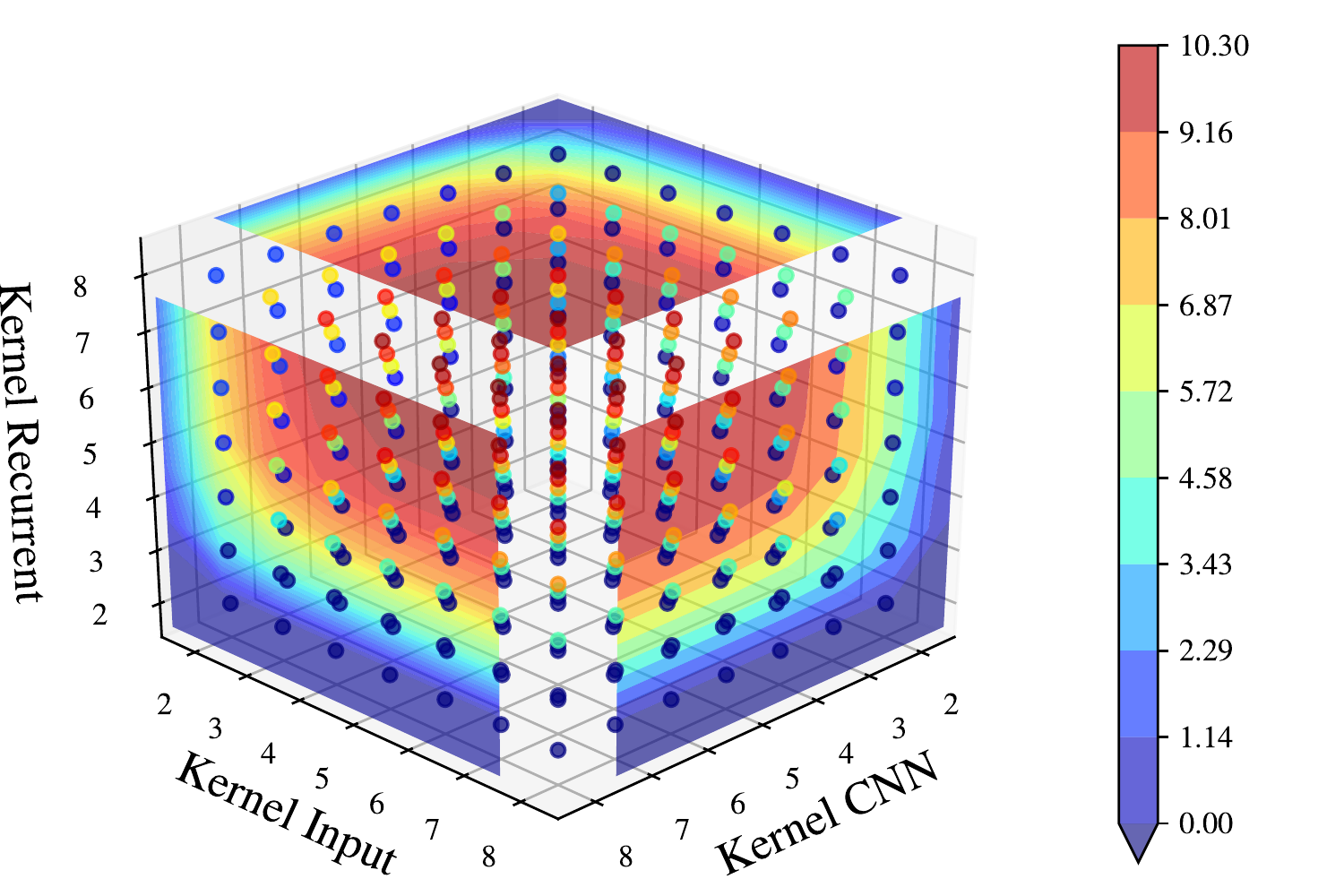}
\caption{Uniform Quantization (Sim1).}  
\end{subfigure}
\hfill
\begin{subfigure}{.32\textwidth}
    \centering
\includegraphics[width=0.99\textwidth]{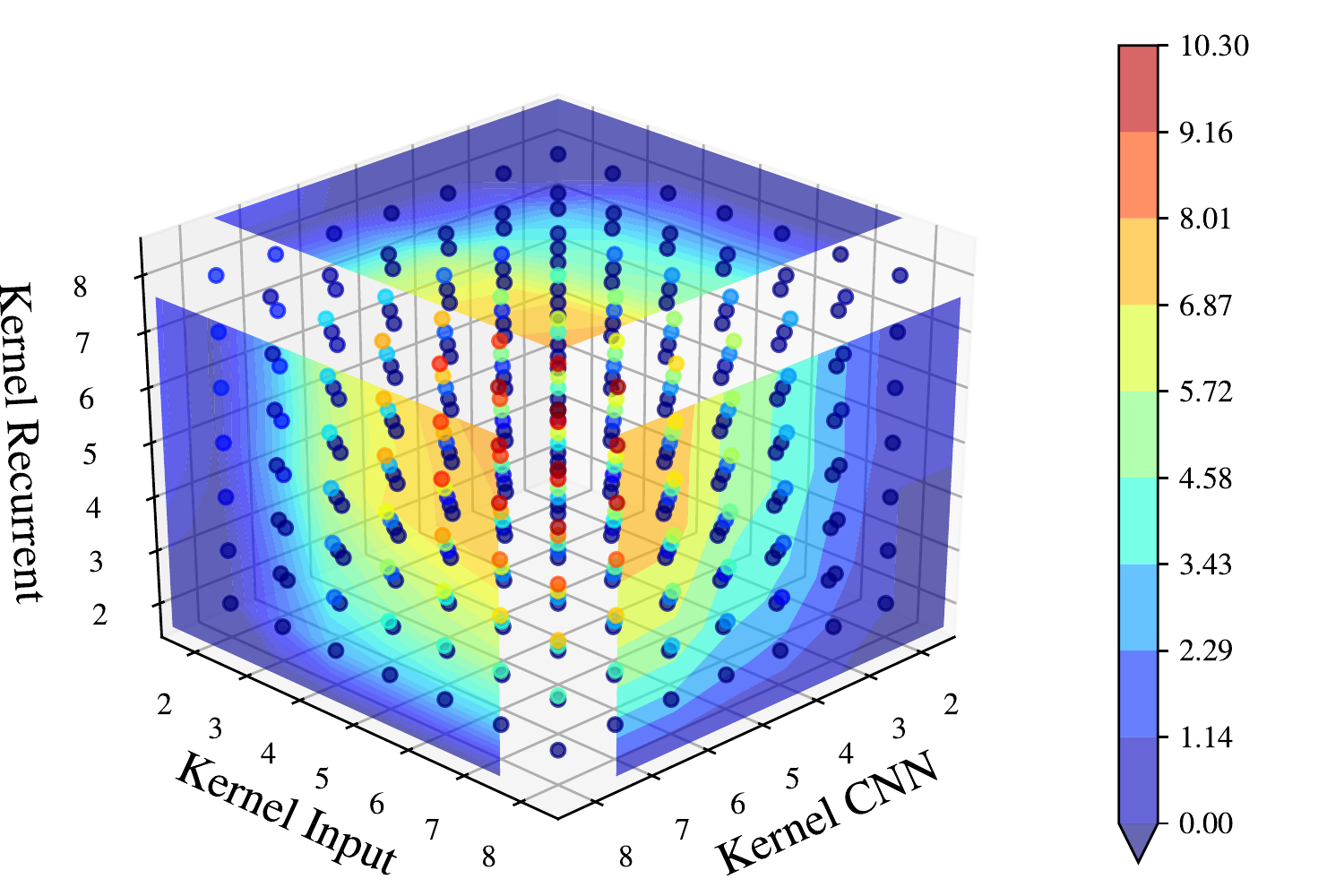}
\caption{APoT Original (Sim1).}
\end{subfigure}
\hfill
\begin{subfigure}{.32\textwidth}
    \centering
\includegraphics[width=0.99\textwidth]{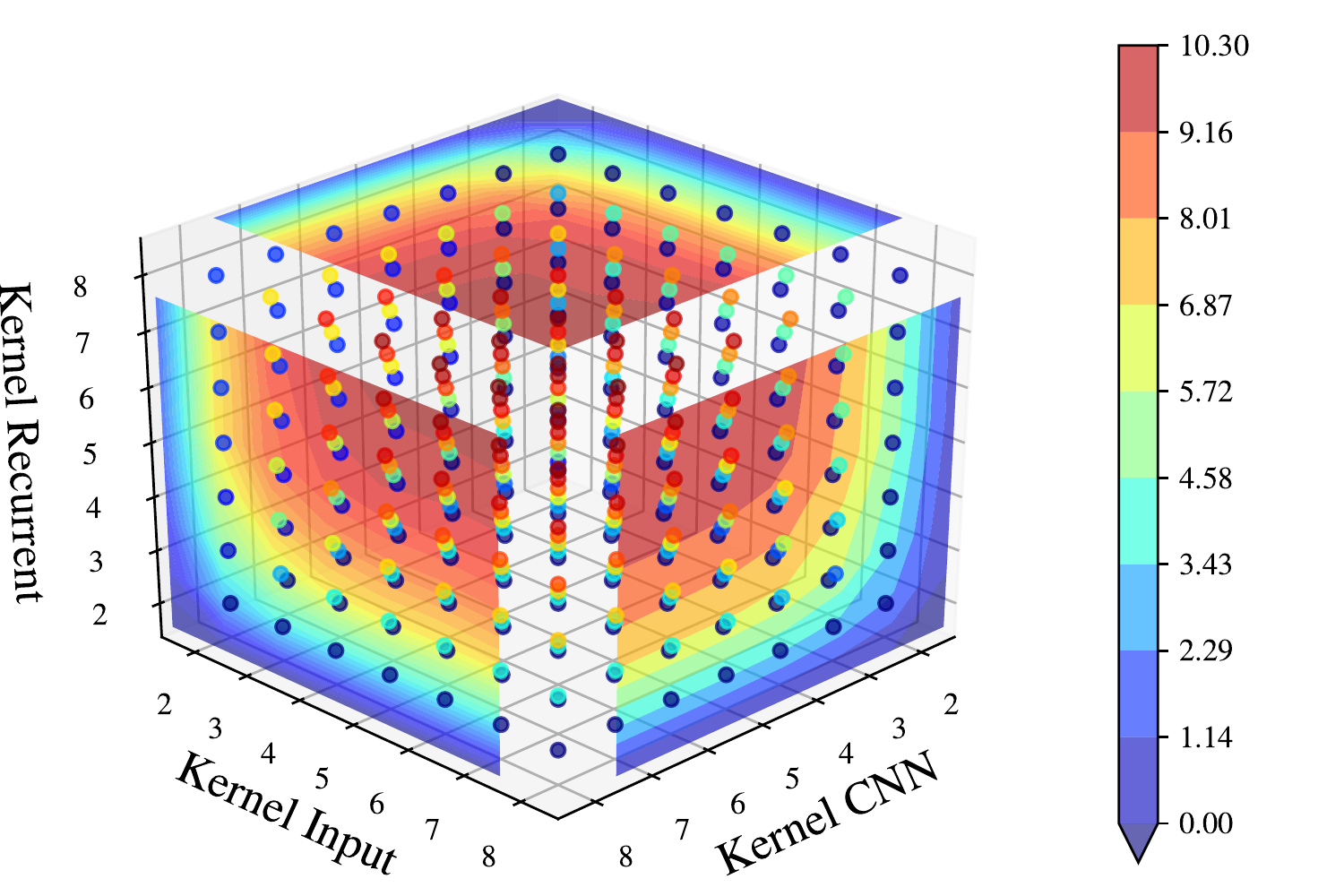}
\caption{Mixed Quantization (Sim1).}
\end{subfigure}
\medskip
\begin{subfigure}{.32\textwidth}
  \centering
\includegraphics[width=0.99\textwidth]{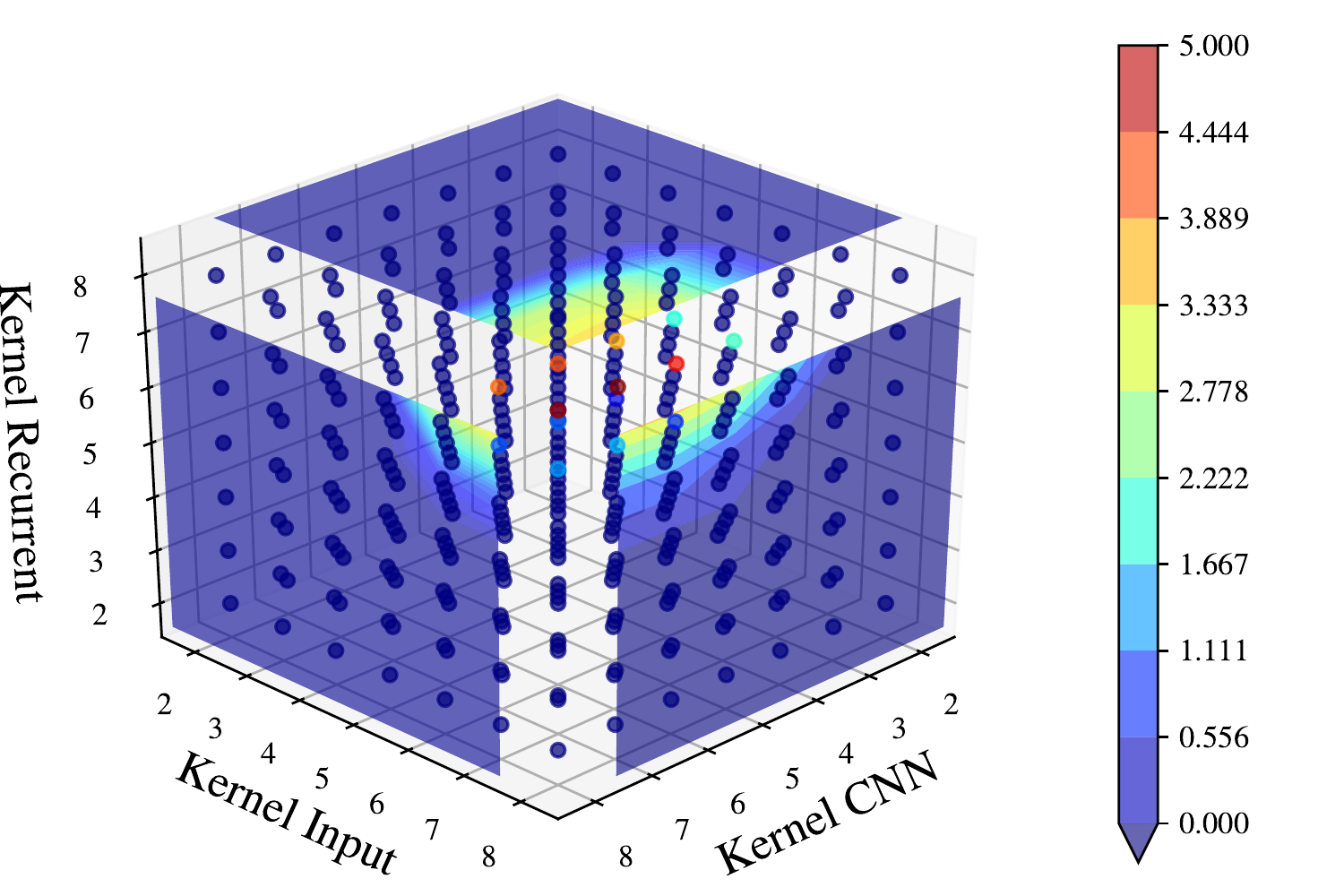}
\caption{Uniform Quantization (Sim2).}  
\end{subfigure}
\hfill
\begin{subfigure}{.32\textwidth}
    \centering
\includegraphics[width=0.99\textwidth]{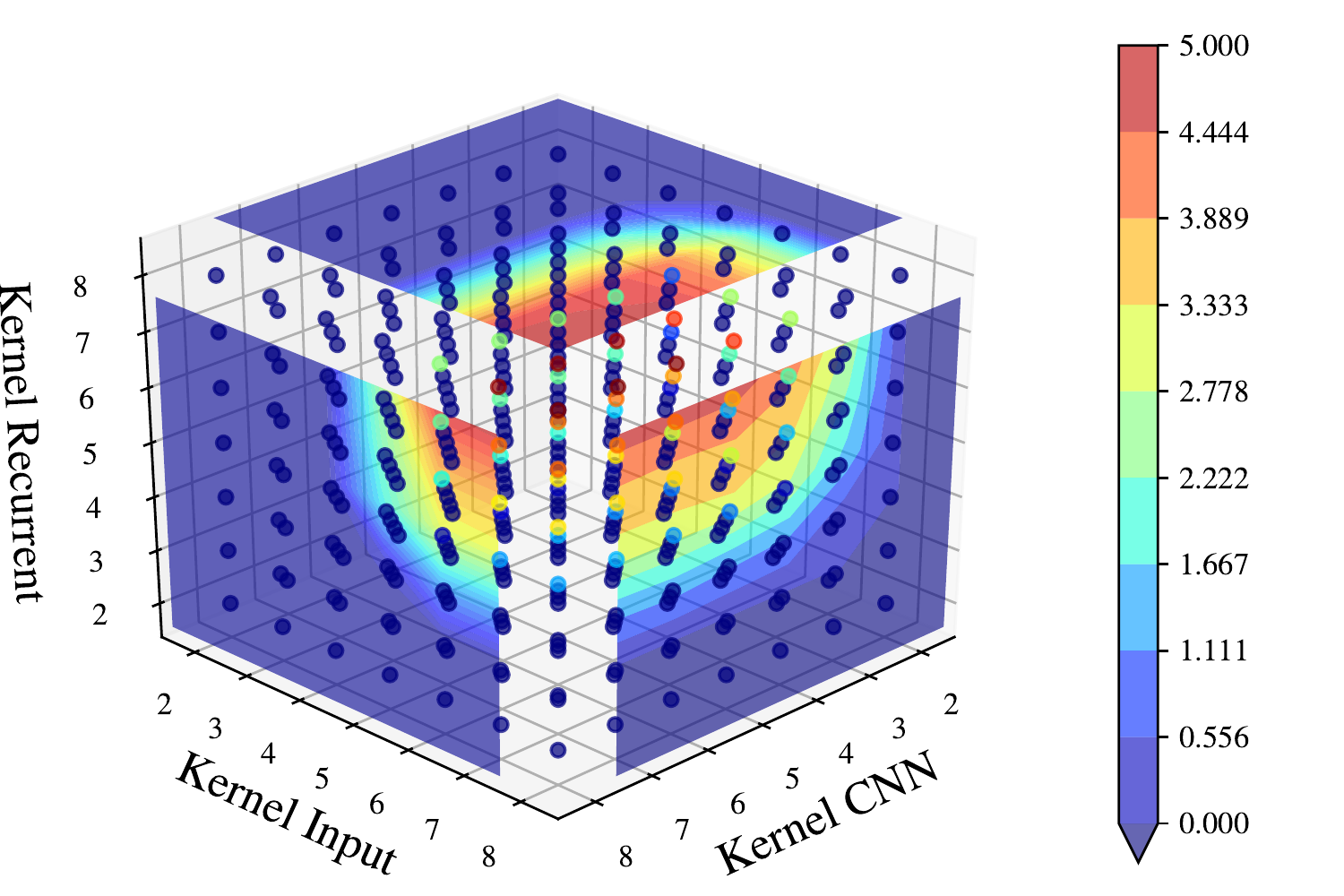}
\caption{APoT Original (Sim2).}
\end{subfigure}
\hfill
\begin{subfigure}{.32\textwidth}
    \centering
\includegraphics[width=0.99\textwidth]{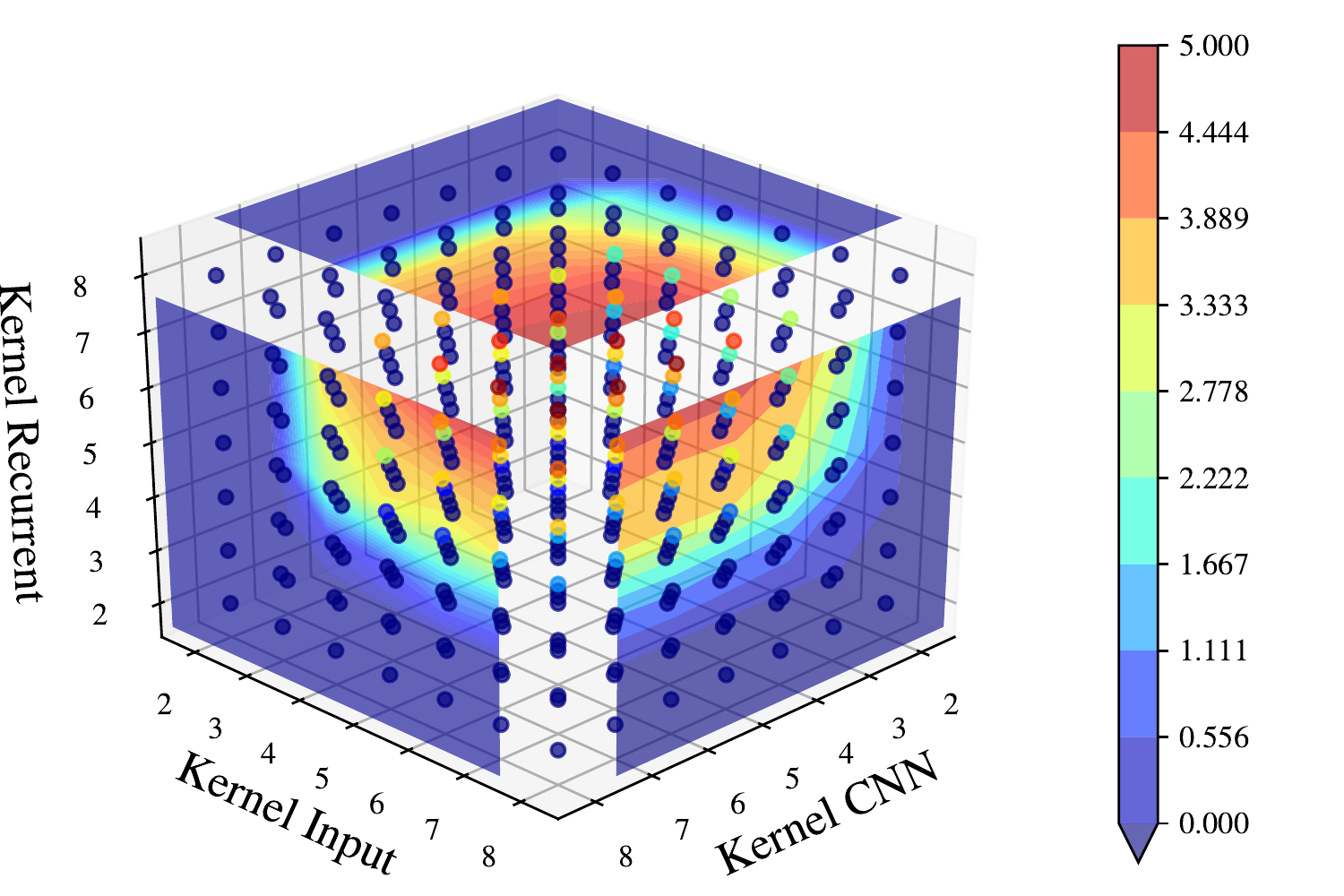}
\caption{Mixed Quantization (Sim2).}
\end{subfigure}
\medskip
\begin{subfigure}{.32\textwidth}
  \centering
\includegraphics[width=0.99\textwidth]{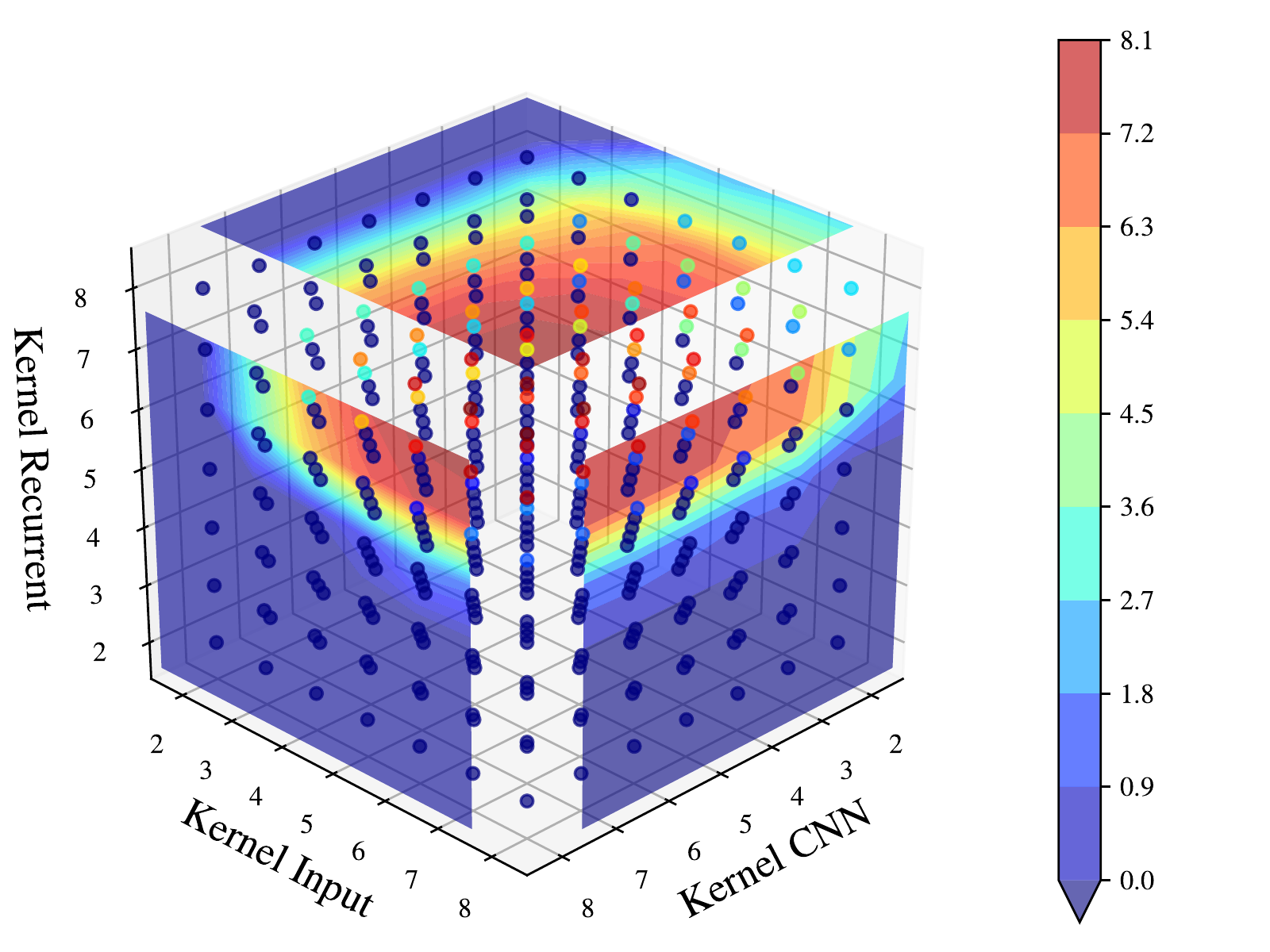}
\caption{Uniform Quantization (Exp).}  
\end{subfigure}
\hfill
\begin{subfigure}{.32\textwidth}
    \centering
\includegraphics[width=0.99\textwidth]{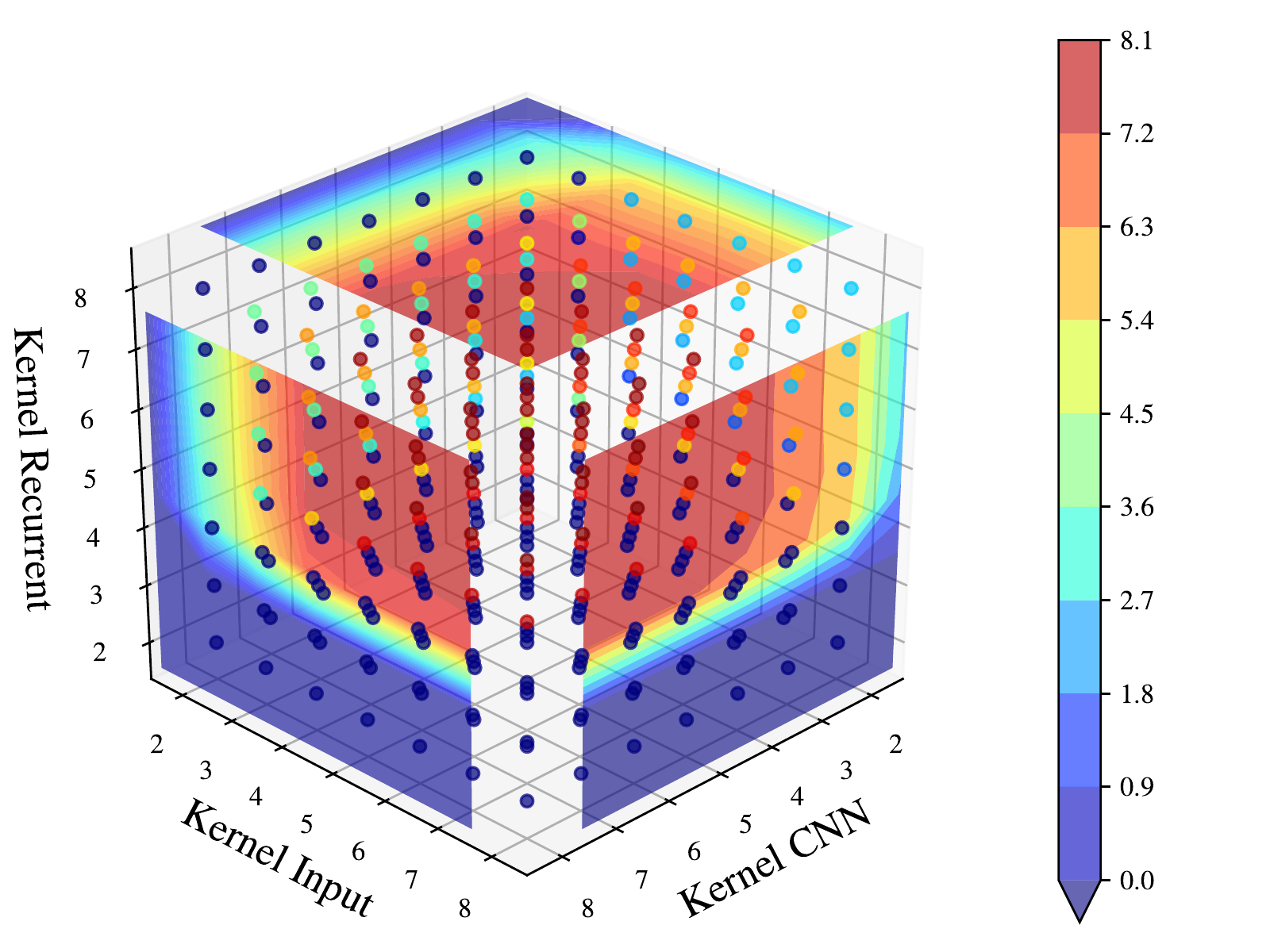}
\caption{APoT Original (Exp).}
\end{subfigure}
\hfill
\begin{subfigure}{.32\textwidth}
    \centering
\includegraphics[width=0.99\textwidth]{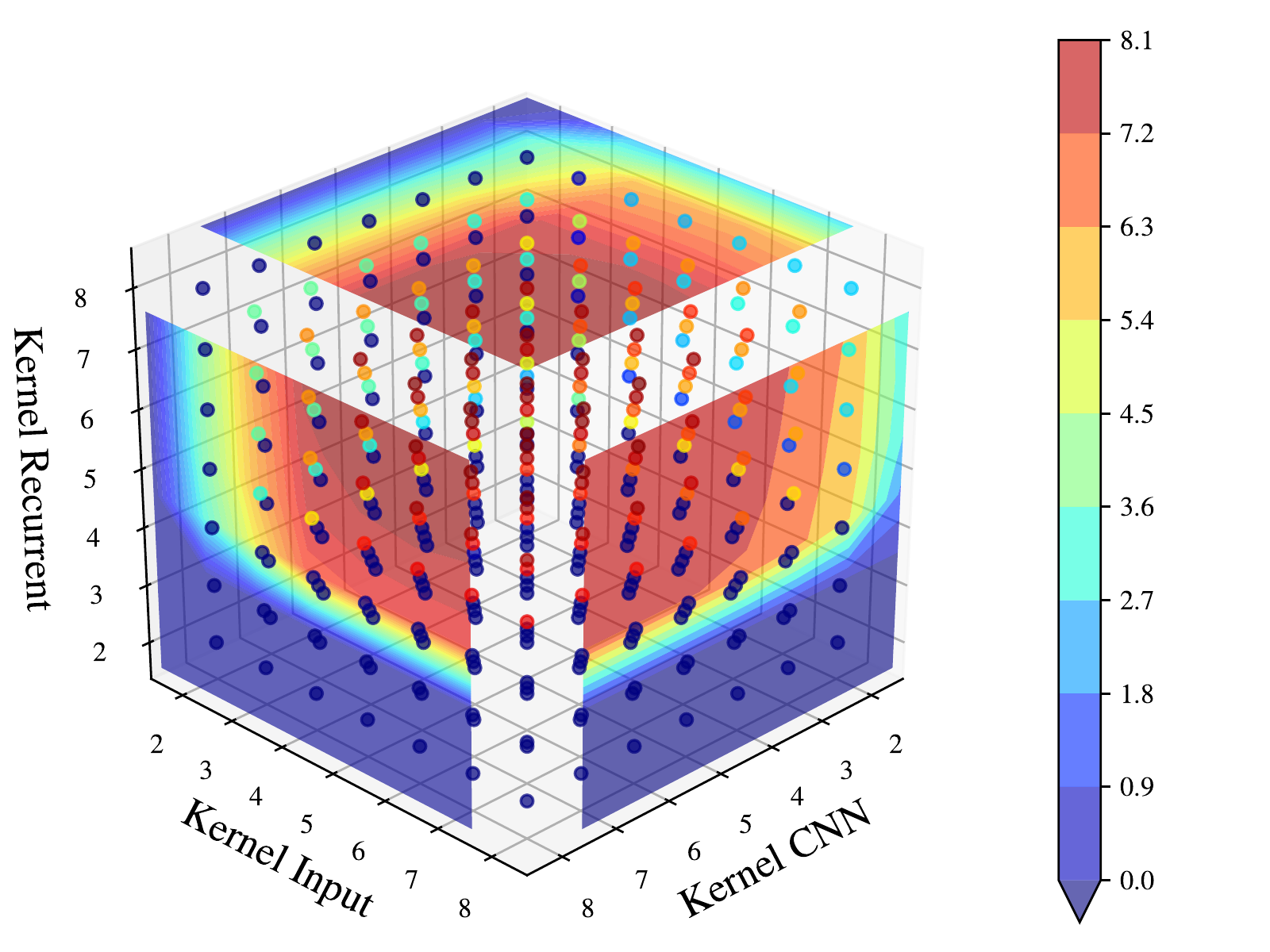}
\caption{Mixed Quantization (Exp).}
\end{subfigure}
\caption{Performance of the Post Training Quantization (Heterogeneous Approach).}

\label{fig:Quantiztion_cube}
\end{figure*}

Nevertheless, and as was mentioned in Sec.~\ref{sec:complexity metric}, when comparing the use of different quantization strategies and bitwidth precisions, the BoPs metric can not be recommended, as it, actually, does not account for the effect resulting from different quantization strategies. To have a better metric, the NABS metric ought to be used, as it allows us to compare the result of the model compression in terms of the number of additions and bit shifts. Figs.~\ref{fig:NABS_11} and~\ref{fig:NABS_12} show the NABS as a function of $b_{w}$ (the bit-precision) for the different quantization strategies employed in this work. As expected, the uniform quantization technique leads to the highest complexity, whereas the PoT quantization gives the smallest one. The APoT quantization leads to a complexity in-between the uniform and PoT approaches. In the APoT case, the complexity depends not only on $b_{w}$, but also on the number of additive terms that are considered. Interestingly, the least complex APoT strategies, i.e., those with the smaller number of additive terms, are the ones leading to better performance for the low bit precision region, see Fig.~\ref{fig:result_PTQ}. It is also interesting to note that, when we reduce the bit precision to 5 bits, which already leads to high-performance degradation, the NABS using uniform quantization becomes the same as that when using the APoT with 4 additive terms. If the bit precision is further reduced to 4 bits, the NABS using uniform quantization is the same as that when using the APoT with 3 additive terms. In the same way, if the bit precision is reduced to 3 bits, the NABS using uniform quantization is the same as that when using the APoT with 2 additive terms and, finally, if the bit precision is reduced to 2 bits, the NABS using uniform quantization is the same as that when using the PoT quantization.


\subsubsection{PTQ heterogeneous approach} \label{sec:PTQ_Heterog_subsec}
When using the heterogeneous approach, the bit precision and quantization method are allowed to vary in different parts of the NN structure. For simplicity, here we only consider the uniform quantization, the original APoT, and the mix, where different types of quantization are used throughout the NN structure.

Fig.~\ref{fig:Quantiztion_cube} depicts 3D plots with the Q-factor as a function of i) the bitwidth of the input and recurrent kernel of the LSTM layer, and ii) the bitwidth of the filter kernel of the CNN layer. A gradient of colors is used, with the warmer colors corresponding to the higher Q-factor, so we can identify which combination of $b_w$ values gives the best performance.

Same as in the homogeneous approach (see Fig.~\ref{fig:result_PTQ}), the uniform quantization is the most interesting solution for the Sim1 transmission scenario, whereas the APoT leads to better performance for the Sim2 and Exp scenarios.

\begin{figure*}[ht!] 
\begin{subfigure}{.33\textwidth}
    \centering
 \begin{tikzpicture}[scale=0.7]
    \begin{axis} [ylabel={BER}, 
        xlabel={Bitwidth of Weights [bits]},
        ylabel={Q-Factor [dB]},
        grid=both,  
        xmin=2, xmax=6,
    	xtick={6, ..., 2},
          x axis line style={-},
          x post scale=-1,
    	ymin=5, ymax=10.5,
        legend style={legend pos=south west, legend cell align=left,fill=white, fill opacity=0.6, draw opacity=1,text opacity=1},
    	grid style={dashed}]
        ]
    \addplot[color=blue, mark=square, very thick]     coordinates {
    (2,5.2)(3,6.39)(4,7.9)(5,9.46) (6,10.1)    
    };
    \addlegendentry{Uniform};
    
    \addplot[color=yellow, mark=*, very thick]   
    coordinates {
    (2,5.2)(3,6.39)(4,7.6)(5,7.65) (6,7.79)    
    };
    \addlegendentry{PoT};

    \addplot[color=green, mark=o, very thick]     coordinates {
    (2,5.2)(3,6.39)(4,8.33)(5,9.24) (6,9.24)    
 
    };
    \addlegendentry{APoT - 2 Terms};

            \addplot [color=cyan, mark=x, very thick]    coordinates {
    (2,5.2)(3,6.39)(4,7.9)(5,9.21) (6,9.79)    
    };
    \addlegendentry{APoT - 3 Terms};
    
                \addplot [color=purple, mark=+, very thick]    coordinates {
    (2,5.2)(3,6.39)(4,7.9)(5,9.46) (6,10.1)    

    };
    \addlegendentry{APoT - 4 Terms};

                \addplot [color=orange, mark=triangle, very thick]    coordinates {
    (2,5.2)(3,6.39)(4,8.52)(5,9.24) (6,9.84)    
    };
    \addlegendentry{APoT - Original};
    
        \addplot[color=red, dashed, very thick]    coordinates {
    (2,10.31)(3,10.31)(4,10.31)(5,10.31) (6,10.31) 
    };
    \addlegendentry{Non-Quantized Ref};
    
    \end{axis}
    \end{tikzpicture}
    \caption{Single Channel-DP 30GBd; 64QAM;\\20$\times$50km SSMF link (Sim1).}
    \label{fig:res_a1} 
\end{subfigure}\hfill
\begin{subfigure}{.33\textwidth}
    \centering
 \begin{tikzpicture}[scale=0.7]
    \begin{axis} [ylabel={BER}, 
        xlabel={Bitwidth of Weights [bits]},
        ylabel={Q-Factor [dB]},
        grid=both,  
        xmin=2, xmax=6,
    	xtick={6, ..., 2},
          x axis line style={-},
          x post scale=-1,
    	ymin=3, ymax=5.3,
        legend style={legend pos=south west, legend cell align=left,fill=white, fill opacity=0.6, draw opacity=1,text opacity=1},
    	grid style={dashed}]
        ]
    \addplot[color=blue, mark=square, very thick]     coordinates {
    (2,3.34)(3,3.656)(4,3.795)(5,4.07) (6,4.71)    
    };
    \addlegendentry{Uniform};
    
    \addplot[color=yellow, mark=*, very thick]   
    coordinates {
    (2,3.29)(3,3.656)(4,4.450165190848647)(5,4.401500813707454) (6,4.356193984141446)   
    };
    \addlegendentry{PoT};

    \addplot[color=green, mark=o, very thick]     coordinates {
    (2,3.36)(3,3.656)(4,4.41)(5,4.73) (6,4.841)     
 
    };
    \addlegendentry{APoT - 2 Terms};

            \addplot [color=cyan, mark=x, very thick]    coordinates {
    (2,3.31)(3,3.656)(4,3.873)(5,4.27) (6,4.844)    
    };
    \addlegendentry{APoT - 3 Terms};
    
                \addplot [color=purple, mark=+, very thick]    coordinates {
    (2,3.34)(3,3.656)(4,3.795)(5,4.07) (6,4.84)    

    };
    \addlegendentry{APoT - 4 Terms};

                \addplot [color=orange, mark=triangle, very thick]    coordinates {
    (2,3.217)(3,3.656)(4,4.32)(5,4.75) (6,4.94)     
    };
    \addlegendentry{APoT - Original};
    
        \addplot[color=red, dashed, very thick]    coordinates {
    (2,5.08)(3,5.08)(4,5.08)(5,5.08) (6,5.08) 
    };
    \addlegendentry{Non-Quantized Ref};
    
    \end{axis}
    \end{tikzpicture}
    \caption{Single Channel-DP 34.4GBd; 64QAM;\\9$\times$110km SSMF link (Sim2).}    \label{fig:res_b1} 
\end{subfigure}\hfill
\begin{subfigure}{.33\textwidth}
    \centering
 \begin{tikzpicture}[scale=0.7]
    \begin{axis} [ylabel={BER}, 
        xlabel={Bitwidth of Weights [bits]},
        ylabel={Q-Factor [dB]},
        grid=both,  
        xmin=2, xmax=6,
    	xtick={6, ..., 2},
          x axis line style={-},
          x post scale=-1,
    	ymin=7.05, ymax=8.2,
        legend style={legend pos=south west, legend cell align=left,fill=white, fill opacity=0.6, draw opacity=1,text opacity=1},
    	grid style={dashed}]
        ]
    \addplot[color=blue, mark=square, very thick]     coordinates {
    (2,7.12)(3,7.56)(4,7.68)(5,7.85) (6,7.88)
    };
    \addlegendentry{Uniform};
    
    \addplot[color=yellow, mark=*, very thick]   
    coordinates {
    (2,7.1)(3,7.52)(4,7.65)(5,7.77) (6,7.8)
    };
    \addlegendentry{PoT};

    \addplot[color=green, mark=o, very thick]     coordinates {
    (2,7.15)(3,7.58)(4,7.75)(5,7.89) (6,8)
 
    };
    \addlegendentry{APoT - 2 Terms};

    \addplot [color=cyan, mark=x, very thick]    coordinates {
    (2,7.12)(3,7.56)(4,7.66)(5,7.83) (6,7.92)
    };
    \addlegendentry{APoT - 3 Terms};
    
    \addplot [color=purple, mark=+, very thick]    coordinates {
    (2,7.12)(3,7.56)(4,7.68)(5,7.85) (6,7.91)

    };
    \addlegendentry{APoT - 4 Terms};

    \addplot [color=orange, mark=triangle, very thick]    coordinates {
    (2,7.15)(3,7.58)(4,7.78)(5,7.97) (6,8.1)
    };
    \addlegendentry{APoT - Original};
    
    \addplot[color=red, dashed, very thick]    coordinates {
    (2,8.1)(3,8.1)(4,8.1)(5,8.1) (6,8.1)
    };
    \addlegendentry{Non-Quantized Ref};
    
    \end{axis}
    \end{tikzpicture}
    \caption{Single Channel-DP 34.4GBd; 64QAM (PS-8bits/4D symbol); 9$\times$110km SSMF link (Exp).}    \label{fig:res_c1} 
\end{subfigure}

  \caption{Performance of the Quantization Aware Training (Homogeneous Approach).}
  \label{fig:result_QAT_Homog} 
\end{figure*}
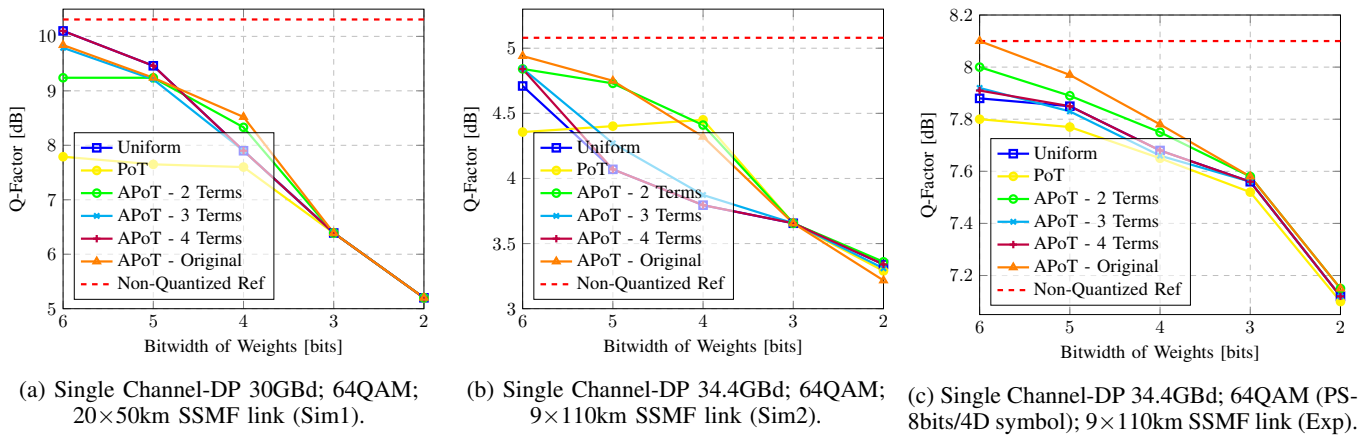

However, when using heterogeneous quantization, we observe that lower complexity can be achieved. For example, considering the Sim1 transmission, and in order to achieve the same optical performance as we have when using the 1 STpS DBP, we may quantize all weights with 6 bits (homogeneous quantization), or we may further reduce the recurrent kernel and CNN kernel to 5 bits using heterogeneous quantization without any significant degradation in optical performance. Similar results can be observed in the Sim2 and Exp transmission scenarios.

In addition to using different bit precision in different parts of the NN, we can also use different types of quantization to improve the performance. For this objective, we have used a grid search, testing different combinations of quantization types. The result of this optimization is referred to in Fig.~\ref{fig:Quantiztion_cube} as Mixed Quantization. By following such an approach, we observe an improvement in the hot area of the Sim 1 results when we quantize the input and CNN kernels with uniform quantization and the recurrent kernel with APoT using the original terms; for Sim2, we quantized the input with the 3 terms APoT and the recurrent and CNN kernels with the original APoT. Unfortunately, no significant optical improvement was observed when compared to using just the original APoT. In this case, just an improvement of 0.1dB in Q-factor is achieved in the mix quantization, where we quantize the input and CNN kernels with the original APoT and the recurrent kernel with the 2 terms APoT, compared to the case with all weights quantized with the original APoT.

\subsubsection{QAT homogeneous approach}
 We now evaluate the potential of implementing quantization during the training phase of the NN to mitigate the error introduced by the low bit precision of weights. Since QAT leads to at least as good performance as PTQ and the results depicted in Fig.~\ref{fig:result_QAT_Homog} show that the optical performance is highly impacted when the bitwidth decreases below 6 bits, we will focus the QAT analysis in this region (between 6 and 2 bits).

Fig.~\ref{fig:result_QAT_Homog} depicts the Q-factor as a function of the bit precision for the three considered transmission scenarios. The considered quantization techniques are the Uniform, PoT, APoT with 2, 3, and 4 additive terms and the original version of APoT that can be found in Ref.~\cite{li2019additive}. In order to better illustrate the impact of QAT, we compare the results in Figs.~\ref{fig:result_PTQ} and~\ref{fig:result_QAT_Homog}.
As an example, let us assume that all weights are quantized equally with 4 bits. For Sim1, Fig. \ref{fig:result_PTQ} (a) shows that the uniform quantization provided the best performance with a Q-factor close to 4 dB, whereas Fig.~\ref{fig:result_QAT_Homog} (a) shows that using the APoT original and following a QAT strategy enables reaching Q-factor values close to 8.5~dB. Similar conclusions can be drawn in Sim2 and Exp transmission scenarios: Fig. \ref{fig:result_QAT_Homog} (b) and (c) show that optical performance is highly impacted when weights are quantized to 4 bits whereas, when implementing QAT, the original APoT provides Q-factor values close to 4.5~dB and 7.8~dB, respectively. These results demonstrate the huge positive impact of QAT on the compression of the NN model. Moreover, when further reducing the bitwidth, the optical performance degradation is not as drastic as in the case of PTQ. 

The original APoT quantization technique leads to very good optical performance in most of the cases depicted in Fig.~\ref{fig:result_QAT_Homog}. This good performance is a direct consequence of performing quantization during the training phase. Indeed, in this case, the weights remaining after the pruning are no longer underrepresented, but rather adjusted to the non-uniform levels of quantization, leading to the good performance.  
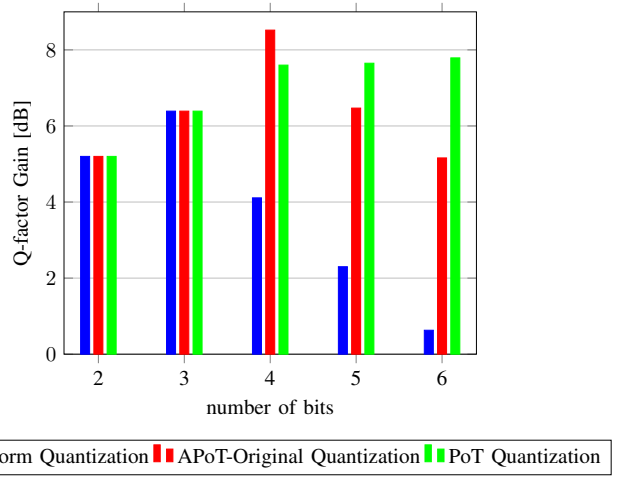
\begin{figure}[ht!] 
 \centering
\begin{tikzpicture}[scale=0.8]

 \begin{axis}[
 ybar=5*\pgflinewidth,
 symbolic x coords={2,3,4,5,6,7,8,9,10,11,12},
 bar width=0.15cm,
 enlarge x limits=0.1,
 ymin=0,ymax=9,
 xtick=data,
 ymajorgrids = true,
 ylabel = {\textcolor{black}{Q-factor Gain [dB]}} ,
 xlabel = {\textcolor{black}{number of bits}} ,
 legend style={at={(0.5,-0.25)},
anchor=north,legend columns=-1},
 ]

 \addplot[style={blue!100,fill=blue!100,mark=none}]
 coordinates {(2, 5.2) (3, 6.39)(4, 4.11)(5, 2.3) (6,0.63) };
 \addplot[style={red!100,fill=red!100,mark=none}]
 coordinates {(2, 5.2) (3, 6.39)(4, 8.52)(5, 6.47) (6,5.16) };
 \addplot[style={green!100,fill=green!100,mark=none}]
 coordinates {(2, 5.2) (3, 6.39)(4, 7.6)(5, 7.65) (6,7.79) };
\legend{Uniform Quantization, APoT-Original Quantization, PoT Quantization}
 \end{axis}
 
\end{tikzpicture}
 \caption{Gain achieved by using Quantization Aware Training vs Post Training Quantization in the SC-DP 30GBd; 64QAM; 20$\times$50km SSMF link (Sim1) dataset.}
 \label{fig:qat_ptq_comparisson}
\end{figure}

 Nevertheless, the difference between PoT and APoT is no longer as significant as the one observed in Fig.~\ref{fig:result_PTQ}. 
 To highlight the Q-factor gains that training gives, we summarized in Fig.~\ref{fig:qat_ptq_comparisson} the Q-factor difference between the PTQ and QAT both homogeneous for the Sim 1 case. In this case, we see that APOT original and POT have benefit from the extra training after quantization, mitigating almost completely the impacts of such quantizations for certain number of bits (e.g. 6 bits). As a consequence, no universal conclusion can be drawn about which is the best quantization technique for QAT. In general, the original APoT and the APoT with 2 terms were the two techniques that performed the best for the range of bitwidths studied and transmission scenarios. But, since the APOT with 2 terms uses only one adder and bit shift for each multiplication, it is probably the best choice in terms of trade-off between optical performance and computational complexity.

\begin{table*}[th!]
\centering
\caption{Results obtained using the Bayesian Optimizer. The performance results of both QAT and PTQ are depicted to highlight the benefit of QAT.}
\label{table:QAT:heterogeneous}
\begin{tabular}{|c|ccc|ccc|c|c|c|c|c|}
\hline
\multirow{2}{*}{Scenario} & \multicolumn{3}{c|}{$b_w$}                                         & \multicolumn{3}{c|}{Quant. Type}                                  & \multirow{2}{*}{\begin{tabular}[c]{@{}c@{}}Learning\\ Rate\end{tabular}}  & \multirow{2}{*}{\begin{tabular}[c]{@{}c@{}}Mini-Batch\\  Size\end{tabular}} & \multirow{2}{*}{\begin{tabular}[c]{@{}c@{}}Q-factor\\ PTQ\end{tabular}} & \multirow{2}{*}{\begin{tabular}[c]{@{}c@{}}Q-factor\\ QAT\end{tabular}} & \multirow{2}{*}{\begin{tabular}[c]{@{}c@{}}Q-factor\\ w/o Quant.\end{tabular}} \\ \cline{2-7}
                          & \multicolumn{1}{c|}{Input} & \multicolumn{1}{c|}{Recurrent} & CNN & \multicolumn{1}{c|}{Input} & \multicolumn{1}{c|}{Recurrent} & CNN &                     &                                                                               &                                                                         &                                                                         &                                                                                \\ \hline
Sim1                      & \multicolumn{1}{c|}{$3$}   & \multicolumn{1}{c|}{$5$}       & $4$ & \multicolumn{1}{c|}{ APOT Orig.}   & \multicolumn{1}{c|}{ APOT Orig.}       & Uniform & $0.00114$                 & $4347$                                                                           & $0.18$ dB                                                                    & $8.6$ dB                                                                    & $10.31$ dB                                                                           \\ \hline
Sim2                      & \multicolumn{1}{c|}{$4$}   & \multicolumn{1}{c|}{$3$}       & $4$ & \multicolumn{1}{c|}{APOT 2 Terms}   & \multicolumn{1}{c|}{APOT Orig.}       & APOT Orig. & $0.00132$                 & $6568$                                                                           & $0$ dB                                                                    & $4.5$  dB                                                                   & $5.1$ dB                                                                           \\ \hline
Exp                       & \multicolumn{1}{c|}{$2$}   & \multicolumn{1}{c|}{$3$}       & $2$ & \multicolumn{1}{c|}{APOT Orig.}   & \multicolumn{1}{c|}{APOT Orig.}       & APOT Orig. & $0.00085$                 & $5253$                                                                           & $0$ dB                                                                    & $7.41$  dB                                                                   & $8.1$ dB                                                                           \\ \hline
\end{tabular}
\end{table*}

We would like to stress that the training of such NN structures is unstable. Consequently, the model needs to be monitored during training. In this work, early stopping was not used for QAT. Instead, the quantized NN structure was trained for 5000 epochs, with the intermediate NN models leading to the best Q-factor being saved and used as the final NN. As described in Ref.~\cite{li2017training}, the training phase of the quantized model can suffer from learning problems (e.g. exploration vs. exploitation trade-offs). As suggested in that reference, we also used large mini-batch sizes ($\geq 4000$), since ``this shrinks the variance of the gradient distribution without changing the mean and concentrates more of the gradient distribution towards downhill directions, making the algorithm more greedy''. As a result, we emphasize that when performing the QAT, the training hyperparameters must be properly set, and the training will most likely require a higher number of epochs as the bitwidth of the weights is reduced. 

\subsubsection{QAT heterogeneous approach}
In this compression technique, and as described in Sec. \ref{sec:Compression},  the BO is used to determine the ideal bitwidth per layer as well as the type of quantization in each layer (and other hyperparameters, like the learning rate), seeking to improve the overall performance.

Similarly to the PTQ case, when going from the homogeneous to the heterogeneous approach, the bitwidths and quantization types employed can be different in different parts of the NN architecture. Like in Sec.\ref{sec:PTQ_Heterog_subsec}, the performance of the heterogeneous approach is evaluated by considering the different bit precision of the input kernel of the LSTM layer, the recurrent kernel of the LSTM layer, and the filter kernel of the CNN layer.  The values obtained by the BO can be found in Table \ref{table:QAT:heterogeneous} for the considered transmission scenarios as well as the Q-factor achieved when i) the NN model is not quantized (w/o Quant.), ii) the model is only quantized (PTQ), and iii) the model is simultaneously quantized and fine-tuned (QAT). This table shows that, at the low levels of bit precision, the PTQ corrupted the NN model completely. Nevertheless, the QAT adapts the weights in such a way that only a small degradation of performance is observed.

To conclude, we evaluate the complexity of the different approaches. We do this in terms of NABS insofar as we employ simultaneously different bit precision and quantization techniques (see \ref{sec:complexity metric}). Our reference complexity is the ``traditional'' uniform quantization with 8 bits in all layers, which we compare against the heterogeneous structures depicted in Table \ref{table:QAT:heterogeneous}.  For Sim1, the reference complexity is 28.6M NABSs while, for the heterogeneous architecture, the complexity is  10.9M NABSs, which translates into a complexity reduction of $\approx 62\%$.
Similarly, for Sim2 and Exp, the reference complexity is $\approx31$M NABSs, whereas after heterogeneous QAT it is $\approx7.9$M and  $\approx7.4$M NABSs, for the two cases, respectively, representing a reduction of $\approx 76\%$.

\section{Conclusions, Open Problems and Research Directions}\label{sec:Conclusion}

In this paper, a full-scale study focusing on the reduction of the computational complexity of NN-based solutions was presented, evaluating them in the context of coherent transmission equalization. 

First, we demonstrate the complexity bottleneck resulting from recovering one symbol at a time, showing the computational complexity benefit resulting from changing the NN structure to recover multiple symbols instead.

Then, we introduced the first compression method evaluated in the paper: the pruning strategy. We provided examples of the three most well-known pruning techniques: fine-tuning, weight-rewinding, and learning rate rewinding. We explained their theoretical foundation and proposed a new strategy, which results from combining fine-tuning with BO to improve the learning of the hyperparameters of the pruning.
Later, we demonstrated that, at the cost of making the model's training more difficult, the latter leads to smaller performance degradation and more sparsity when compared to the former.

Next, we present the second compression technique, known as weight clustering (or weight sharing). We demonstrated its application in both recurrent and feedforward layers, emphasizing its goal of reducing the number of effective weights and effective multiplications required by the model. This is achieved by having multiple connections that share the same weight and then fine-tuning those shared weights. In addition, we demonstrated the advantages of using the BO to determine the number of clusters per layer and the training hyperparameters for the fine-tuning phase.

Afterward, we provided a comprehensive overview of the various aspects of quantization in neural networks. The difference between post-training quantization and quantization-aware training was discussed, as well as how the Bayesian optimizer can aid in the design process. In addition, the use of different quantization types, such as uniform, APOT, and POT quantizations, in the field of optical channel equalization, was examined. It is challenging to compare the computational complexity of two different NN structures with different quantization types. As a result, we've covered several computational metrics and discussed when they're useful and how to calculate them for our NN equalizer.

Finally, we evaluated the performance of the different compression techniques considering three different transmission setups, comparing the Q-factor versus computational complexity in all scenarios. As the most relevant result, we observed that when using weight clustering and pruning, as a nonuniform quantization step, for the Sim1 transmission, we presented a Q-factor gain of 1.6~dB compared to the CDC  in the case of 3 clustered weights with the cost of increasing the complexity by 182\%, and a Q-factor gain of 0.6~dB at the expense of a 61\% increase in complexity in the case of 2 clustered weights. This result represents a big step forward in reaching commercial implementation since we are approaching the computational complexity of the existing CDC block in the DSP chain.  

Next, we describe some  open  problems  in  the design  of low complexity NN-based equalizers,  with  the  aim  of  spurring more  research  effort  on  advancing  the  design  of  machine learning solution in optical communication systems.

1) \textbf{The parallelization problem}. Training and evaluating each node can be very time-consuming in large NNs. This is unquestionably a bottleneck in the development of a high-speed NN design suitable for optical communication applications. A possible solution is to parallelize such models when implementing them in hardware. This topic was partially covered in Ref.~\cite{misra2010artificial} for feed-forward layers. If recurrent layers, like the LSTM layer proposed in this work, are in demand for future industrial applications, it would be interesting to investigate how to parallelize them in hardware implementations.

2) \textbf{Knowledge distillation}. This is another possible type of compression that was not investigated in this work, but that is receiving increasing attention from the community\cite{gou2021knowledge}.  The idea behind knowledge distillation is to train a distilled NN model that has many layers and is truly computationally complex, and then use it to train a more compact NN model. An evaluation of the possible benefit of this technique in the design of NN-based equalizers is an interesting direction for future work.

3) \textbf{Meta Learning Based Compression}.  In  Ref.~\cite{ye2021hybrid}, the authors have jointly considered network pruning and quantization in an end-to-end meta-learning framework. Other papers, e.g. Ref.~\cite{chen2019metaquant,liu2019metapruning}, used meta-learning to learn how to quantize or how to prune the NN structure. We see this as a potentially good alternative to the BO technique used in this paper, which,  perhaps, could provide the same or even better solutions, but in a faster manner.

4) \textbf{Stabilization of the quantization training} for different transmission scenarios. The effectiveness of quantization, as mentioned in this work, is highly dependent on the difficulty of the transmission equalization task and the learning process. Several authors have already discussed the challenges and possible solutions of the training process for the quantized NN models \cite{li2017training, bartan2021training,zhuang2020training}. To fully understand this application for future industrial applications, e.g., in the optical communications field, a deeper investigation into how to make this training process more stable and faster, independently of the transmission setup, is required.

5) \textbf{Flexibility study after compression}. In Refs.~\cite{9523752,9748452,9226070}, it can be observed that by using techniques such as transfer learning and domain randomization, an NN-based equalizer can operate in multiple distances, modulation formats, launch powers, and symbol rates. However, in the context of this study, the following question naturally arises: Can the NN equalizer keep its re-usability and flexibility if its representability capacity is drastically lowered by compression approaches (such as pruning and weight sharing)?

Some works in the machine learning field \cite{xiao2022optimizations,myung2022pac, liu2021transtailor,gordon2020compressing} have presented some of the good and bad aspects on the NN flexibility when compression is applied in the NN model, but a deeper report for the channel equalization task is also required because flexibility is a key feature desired by the telecommunications industry.

\bibliographystyle{IEEEtran}
\bibliography{references}

\begin{thebibliography}{100}
\providecommand{\url}[1]{#1}
\csname url@samestyle\endcsname
\providecommand{\newblock}{\relax}
\providecommand{\bibinfo}[2]{#2}
\providecommand{\BIBentrySTDinterwordspacing}{\spaceskip=0pt\relax}
\providecommand{\BIBentryALTinterwordstretchfactor}{4}
\providecommand{\BIBentryALTinterwordspacing}{\spaceskip=\fontdimen2\font plus
\BIBentryALTinterwordstretchfactor\fontdimen3\font minus
  \fontdimen4\font\relax}
\providecommand{\BIBforeignlanguage}[2]{{%
\expandafter\ifx\csname l@#1\endcsname\relax
\typeout{** WARNING: IEEEtran.bst: No hyphenation pattern has been}%
\typeout{** loaded for the language `#1'. Using the pattern for}%
\typeout{** the default language instead.}%
\else
\language=\csname l@#1\endcsname
\fi
#2}}
\providecommand{\BIBdecl}{\relax}
\BIBdecl

\bibitem{akc16}
E.~Agrell, M.~Karlsson, A.~R. Chraplyvy, D.~J. Richardson, P.~M. Krummrich,
  P.~Winzer, K.~Roberts, J.~K. Fischer, S.~J. Savory, B.~J. Eggleton,
  M.~Secondini, F.~R. Kschischang, A.~Lord, J.~Prat, I.~Tomkos, J.~E. Bowers,
  S.~Srinivasan, M.~Brandt-Pearce, and N.~Gisin, ``Roadmap of optical
  communications,'' \emph{Journal of Optics}, vol.~18, no.~6, p. 063002, may
  2016.

\bibitem{winzer2018fiber}
P.~J. Winzer, D.~T. Neilson, and A.~R. Chraplyvy, ``Fiber-optic transmission
  and networking: the previous 20 and the next 20 years,'' \emph{Optics
  Express}, vol.~26, no.~18, pp. 24\,190--24\,239, 2018.

\bibitem{Cartledge:17}
J.~C. Cartledge, F.~P. Guiomar, F.~R. Kschischang, G.~Liga, and M.~P. Yankov,
  ``Digital signal processing for fiber nonlinearities
  \&\#x0005b;invited\&\#x0005d;,'' \emph{Opt. Express}, vol.~25, no.~3, pp.
  1916--1936, Feb 2017.

\bibitem{musumeci2018overview}
F.~Musumeci, C.~Rottondi, A.~Nag, I.~Macaluso, D.~Zibar, M.~Ruffini, and
  M.~Tornatore, ``An overview on application of machine learning techniques in
  optical networks,'' \emph{IEEE Communications Surveys \& Tutorials}, vol.~21,
  no.~2, pp. 1383--1408, 2018.

\bibitem{hager2018nonlinear}
C.~H{\"a}ger and H.~D. Pfister, ``Nonlinear interference mitigation via deep
  neural networks,'' in \emph{2018 Optical Fiber Communications Conference and
  Exposition (OFC)}.\hskip 1em plus 0.5em minus 0.4em\relax IEEE, 2018, pp.
  1--3.

\bibitem{hager2020physics}
------, ``Physics-based deep learning for fiber-optic communication systems,''
  \emph{IEEE Journal on Selected Areas in Communications}, vol.~39, no.~1, pp.
  280--294, 2020.

\bibitem{freire2021performance}
P.~J. Freire, Y.~Osadchuk, B.~Spinnler, A.~Napoli, W.~Schairer, N.~Costa, J.~E.
  Prilepsky, and S.~K. Turitsyn, ``Performance versus complexity study of
  neural network equalizers in coherent optical systems,'' \emph{Journal of
  Lightwave Technology}, vol.~39, no.~19, pp. 6085--6096, 2021.

\bibitem{freire2022neural}
P.~J. Freire, A.~Napoli, B.~Spinnler, N.~Costa, S.~K. Turitsyn, and J.~E.
  Prilepsky, ``Neural networks-based equalizers for coherent optical
  transmission: Caveats and pitfalls,'' \emph{IEEE Journal of Selected Topics
  in Quantum Electronics}, vol.~28, no.~4, pp. 1--23, 2022.

\bibitem{nevin2021machine}
J.~W. Nevin, S.~Nallaperuma, N.~A. Shevchenko, X.~Li, M.~S. Faruk, and S.~J.
  Savory, ``Machine learning for optical fiber communication systems: An
  introduction and overview,'' \emph{APL Photonics}, vol.~6, no.~12, p. 121101,
  2021.

\bibitem{deligiannidis2020compensation}
S.~{Deligiannidis}, A.~{Bogris}, C.~{Mesaritakis}, and Y.~{Kopsinis},
  ``Compensation of fiber nonlinearities in digital coherent systems leveraging
  long short-term memory neural networks,'' \emph{Journal of Lightwave
  Technology}, vol.~38, no.~21, pp. 5991--5999, 2020.

\bibitem{deligiannidis2021performance}
S.~Deligiannidis, C.~Mesaritakis, and A.~Bogris, ``Performance and complexity
  analysis of bi-directional recurrent neural network models versus volterra
  nonlinear equalizers in digital coherent systems,'' \emph{Journal of
  Lightwave Technology}, vol.~39, no.~18, pp. 5791--5798, 2021.

\bibitem{sidelnikov2018equalization}
O.~Sidelnikov, A.~Redyuk, and S.~Sygletos, ``Equalization performance and
  complexity analysis of dynamic deep neural networks in long haul transmission
  systems,'' \emph{Optics Express}, vol.~26, no.~25, pp. 32\,765--32\,776,
  2018.

\bibitem{ibnkahla2000applications}
M.~Ibnkahla, ``Applications of neural networks to digital communications--a
  survey,'' \emph{Signal processing}, vol.~80, no.~7, pp. 1185--1215, 2000.

\bibitem{HORNIK1989359}
\BIBentryALTinterwordspacing
K.~Hornik, M.~Stinchcombe, and H.~White, ``Multilayer feedforward networks are
  universal approximators,'' \emph{Neural Networks}, vol.~2, no.~5, pp.
  359--366, 1989. [Online]. Available:
  \url{https://www.sciencedirect.com/science/article/pii/0893608089900208}
\BIBentrySTDinterwordspacing

\bibitem{schafer2007recurrent}
A.~M. Sch{\"a}fer and H.-G. Zimmermann, ``Recurrent neural networks are
  universal approximators,'' \emph{International journal of neural systems},
  vol.~17, no.~04, pp. 253--263, 2007.

\bibitem{Goodfellow-et-al-2016}
I.~Goodfellow, Y.~Bengio, and A.~Courville, \emph{Deep Learning}.\hskip 1em
  plus 0.5em minus 0.4em\relax MIT Press, 2016,
  \url{http://www.deeplearningbook.org}.

\bibitem{jabbar2015methods}
H.~Jabbar and R.~Z. Khan, ``Methods to avoid over-fitting and under-fitting in
  supervised machine learning (comparative study),'' \emph{Computer Science,
  Communication and Instrumentation Devices}, vol.~70, 2015.

\bibitem{sang2022low}
B.~Sang, W.~Zhou, Y.~Tan, M.~Kong, C.~Wang, M.~Wang, L.~Zhao, J.~Zhang, and
  J.~Yu, ``Low complexity neural network equalization based on multi-symbol
  output technique for 200+ gbps im/dd short reach optical system,''
  \emph{Journal of Lightwave Technology}, vol.~40, no.~9, pp. 2890--2900, 2022.

\bibitem{agrawal21}
\BIBentryALTinterwordspacing
G.~P. Agrawal, \emph{Fiber-Optic Communication Systems}, 5th~ed.\hskip 1em plus
  0.5em minus 0.4em\relax Wiley, 2021. [Online]. Available:
  \url{https://www.wiley.com/en-us/Fiber+Optic+Communication+Systems%2C+5th+Edition-p-9781119737360#description-section}
\BIBentrySTDinterwordspacing

\bibitem{kodama1985optical}
Y.~Kodama, ``Optical solitons in a monomode fiber,'' \emph{Journal of
  Statistical Physics}, vol.~39, no.~5, pp. 597--614, 1985.

\bibitem{ferrari2020assessment}
A.~Ferrari, A.~Napoli, J.~K. Fischer, N.~Costa, A.~D’Amico, J.~Pedro,
  W.~Forysiak, E.~Pincemin, A.~Lord, A.~Stavdas \emph{et~al.}, ``Assessment on
  the achievable throughput of multi-band itu-t g. 652. d fiber transmission
  systems,'' \emph{Journal of Lightwave Technology}, vol.~38, no.~16, pp.
  4279--4291, 2020.

\bibitem{shannon1948mathematical}
C.~E. Shannon, ``A mathematical theory of communication,'' \emph{The Bell
  system technical journal}, vol.~27, no.~3, pp. 379--423, 1948.

\bibitem{essiambre2010capacity}
R.-J. Essiambre, G.~Kramer, P.~J. Winzer, G.~J. Foschini, and B.~Goebel,
  ``Capacity limits of optical fiber networks,'' \emph{Journal of Lightwave
  Technology}, vol.~28, no.~4, pp. 662--701, 2010.

\bibitem{sinkin2003optimization}
O.~V. Sinkin, R.~Holzl{\"o}hner, J.~Zweck, and C.~R. Menyuk, ``Optimization of
  the split-step fourier method in modeling optical-fiber communications
  systems,'' \emph{Journal of lightwave technology}, vol.~21, no.~1, p.~61,
  2003.

\bibitem{millar2010mitigation}
D.~S. Millar, S.~Makovejs, C.~Behrens, S.~Hellerbrand, R.~I. Killey, P.~Bayvel,
  and S.~J. Savory, ``Mitigation of fiber nonlinearity using a digital coherent
  receiver,'' \emph{IEEE Journal of Selected Topics in Quantum Electronics},
  vol.~16, no.~5, pp. 1217--1226, 2010.

\bibitem{agazzi2004impact}
O.~E. Agazzi and V.~Gopinathan, ``The impact of nonlinearity on electronic
  dispersion compensation of optical channels,'' in \emph{Optical Fiber
  Communication Conference}.\hskip 1em plus 0.5em minus 0.4em\relax Optica
  Publishing Group, 2004, p. TuG6.

\bibitem{savory2007imdd}
S.~Savory, Y.~Benlachtar, R.~Killey, P.~Bayvel, G.~Bosco, P.~Poggiolini,
  J.~Prat, and M.~Omella, ``Imdd transmission over 1,040 km of standard
  single-mode fiber at 10gbit/s using a one-sample-per-bit reduced-complexity
  mlse receiver,'' in \emph{OFC/NFOEC 2007-2007 Conference on Optical Fiber
  Communication and the National Fiber Optic Engineers Conference}.\hskip 1em
  plus 0.5em minus 0.4em\relax IEEE, 2007, pp. 1--3.

\bibitem{kupfer2008measurement}
T.~Kupfer, C.~Dorschky, M.~Ene, and S.~Langenbach, ``Measurement of the
  performance of 16-states mlse digital equalizer with different optical
  modulation formats,'' in \emph{Optical Fiber Communication Conference}.\hskip
  1em plus 0.5em minus 0.4em\relax Optica Publishing Group, 2008, p. PDP13.

\bibitem{benedetto1979modeling}
S.~Benedetto, E.~Biglieri, and R.~Daffara, ``Modeling and performance
  evaluation of nonlinear satellite links-a volterra series approach,''
  \emph{IEEE Transactions on Aerospace and Electronic Systems}, no.~4, pp.
  494--507, 1979.

\bibitem{guiomar2012mitigation}
F.~P. Guiomar, J.~D. Reis, A.~L. Teixeira, and A.~N. Pinto, ``Mitigation of
  intra-channel nonlinearities using a frequency-domain volterra series
  equalizer,'' \emph{Optics express}, vol.~20, no.~2, pp. 1360--1369, 2012.

\bibitem{cho2022volterra}
J.~Cho and S.~T. Le, ``Volterra equalization to compensate for transceiver
  nonlinearity: Performance and pitfalls,'' in \emph{2022 Optical Fiber
  Communications Conference and Exhibition (OFC)}.\hskip 1em plus 0.5em minus
  0.4em\relax IEEE, 2022, pp. 1--3.

\bibitem{ip2008compensation}
E.~Ip and J.~M. Kahn, ``Compensation of dispersion and nonlinear impairments
  using digital backpropagation,'' \emph{Journal of Lightwave Technology},
  vol.~26, no.~20, pp. 3416--3425, 2008.

\bibitem{napoli2014reduced}
A.~Napoli, Z.~Maalej, V.~A. Sleiffer, M.~Kuschnerov, D.~Rafique, E.~Timmers,
  B.~Spinnler, T.~Rahman, L.~D. Coelho, and N.~Hanik, ``Reduced complexity
  digital back-propagation methods for optical communication systems,''
  \emph{Journal of Lightwave Technology}, vol.~32, no.~7, pp. 1351--1362, 2014.

\bibitem{zhu2012nonlinearity}
L.~Zhu and G.~Li, ``Nonlinearity compensation using dispersion-folded digital
  backward propagation,'' \emph{Optics express}, vol.~20, no.~13, pp.
  14\,362--14\,370, 2012.

\bibitem{rafique2011compensation}
D.~Rafique, M.~Mussolin, M.~Forzati, J.~M{\aa}rtensson, M.~N. Chugtai, and
  A.~D. Ellis, ``Compensation of intra-channel nonlinear fibre impairments
  using simplified digital back-propagation algorithm,'' \emph{Optics express},
  vol.~19, no.~10, pp. 9453--9460, 2011.

\bibitem{freire2020complex}
P.~J. Freire, V.~Neskornuik, A.~Napoli, B.~Spinnler, N.~Costa, G.~Khanna,
  E.~Riccardi, J.~E. Prilepsky, and S.~K. Turitsyn, ``Complex-valued neural
  network design for mitigation of signal distortions in optical links,''
  \emph{Journal of Lightwave Technology}, vol.~39, no.~6, pp. 1696--1705, 2021.

\bibitem{alqahtani2021literature}
\BIBentryALTinterwordspacing
A.~Alqahtani, X.~Xie, and M.~W. Jones, ``Literature review of deep network
  compression,'' \emph{Informatics}, vol.~8, no.~4, 2021. [Online]. Available:
  \url{https://www.mdpi.com/2227-9709/8/4/77}
\BIBentrySTDinterwordspacing

\bibitem{deng2020model}
L.~Deng, G.~Li, S.~Han, L.~Shi, and Y.~Xie, ``Model compression and hardware
  acceleration for neural networks: A comprehensive survey,'' \emph{Proceedings
  of the IEEE}, vol. 108, no.~4, pp. 485--532, 2020.

\bibitem{liang2021pruning}
T.~Liang, J.~Glossner, L.~Wang, S.~Shi, and X.~Zhang, ``Pruning and
  quantization for deep neural network acceleration: A survey,''
  \emph{Neurocomputing}, vol. 461, pp. 370--403, 2021.

\bibitem{9043731}
L.~Deng, G.~Li, S.~Han, L.~Shi, and Y.~Xie, ``Model compression and hardware
  acceleration for neural networks: A comprehensive survey,'' \emph{Proceedings
  of the IEEE}, vol. 108, no.~4, pp. 485--532, 2020.

\bibitem{ron2022experimental}
D.~A. Ron, P.~J. Freire, J.~E. Prilepsky, M.~Kamalian-Kopae, A.~Napoli, and
  S.~K. Turitsyn, ``Experimental implementation of a neural network optical
  channel equalizer in restricted hardware using pruning and quantization,''
  \emph{Scientific Reports}, vol.~12, no.~1, pp. 1--14, 2022.

\bibitem{blalock2020state}
D.~Blalock, J.~J.~G. Ortiz, J.~Frankle, and J.~Guttag, ``What is the state of
  neural network pruning?'' \emph{arXiv preprint arXiv:2003.03033}, 2020.

\bibitem{liu2018rethinking}
Z.~Liu, M.~Sun, T.~Zhou, G.~Huang, and T.~Darrell, ``Rethinking the value of
  network pruning,'' \emph{arXiv preprint arXiv:1810.05270}, 2018.

\bibitem{augasta2013pruning}
M.~Augasta and T.~Kathirvalavakumar, ``Pruning algorithms of neural
  networks—a comparative study,'' \emph{Open Computer Science}, vol.~3,
  no.~3, pp. 105--115, 2013.

\bibitem{vadera2020methods}
S.~Vadera and S.~Ameen, ``Methods for pruning deep neural networks,''
  \emph{arXiv preprint arXiv:2011.00241}, 2020.

\bibitem{han2015deep}
S.~Han, H.~Mao, and W.~J. Dally, ``Deep compression: Compressing deep neural
  networks with pruning, trained quantization and huffman coding,'' \emph{arXiv
  preprint arXiv:1510.00149}, 2015.

\bibitem{frankle2018lottery}
J.~Frankle and M.~Carbin, ``The lottery ticket hypothesis: Finding sparse,
  trainable neural networks,'' \emph{arXiv preprint arXiv:1803.03635}, 2018.

\bibitem{tung2018deep}
F.~Tung and G.~Mori, ``Deep neural network compression by in-parallel
  pruning-quantization,'' \emph{IEEE transactions on pattern analysis and
  machine intelligence}, vol.~42, no.~3, pp. 568--579, 2018.

\bibitem{renda2020comparing}
A.~Renda, J.~Frankle, and M.~Carbin, ``Comparing rewinding and fine-tuning in
  neural network pruning,'' \emph{arXiv preprint arXiv:2003.02389}, 2020.

\bibitem{zhu2017prune}
M.~Zhu and S.~Gupta, ``To prune, or not to prune: exploring the efficacy of
  pruning for model compression,'' \emph{arXiv preprint arXiv:1710.01878},
  2017.

\bibitem{chuang2019sparse}
C.-Y. Chuang, W.-F. Chang, C.-C. Wei, C.-J. Ho, C.-Y. Huang, J.-W. Shi,
  L.~Henrickson, Y.-K. Chen, and J.~Chen, ``Sparse volterra nonlinear equalizer
  by employing pruning algorithm for high-speed pam-4 850-nm vcsel optical
  interconnect,'' in \emph{Optical Fiber Communication Conference}.\hskip 1em
  plus 0.5em minus 0.4em\relax Optical Society of America, 2019, pp. M1F--2.

\bibitem{huang201893}
W.-J. Huang, W.-F. Chang, C.-C. Wei, J.-J. Liu, Y.-C. Chen, K.-L. Chi, C.-L.
  Wang, J.-W. Shi, and J.~Chen, ``93\% complexity reduction of volterra
  nonlinear equalizer by l1-regularization for 112-gbps pam-4 850-nm vcsel
  optical interconnect,'' in \emph{2018 Optical Fiber Communications Conference
  and Exposition (OFC)}.\hskip 1em plus 0.5em minus 0.4em\relax IEEE, 2018, pp.
  1--3.

\bibitem{6647643}
F.~P. Guiomar, S.~B. Amado, N.~J. Muga, J.~D. Reis, A.~L. Teixeira, and A.~N.
  Pinto, ``Simplified volterra series nonlinear equalizer by intra-channel
  cross-phase modulation oriented pruning,'' in \emph{39th European Conference
  and Exhibition on Optical Communication (ECOC 2013)}, 2013, pp. 1--3.

\bibitem{zhang2019field}
S.~Zhang, F.~Yaman, K.~Nakamura, T.~Inoue, V.~Kamalov, L.~Jovanovski,
  V.~Vusirikala, E.~Mateo, Y.~Inada, and T.~Wang, ``Field and lab experimental
  demonstration of nonlinear impairment compensation using neural networks,''
  \emph{Nature communications}, vol.~10, no.~1, pp. 1--8, 2019.

\bibitem{melek2020nonlinearity}
M.~M. Melek and D.~Yevick, ``Nonlinearity mitigation with a perturbation based
  neural network receiver,'' \emph{Optical and Quantum Electronics}, vol.~52,
  no.~10, pp. 1--10, 2020.

\bibitem{kumar2021deep}
O.~S. Kumar, L.~Lampe, S.~Luo, M.~Jana, J.~Mitra, and C.~Li, ``Deep neural
  network assisted second-order perturbation-based nonlinearity compensation,''
  in \emph{Signal Processing in Photonic Communications}.\hskip 1em plus 0.5em
  minus 0.4em\relax Optical Society of America, 2021, pp. SpF2E--2.

\bibitem{li2021high}
M.~Li, W.~Zhang, Q.~Chen, and Z.~He, ``High-throughput hardware deployment of
  pruned neural network based nonlinear equalization for 100-gbps short-reach
  optical interconnect,'' \emph{Optics Letters}, vol.~46, no.~19, pp.
  4980--4983, 2021.

\bibitem{wan2018nonlinear}
Z.~Wan, J.~Li, L.~Shu, M.~Luo, X.~Li, S.~Fu, and K.~Xu, ``Nonlinear
  equalization based on pruned artificial neural networks for 112-gb/s ssb-pam4
  transmission over 80-km ssmf,'' \emph{Optics express}, vol.~26, no.~8, pp.
  10\,631--10\,642, 2018.

\bibitem{zhang2020compressed}
W.~Zhang, L.~Ge, Y.~Zhang, C.~Liang, and Z.~He, ``Compressed nonlinear
  equalizers for 112-gbps optical interconnects: Efficiency and stability,''
  \emph{Sensors}, vol.~20, no.~17, p. 4680, 2020.

\bibitem{wang2021low}
L.~Wang, X.~Zeng, J.~Wang, D.~Gao, and M.~Bai, ``Low-complexity nonlinear
  equalizer based on artificial neural network for 112 gbit/s pam-4
  transmission using dml,'' \emph{Optical Fiber Technology}, vol.~67, p.
  102724, 2021.

\bibitem{ge2020compressed}
L.~Ge, W.~Zhang, C.~Liang, and Z.~He, ``Compressed neural network equalization
  based on iterative pruning algorithm for 112-gbps vcsel-enabled optical
  interconnects,'' \emph{Journal of Lightwave Technology}, vol.~38, no.~6, pp.
  1323--1329, 2020.

\bibitem{reza2018nonlinear}
A.~G. Reza and J.-K.~K. Rhee, ``Nonlinear equalizer based on neural networks
  for pam-4 signal transmission using dml,'' \emph{IEEE Photonics Technology
  Letters}, vol.~30, no.~15, pp. 1416--1419, 2018.

\bibitem{Koike2021}
T.~Koike-Akino, Y.~Wang, K.~Kojima, K.~Parsons, and T.~Yoshida,
  ``Zero-multiplier sparse dnn equalization for fiber-optic qam systems with
  probabilistic amplitude shaping,'' in \emph{2021 European Conference on
  Optical Communications (ECOC)}.\hskip 1em plus 0.5em minus 0.4em\relax IEEE,
  2021, pp. 1--4.

\bibitem{weerts2020importance}
H.~J. Weerts, A.~C. Mueller, and J.~Vanschoren, ``Importance of tuning
  hyperparameters of machine learning algorithms,'' \emph{arXiv preprint
  arXiv:2007.07588}, 2020.

\bibitem{he2018amc}
Y.~He, J.~Lin, Z.~Liu, H.~Wang, L.-J. Li, and S.~Han, ``Amc: Automl for model
  compression and acceleration on mobile devices,'' in \emph{Proceedings of the
  European conference on computer vision (ECCV)}, 2018, pp. 784--800.

\bibitem{cho2020basic}
H.~Cho, Y.~Kim, E.~Lee, D.~Choi, Y.~Lee, and W.~Rhee, ``Basic enhancement
  strategies when using bayesian optimization for hyperparameter tuning of deep
  neural networks,'' \emph{IEEE Access}, vol.~8, pp. 52\,588--52\,608, 2020.

\bibitem{snoek2012practical}
J.~Snoek, H.~Larochelle, and R.~P. Adams, ``Practical bayesian optimization of
  machine learning algorithms,'' \emph{Advances in neural information
  processing systems}, vol.~25, 2012.

\bibitem{frazier2018tutorial}
P.~I. Frazier, ``A tutorial on bayesian optimization,'' \emph{arXiv preprint
  arXiv:1807.02811}, 2018.

\bibitem{wang2020compressing}
L.-N. Wang, W.~Liu, X.~Liu, G.~Zhong, P.~P. Roy, J.~Dong, and K.~Huang,
  ``Compressing deep networks by neuron agglomerative clustering,''
  \emph{Sensors}, vol.~20, no.~21, p. 6033, 2020.

\bibitem{son2018clustering}
S.~Son, S.~Nah, and K.~M. Lee, ``Clustering convolutional kernels to compress
  deep neural networks,'' in \emph{Proceedings of the European Conference on
  Computer Vision (ECCV)}, 2018, pp. 216--232.

\bibitem{wu2018deep}
J.~Wu, Y.~Wang, Z.~Wu, Z.~Wang, A.~Veeraraghavan, and Y.~Lin, ``Deep k-means:
  Re-training and parameter sharing with harder cluster assignments for
  compressing deep convolutions,'' in \emph{International Conference on Machine
  Learning}.\hskip 1em plus 0.5em minus 0.4em\relax PMLR, 2018, pp. 5363--5372.

\bibitem{lee2021cluster}
J.~H. Lee, J.~Yun, S.~J. Hwang, and E.~Yang, ``Cluster-promoting quantization
  with bit-drop for minimizing network quantization loss,'' in
  \emph{Proceedings of the IEEE/CVF International Conference on Computer
  Vision}, 2021, pp. 5370--5379.

\bibitem{Clsteringtensorflow}
Github implementation of the weights clustering algorithm in tensorflow.
  \url{https://github.com/tensorflow/model-optimization/blob/v0.7.2/tensorflow_model_optimization/python/core/clustering/keras/clustering_algorithm.py#L24-L194}.
  Accessed: 2022-05-30.

\bibitem{cho2021dkm}
M.~Cho, K.~A. Vahid, S.~Adya, and M.~Rastegari, ``Dkm: Differentiable k-means
  clustering layer for neural network compression,'' \emph{arXiv preprint
  arXiv:2108.12659}, 2021.

\bibitem{gholami2021survey}
A.~Gholami, S.~Kim, Z.~Dong, Z.~Yao, M.~W. Mahoney, and K.~Keutzer, ``A survey
  of quantization methods for efficient neural network inference,'' \emph{arXiv
  preprint arXiv:2103.13630}, 2021.

\bibitem{cheng2017survey}
Y.~Cheng, D.~Wang, P.~Zhou, and T.~Zhang, ``A survey of model compression and
  acceleration for deep neural networks,'' \emph{arXiv preprint
  arXiv:1710.09282}, 2017.

\bibitem{weng2021neural}
O.~Weng, ``Neural network quantization for efficient inference: A survey,''
  \emph{arXiv preprint arXiv:2112.06126}, 2021.

\bibitem{bai2021towards}
H.~Bai, L.~Hou, L.~Shang, X.~Jiang, I.~King, and M.~R. Lyu, ``Towards efficient
  post-training quantization of pre-trained language models,'' \emph{arXiv
  preprint arXiv:2109.15082}, 2021.

\bibitem{alvarez2016efficient}
R.~Alvarez, R.~Prabhavalkar, and A.~Bakhtin, ``On the efficient representation
  and execution of deep acoustic models,'' \emph{arXiv preprint
  arXiv:1607.04683}, 2016.

\bibitem{duarte2018fast}
J.~Duarte, S.~Han, P.~Harris, S.~Jindariani, E.~Kreinar, B.~Kreis, J.~Ngadiuba,
  M.~Pierini, R.~Rivera, N.~Tran \emph{et~al.}, ``Fast inference of deep neural
  networks in fpgas for particle physics,'' \emph{Journal of Instrumentation},
  vol.~13, no.~07, p. P07027, 2018.

\bibitem{coelho2021automatic}
C.~N. Coelho, A.~Kuusela, S.~Li, H.~Zhuang, J.~Ngadiuba, T.~K. Aarrestad,
  V.~Loncar, M.~Pierini, A.~A. Pol, and S.~Summers, ``Automatic heterogeneous
  quantization of deep neural networks for low-latency inference on the edge
  for particle detectors,'' \emph{Nature Machine Intelligence}, pp. 1--12,
  2021.

\bibitem{goyal2021fixed}
R.~Goyal, J.~Vanschoren, V.~Van~Acht, and S.~Nijssen, ``Fixed-point
  quantization of convolutional neural networks for quantized inference on
  embedded platforms,'' \emph{arXiv preprint arXiv:2102.02147}, 2021.

\bibitem{zhou2017incremental}
A.~Zhou, A.~Yao, Y.~Guo, L.~Xu, and Y.~Chen, ``Incremental network
  quantization: Towards lossless cnns with low-precision weights,'' \emph{arXiv
  preprint arXiv:1702.03044}, 2017.

\bibitem{li2019additive}
Y.~Li, X.~Dong, and W.~Wang, ``Additive powers-of-two quantization: An
  efficient non-uniform discretization for neural networks,'' \emph{arXiv
  preprint arXiv:1909.13144}, 2019.

\bibitem{prasanna2019deep}
S.~Prasanna, ``Deep learning deployment with nvidia tensorrt,'' \emph{NVIDIA
  Deep Learning Institute, New York}, 2019.

\bibitem{DNNDK}
Xilinx. dnndk user guide.
  \url{:https://www.xilinx.com/support/documentation/sw_manuals/ai_inference/v1_6/ug1327-dnndk-user-guide.pdf.47}.
  Accessed: 2022-05-30.

\bibitem{miyashita2016convolutional}
D.~Miyashita, E.~H. Lee, and B.~Murmann, ``Convolutional neural networks using
  logarithmic data representation,'' \emph{arXiv preprint arXiv:1603.01025},
  2016.

\bibitem{przewlocka2022power}
D.~Przewlocka-Rus, S.~S. Sarwar, H.~E. Sumbul, Y.~Li, and B.~De~Salvo,
  ``Power-of-two quantization for low bitwidth and hardware compliant neural
  networks,'' \emph{arXiv preprint arXiv:2203.05025}, 2022.

\bibitem{nayak2019bit}
P.~Nayak, D.~Zhang, and S.~Chai, ``Bit efficient quantization for deep neural
  networks,'' in \emph{2019 Fifth Workshop on Energy Efficient Machine Learning
  and Cognitive Computing-NeurIPS Edition (EMC2-NIPS)}.\hskip 1em plus 0.5em
  minus 0.4em\relax IEEE, 2019, pp. 52--56.

\bibitem{habi2021hptq}
H.~V. Habi, R.~Peretz, E.~Cohen, L.~Dikstein, O.~Dror, I.~Diamant, R.~H.
  Jennings, and A.~Netzer, ``Hptq: Hardware-friendly post training
  quantization,'' \emph{arXiv preprint arXiv:2109.09113}, 2021.

\bibitem{zhang2021training}
X.~Zhang, I.~Colbert, K.~Kreutz-Delgado, and S.~Das, ``Training deep neural
  networks with joint quantization and pruning of weights and activations,''
  \emph{arXiv preprint arXiv:2110.08271}, 2021.

\bibitem{hawks2021ps}
B.~Hawks, J.~Duarte, N.~J. Fraser, A.~Pappalardo, N.~Tran, and Y.~Umuroglu,
  ``Ps and qs: Quantization-aware pruning for efficient low latency neural
  network inference,'' \emph{arXiv preprint arXiv:2102.11289}, 2021.

\bibitem{jacob2018quantization}
B.~Jacob, S.~Kligys, B.~Chen, M.~Zhu, M.~Tang, A.~Howard, H.~Adam, and
  D.~Kalenichenko, ``Quantization and training of neural networks for efficient
  integer-arithmetic-only inference,'' in \emph{Proceedings of the IEEE
  conference on computer vision and pattern recognition}, 2018, pp. 2704--2713.

\bibitem{kaneda2020fpga}
N.~Kaneda, Z.~Zhu, C.-Y. Chuang, A.~Mahadevan, B.~Farah, K.~Bergman,
  D.~Van~Veen, and V.~Houtsma, ``Fpga implementation of deep neural network
  based equalizers for high-speed pon,'' in \emph{Optical Fiber Communication
  Conference}.\hskip 1em plus 0.5em minus 0.4em\relax Optical Society of
  America, 2020, pp. T4D--2.

\bibitem{huang2022low}
X.~Huang, D.~Zhang, X.~Hu, C.~Ye, and K.~Zhang, ``Low-complexity recurrent
  neural network based equalizer with embedded parallelization for
  100-gbit/s/$\lambda$ pon,'' \emph{Journal of Lightwave Technology}, vol.~40,
  no.~5, pp. 1353--1359, 2022.

\bibitem{he2021fiber}
P.~He, F.~Wu, M.~Yang, A.~Yang, P.~Guo, Y.~Qiao, and X.~Xin, ``A fiber
  nonlinearity compensation scheme with complex-valued dimension-reduced neural
  network,'' \emph{IEEE Photonics Journal}, vol.~13, no.~6, pp. 1--7, 2021.

\bibitem{aoudia2019towards}
F.~A. Aoudia and J.~Hoydis, ``Towards hardware implementation of neural
  network-based communication algorithms,'' in \emph{2019 IEEE 20th
  International Workshop on Signal Processing Advances in Wireless
  Communications (SPAWC)}.\hskip 1em plus 0.5em minus 0.4em\relax IEEE, 2019,
  pp. 1--5.

\bibitem{xu2019efficient}
W.~Xu, X.~Tan, Y.~Lin, X.~You, C.~Zhang, and Y.~Be’ery, ``On the efficient
  design of neural networks in communication systems,'' in \emph{2019 53rd
  Asilomar Conference on Signals, Systems, and Computers}.\hskip 1em plus 0.5em
  minus 0.4em\relax IEEE, 2019, pp. 522--526.

\bibitem{matsumoto1998mersenne}
M.~Matsumoto and T.~Nishimura, ``Mersenne twister: a 623-dimensionally
  equidistributed uniform pseudo-random number generator,'' \emph{ACM
  Transactions on Modeling and Computer Simulation (TOMACS)}, vol.~8, no.~1,
  pp. 3--30, 1998.

\bibitem{kuschnerov2010data}
M.~Kuschnerov, M.~Chouayakh, K.~Piyawanno, B.~Spinnler, E.~De~Man,
  P.~Kainzmaier, M.~S. Alfiad, A.~Napoli, and B.~Lankl, ``Data-aided versus
  blind single-carrier coherent receivers,'' \emph{IEEE Photonics Journal},
  vol.~2, no.~3, pp. 387--403, 2010.

\bibitem{gulli2017deep}
A.~Gulli and S.~Pal, \emph{Deep learning with Keras}.\hskip 1em plus 0.5em
  minus 0.4em\relax Packt Publishing Ltd, 2017.

\bibitem{sidelnikov2021advanced}
O.~Sidelnikov, A.~Redyuk, S.~Sygletos, M.~Fedoruk, and S.~Turitsyn, ``Advanced
  convolutional neural networks for nonlinearity mitigation in long-haul wdm
  transmission systems,'' \emph{Journal of Lightwave Technology}, vol.~39,
  no.~8, pp. 2397--2406, 2021.

\bibitem{lin2014adaptive}
C.-Y. Lin, A.~Napoli, B.~Spinnler, V.~Sleiffer, D.~Rafique, M.~Kuschnerov,
  M.~Bohn, and B.~Schmauss, ``Adaptive digital back-propagation for optical
  communication systems,'' in \emph{Optical Fiber Communication
  Conference}.\hskip 1em plus 0.5em minus 0.4em\relax Optical Society of
  America, 2014, pp. M3C--4.

\bibitem{napoli2014performance}
A.~Napoli, D.~Rafique, B.~Spinnler, M.~Kuschnerov, M.~Noelle, and M.~Bohn,
  ``Performance dependence of single-carrier digital back-propagation on fiber
  types and data rates,'' in \emph{Optical Fiber Communication
  Conference}.\hskip 1em plus 0.5em minus 0.4em\relax Optica Publishing Group,
  2014, pp. W2A--49.

\bibitem{spinnler2010equalizer}
B.~Spinnler, ``Equalizer design and complexity for digital coherent
  receivers,'' \emph{IEEE Journal of Selected Topics in Quantum Electronics},
  vol.~16, no.~5, pp. 1180--1192, 2010.

\bibitem{freire2022computational}
P.~J. Freire, S.~Srivallapanondh, A.~Napoli, J.~E. Prilepsky, and S.~K.
  Turitsyn, ``Computational complexity evaluation of neural network
  applications in signal processing,'' 2022.

\bibitem{jacobsen2007fast}
E.~Jacobsen and P.~Kootsookos, ``Fast, accurate frequency estimators [dsp tips
  \& tricks],'' \emph{IEEE Signal Processing Magazine}, vol.~24, no.~3, pp.
  123--125, 2007.

\bibitem{baskin2021uniq}
C.~Baskin, et~al., ``Uniq: Uniform noise injection for non-uniform quantization
  of neural networks,'' \emph{ACM Transactions on Computer Systems (TOCS)},
  vol.~37, no. 1--4, pp. 1--15, 2021.

\bibitem{sahin2006neural}
S.~Sahin, Y.~Becerikli, and S.~Yazici, ``Neural network implementation in
  hardware using fpgas,'' in \emph{International conference on neural
  information processing}.\hskip 1em plus 0.5em minus 0.4em\relax Springer,
  2006, pp. 1105--1112.

\bibitem{5280233}
A.~Dinu, M.~N. Cirstea, and S.~E. Cirstea, ``Direct neural-network
  hardware-implementation algorithm,'' \emph{IEEE Transactions on Industrial
  Electronics}, vol.~57, no.~5, pp. 1845--1848, 2010.

\bibitem{sun2007fpga}
W.~Sun, M.~J. Wirthlin, and S.~Neuendorffer, ``Fpga pipeline synthesis design
  exploration using module selection and resource sharing,'' \emph{IEEE
  Transactions on Computer-Aided Design of Integrated Circuits and Systems},
  vol.~26, no.~2, pp. 254--265, 2007.

\bibitem{elhoushi2021deepshift}
M.~Elhoushi, Z.~Chen, F.~Shafiq, Y.~H. Tian, and J.~Y. Li, ``Deepshift: Towards
  multiplication-less neural networks,'' in \emph{Proceedings of the IEEE/CVF
  Conference on Computer Vision and Pattern Recognition}, 2021, pp. 2359--2368.

\bibitem{you2020shiftaddnet}
H.~You, X.~Chen, Y.~Zhang, C.~Li, S.~Li, Z.~Liu, Z.~Wang, and Y.~Lin,
  ``Shiftaddnet: A hardware-inspired deep network,'' \emph{arXiv preprint
  arXiv:2010.12785}, 2020.

\bibitem{gentili1995efficient}
P.~Gentili, F.~Piazza, and A.~Uncini, ``Efficient genetic algorithm design for
  power-of-two fir filters,'' in \emph{1995 International conference on
  acoustics, speech, and signal processing}, vol.~2.\hskip 1em plus 0.5em minus
  0.4em\relax IEEE, 1995, pp. 1268--1271.

\bibitem{evans1994efficient}
J.~B. Evans, ``Efficient fir filter architectures suitable for fpga
  implementation,'' \emph{IEEE Transactions on Circuits and Systems II: Analog
  and Digital Signal Processing}, vol.~41, no.~7, pp. 490--493, 1994.

\bibitem{lee2003frequency}
W.~R. Lee, V.~Rehbock, K.~L. Teo, and L.~Caccetta, ``Frequency-response masking
  based fir filter design with power-of-two coefficients and suboptimum pwr,''
  \emph{Journal of Circuits, Systems, and Computers}, vol.~12, no.~05, pp.
  591--599, 2003.

\bibitem{Kahn:01}
E.~Ip and J.~M. Kahn, ``Compensation of dispersion and nonlinear impairments
  using digital backpropagation,'' \emph{J. Lightw. Technol.}, vol.~26, no.~20,
  p. 3416–3425, October 2008.

\bibitem{freire_transfer}
P.~J. Freire, D.~Abode, J.~E. Prilepsky, N.~Costa, B.~Spinnler, A.~Napoli, and
  S.~K. Turitsyn, ``Transfer learning for neural networks-based equalizers in
  coherent optical systems,'' \emph{Journal of Lightwave Technology}, vol.~39,
  no.~21, pp. 6733--6745, 2021.

\bibitem{cho2019probabilistic}
J.~Cho and P.~J. Winzer, ``Probabilistic constellation shaping for optical
  fiber communications,'' \emph{Journal of Lightwave Technology}, vol.~37,
  no.~6, pp. 1590--1607, 2019.

\bibitem{li2017training}
H.~Li, S.~De, Z.~Xu, C.~Studer, H.~Samet, and T.~Goldstein, ``Training
  quantized nets: A deeper understanding,'' \emph{Advances in Neural
  Information Processing Systems}, vol.~30, 2017.

\bibitem{misra2010artificial}
J.~Misra and I.~Saha, ``Artificial neural networks in hardware: A survey of two
  decades of progress,'' \emph{Neurocomputing}, vol.~74, no. 1-3, pp. 239--255,
  2010.

\bibitem{gou2021knowledge}
J.~Gou, B.~Yu, S.~J. Maybank, and D.~Tao, ``Knowledge distillation: A survey,''
  \emph{International Journal of Computer Vision}, vol. 129, no.~6, pp.
  1789--1819, 2021.

\bibitem{ye2021hybrid}
J.~Ye, S.~Zhang, and J.~Wang, ``Hybrid network compression via meta-learning,''
  in \emph{Proceedings of the 29th ACM International Conference on Multimedia},
  2021, pp. 1423--1431.

\bibitem{chen2019metaquant}
S.~Chen, W.~Wang, and S.~J. Pan, ``Metaquant: Learning to quantize by learning
  to penetrate non-differentiable quantization,'' \emph{Advances in Neural
  Information Processing Systems}, vol.~32, 2019.

\bibitem{liu2019metapruning}
Z.~Liu, H.~Mu, X.~Zhang, Z.~Guo, X.~Yang, K.-T. Cheng, and J.~Sun,
  ``Metapruning: Meta learning for automatic neural network channel pruning,''
  in \emph{Proceedings of the IEEE/CVF international conference on computer
  vision}, 2019, pp. 3296--3305.

\bibitem{bartan2021training}
B.~Bartan and M.~Pilanci, ``Training quantized neural networks to global
  optimality via semidefinite programming,'' in \emph{International Conference
  on Machine Learning}.\hskip 1em plus 0.5em minus 0.4em\relax PMLR, 2021, pp.
  694--704.

\bibitem{zhuang2020training}
B.~Zhuang, L.~Liu, M.~Tan, C.~Shen, and I.~Reid, ``Training quantized neural
  networks with a full-precision auxiliary module,'' in \emph{Proceedings of
  the IEEE/CVF Conference on Computer Vision and Pattern Recognition}, 2020,
  pp. 1488--1497.

\bibitem{9523752}
P.~J. Freire, D.~Abode, J.~E. Prilepsky, N.~Costa, B.~Spinnler, A.~Napoli, and
  S.~K. Turitsyn, ``Transfer learning for neural networks-based equalizers in
  coherent optical systems,'' \emph{Journal of Lightwave Technology}, vol.~39,
  no.~21, pp. 6733--6745, 2021.

\bibitem{9748452}
P.~J. Freire, B.~Spinnler, D.~Abode, J.~E. Prilepsky, A.~Ali, N.~Costa,
  W.~Schairer, A.~Napoli, A.~D. Ellis, and S.~K. Turitsyn, ``Domain adaptation:
  the key enabler of neural network equalizers in coherent optical systems,''
  in \emph{2022 Optical Fiber Communications Conference and Exhibition (OFC)},
  2022, pp. 1--3.

\bibitem{9226070}
Z.~Xu, C.~Sun, T.~Ji, J.~H. Manton, and W.~Shieh, ``Feedforward and recurrent
  neural network-based transfer learning for nonlinear equalization in
  short-reach optical links,'' \emph{Journal of Lightwave Technology}, vol.~39,
  no.~2, pp. 475--480, 2021.

\bibitem{xiao2022optimizations}
J.~Xiao, L.~Sun, C.~Liu, and G.~N. Liu, ``Optimizations and investigations for
  transfer learning of iteratively pruned neural network equalizers for data
  center networking,'' \emph{Optics Express}, vol.~30, no.~20, pp.
  36\,358--36\,367, 2022.

\bibitem{myung2022pac}
S.~Myung, I.~Huh, W.~Jang, J.~M. Choe, J.~Ryu, D.~Kim, K.-E. Kim, and C.~Jeong,
  ``Pac-net: A model pruning approach to inductive transfer learning,'' in
  \emph{International Conference on Machine Learning}.\hskip 1em plus 0.5em
  minus 0.4em\relax PMLR, 2022, pp. 16\,240--16\,252.

\bibitem{liu2021transtailor}
B.~Liu, Y.~Cai, Y.~Guo, and X.~Chen, ``Transtailor: Pruning the pre-trained
  model for improved transfer learning,'' in \emph{Proceedings of the AAAI
  Conference on Artificial Intelligence}, vol.~35, no.~10, 2021, pp.
  8627--8634.

\bibitem{gordon2020compressing}
M.~A. Gordon, K.~Duh, and N.~Andrews, ``Compressing bert: Studying the effects
  of weight pruning on transfer learning,'' \emph{arXiv preprint
  arXiv:2002.08307}, 2020.

\end{thebibliography}

\end{document}